\documentclass[fleqn,usenatbib]{mnras}

\usepackage[T1]{fontenc}

\usepackage{graphicx}	
\usepackage{amsmath}	
\usepackage{amssymb}	
\usepackage{rotating}

\usepackage{pdflscape}
\usepackage{soul}

\title[Discovery and properties of open clusters]{
Discovery and astrophysical properties of Galactic open clusters  
in dense stellar fields using {\textit{Gaia}} DR2}

\author[F. A. Ferreira et al.]{
F. A. Ferreira$^{1}$\thanks{E-mail: filipe1906@ufmg.br},
W. J. B. Corradi$^{1,5}$,
F. F. S. Maia$^{2}$,
M. S. Angelo$^{3}$  \newauthor
\ and J. F. C. Santos Jr.$^{1,4}$
\\
$^{1}$Universidade Federal de Minas Gerais, Departamento de F\'isica, Av. Ant\^onio Carlos 6627, 31270-901, Brazil \\
$^{2}$Universidade Federal do Rio de Janeiro, Instituto de F\'isica, 21941-972, Brazil 
\\
$^{3}$Centro Federal de Educa\c c\~ao Tecnol\'ogica de Minas Gerais, Av. Monsenhor Luiz de Gonzaga, 103, 37250-000, Brazil\\
$^{4}$Departamento de Astronom\'ia, Universidad de La Serena, Av. Juan Cisternas 1200, La Serena, Chile \\
$^{5}$Laborat\'orio Nacional de Astrof{\'{\i}}sica, R. Estados Unidos, 154, 37504-364, Itajub\'a, MG, Brazil
}

\date{Accepted XXX. Received YYY; in original form ZZZ}

\pubyear{2020}

\begin{document}
\label{firstpage}
\pagerange{\pageref{firstpage}--\pageref{lastpage}}
\maketitle

\begin{abstract}

We report the discovery of 25 new open clusters resulting from a search in dense low galactic latitude fields. We also provide, for the first time, structural and astrophysical parameters for the new findings and 34 other recently discovered open clusters using \textit{Gaia} DR2 data. The candidates were confirmed by jointly inspecting the vector point diagrams and spatial distribution.
The discoveries were validated by matching near known objects and comparing their mean astrometric parameters with the available literature.
A decontamination algorithm was applied to the three-dimensional astrometric space to derive membership likelihoods for clusters stars. By rejecting stars with low membership likelihoods, we built decontaminated colour-magnitude diagrams and derived the clusters astrophysical parameters by isochrone fitting. The structural parameters were also derived by King-profile fittings over the stellar distributions. The investigated clusters are mainly located within 3\,kpc from the Sun, with ages ranging from 30\,Myr to 3.2\,Gyr and reddening limited to $E(B-V)=2.5$. On average, our cluster sample presents less concentrated structures than \textit{Gaia} DR2 confirmed clusters, since the derived core radii are larger while the tidal radii are not significantly different. Most of them are located in the IV quadrant of the Galactic disc at low latitudes, therefore they are immersed in dense fields characteristic of the inner Milky Way.

\end{abstract}

\begin{keywords}
Galaxy: stellar content -- open clusters and associations: general -- surveys: \textit{Gaia}
\end{keywords}



\section{Introduction}

Open star clusters are important objects to study and probe the history and structure of the Galactic disc. Young open clusters show us how stars form from embedded systems in molecular clouds and the recent disc history \citep{Lada:2003}.
On the other hand, the old open clusters help us to understand the chemical and dynamical evolution of the Galactic disc and to establish an inferior age limit to this structure \citep[e.g.][]{vm80,friel1995,2007A&A...476..217C,2016A&A...585A.150N}.

\indent Efforts have been made to detect new objects from their spatial stellar overdensities with respect to the background \cite[e.g.][]{Froebrich:2007,Kronberger:2006}. On the other hand, the \textit{Gaia} mission \citep{Brown:2018} has acquired high precision proper motions and parallaxes which allows a search for new objects in another parameters space. For example, now the distinction between star cluster and field stars become possible
by analyzing the confined loci of cluster stars in the astrometric space. It is evident an increase on the number of open clusters discovered recently, specially those projected against crowded fields (\citealp{Ryu:2018,cjv18,cjl18,Torrealba:2018}; \citet[][hereafter Sim2019]{sla19}; \citealt{2019MNRAS.483.5508F}).  Before the \textit{Gaia} era, at least 3000 known open clusters have measured parameters in the literature \citep{Dias:2002,kps13}. However, more recently \cite{2019AJ....157...12B} have shown that the actual census of Galactic star clusters, associations and
candidates reaches a number greater than 10000 objects (although many of them are  asterisms, embedded or globular clusters),
showing that there are much more unstudied objects than
the studied ones. In a statistical sense, the more clusters are discovered and properly characterized, the more we are able to understand the Galactic structure and its chemical and dynamical evolution.

The separation between fiducial members of open clusters from field stars in a high density region of the Galactic disc is a challenging task. Specialized automated searches with different algorithms using \textit{Gaia} DR2 
are useful and necessary tools to detect such objects.

 After the \textit{Gaia} second data release, \citet{Cantat:2018} used an unsupervised membership assignment code, UPMASK, to search for open clusters in astrometric space. They obtained cluster membership, mean astrometric parameters and distances for more than 1200 open clusters while also reporting 60 newly discovered open clusters. 
\citet{cjl19} found 53 new open cluster using an unsupervised, density-based, clustering algorithm, DBSCAN. The authors reported the new objects as well as determined their mean astrometric parameters and most probable members. \citet{cks19} found 41 new open clusters located in Perseus arm by using Gaussian mixture models.
The authors applied an unsupervised membership assignment method, UPMASK, to characterize the candidates and then visually inspected colour-magnitude diagrams to validate the detected objects as new open cluster. The mean astrometric parameters and most probable members of these objects were also reported. Sim2019 performed visual inspection of the \textit{Gaia} DR2 astrometric and photometric data, resulting in the discovery of 207 new OCs within 1\,kpc from the sun that has been missed by previous works. The authors also characterized  the newly discovered objects by finding their ages, distances, sizes and mean astrometric parameters. \citet[][hereafter LP2019]{lp19} detected 2443 star clusters using a clustering algorithm in the 5-D astrometric space ($l$, $b$, $\varpi$, $\mu_{\alpha}^{*}$, $\mu_{\delta}$ ). The authors characterized all the found objects by an automated isochrone fitting scheme determining distances, colour excesses, metalicities and also the mean astrometric parameters. The authors associate a classification scheme to the discovered objects that takes account the isochrone fitting quality, assigning 76 of the found objects as high confidence, new open clusters candidates. More recently, \citet[][hereafter CG2020]{cjl20} have discovered 582 new open clusters by applying an unsupervised clustering algorithm to find overdensities in a five-dimensional parameter space ($l$, $b$, $\varpi$, $\mu_{\alpha}^{*}$, $\mu_{\delta}$). 

In a previous work \citep{2019MNRAS.483.5508F}, we have found 3 new open clusters by visually inspecting proper motion diagrams by applying colour and magnitude filters in colour-magnitude diagrams. We characterized the findings by providing astrometric and astrophysical parameters derived from isochrone fitting methods. We follow on applying this methodology to high density regions of the Galactic disc. 
We surveyed 96 Galactic fields finding 59 new open clusters candidates. However, during the analysis, part of this sample was reported by Sim2019, LP2019 and CG2020. Since these authors did not performed a complete characterization of the objects, we determined the astrophysical properties of the sample of 59 open clusters and report the discovery of 25 new clusters among them. We also investigate their nature and how these new objects are disposed in the Galaxy.


This paper is structured as follows: in Sect. 2 the method used to detect the new open clusters is described. In Sect. 3 we discuss the database used to confirm and validate discoveries. 
In Sect. 4 the analysis procedure is presented, including membership assessment, determination of astrophysical
and structural parameters. In Sect. 5, we compare our sample of
open clusters with known open clusters and discuss our main results. The concluding remarks are given in Sect. 6.

\section{Searching for new cluster candidates}
\label{sect:search}

In this work, we used astrometric and photometric  data from the \textit{Gaia} DR2 catalogue \citep{Lindegren:2018,Evans:2018}, which provides positions, proper motions, parallaxes and magnitudes in the $G$, $G_{BP}$ and $G_{RP}$ bands for more than 1.3 billion sources \citep{Brown:2018}.
We applied quality filters in our samples by using the equations (1), (2) and (3) from \cite{Arenou:2018}. This is a recommended basic filter to clean the  data samples from contamination due to double stars, astrometric effects from binary stars and calibration problems. In other words, it assures the best quality of the data for analysis.

\cite{2019MNRAS.483.5508F} discovered three new open clusters when analyzing the region adjacent to NGC\,5999 in  the  nearby Sagittarius arm by applying colour-magnitude filters. 
The present work extends the search for new open clusters to low latitudes {\bf ($-10^{\circ}$ $\lesssim$ b $\lesssim$ $10^{\circ}$)} and to a longitude range similar to that of the VVV survey \citep{Minniti:2010} ($290^{\circ}$ $\lesssim$ l $\lesssim$10$^{\circ}$). Additionally, we have also analysed six fields presented in \cite{2016A&A...585A.150N}, adding newly found clusters to our sample. 
In total, we have surveyed a projected area of $\sim$650 square degrees, divided into 96 fields, as shown in Fig.~\ref{fig:survey_fields}. The full list of fields can be found in Table \ref{tab:survey_fields}, in the appendix.

\begin{figure}
\centering
\includegraphics[width=0.95\linewidth]{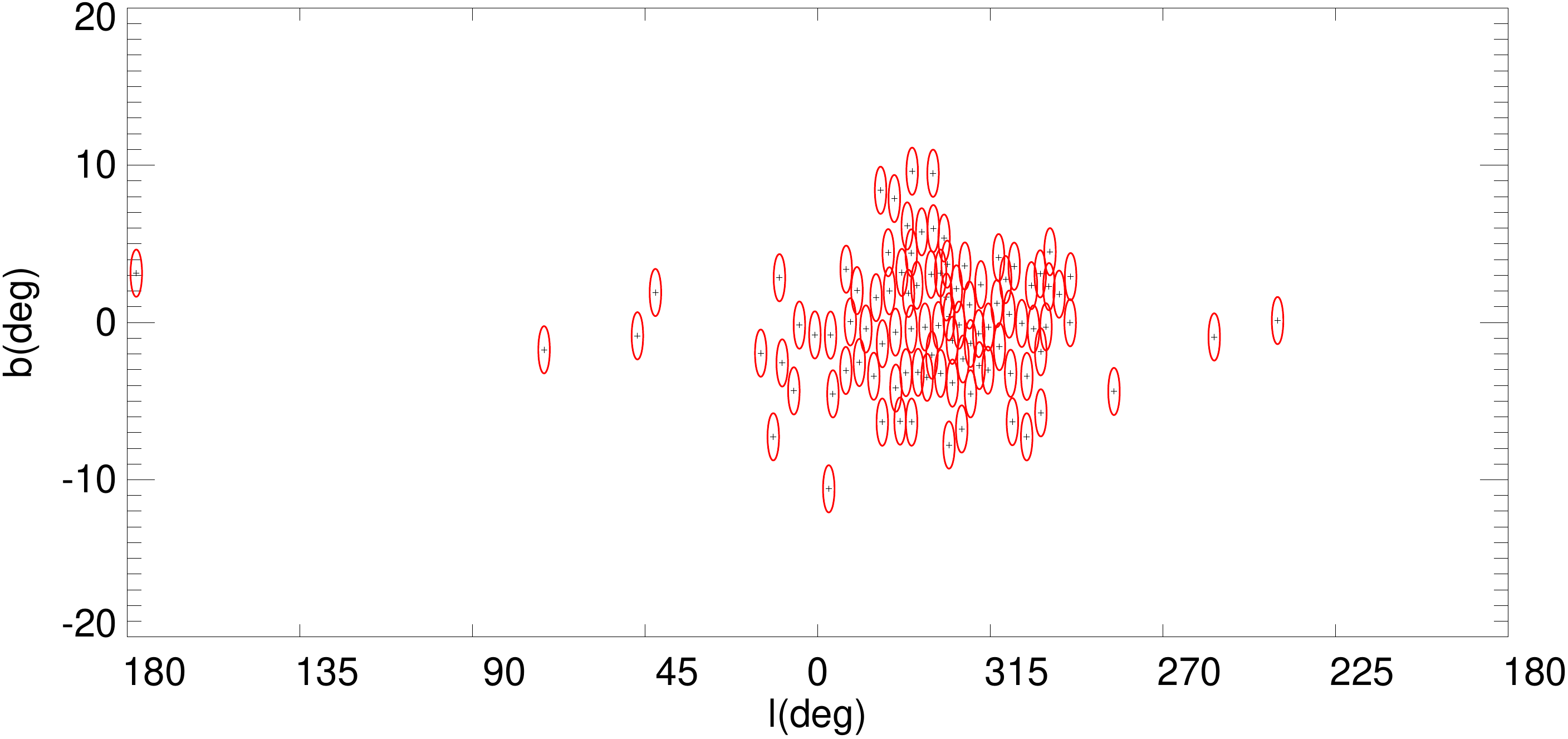} 
\caption{Spatial coverage of the Galactic fields surveyed in this work. The size of the red circles indicates the surveyed regions.}
\label{fig:survey_fields}
\end{figure}

To search for new clusters, we have used the Vizier\footnote[1]{http://vizier.u-strasbg.fr/viz-bin/VizieR} online service to extract \textit{Gaia} DR2 data in regions of 1.5 degrees radius around each analyzed field. Given a tidal radius of $\sim 10 pc$ the cluster would extrapolate the adopted region size, when located closer than $\sim 380 pc$ from the Sun. Thus, our method would not detect such objects. Most of our targets are farther than this, at a distance range of $\sim 1-3 kpc$ (see Sect. ~\ref{sect:sec6}).

Given the data inside the extraction area, we adopted the following steps:
\begin{enumerate}
    \item divide the region in 4 sectors of same area (Sect.~\ref{sect:spatial_filter});
    \item apply colour and magnitude filters in each sector (Sect.~\ref{sect:colour_filter});
    \item search for overdensities in filtered VPD and sky chart of each sector (Sect.~\ref{sect:over_search});
    \item analyze the spatial bonds, proper motions and parallaxes of the stars in overdensities (Sect.~\ref{sect:analysis}).
\end{enumerate}

\subsection{The spatial filter}
\label{sect:spatial_filter}

\begin{figure}
\centering
\includegraphics[width=0.8\linewidth]{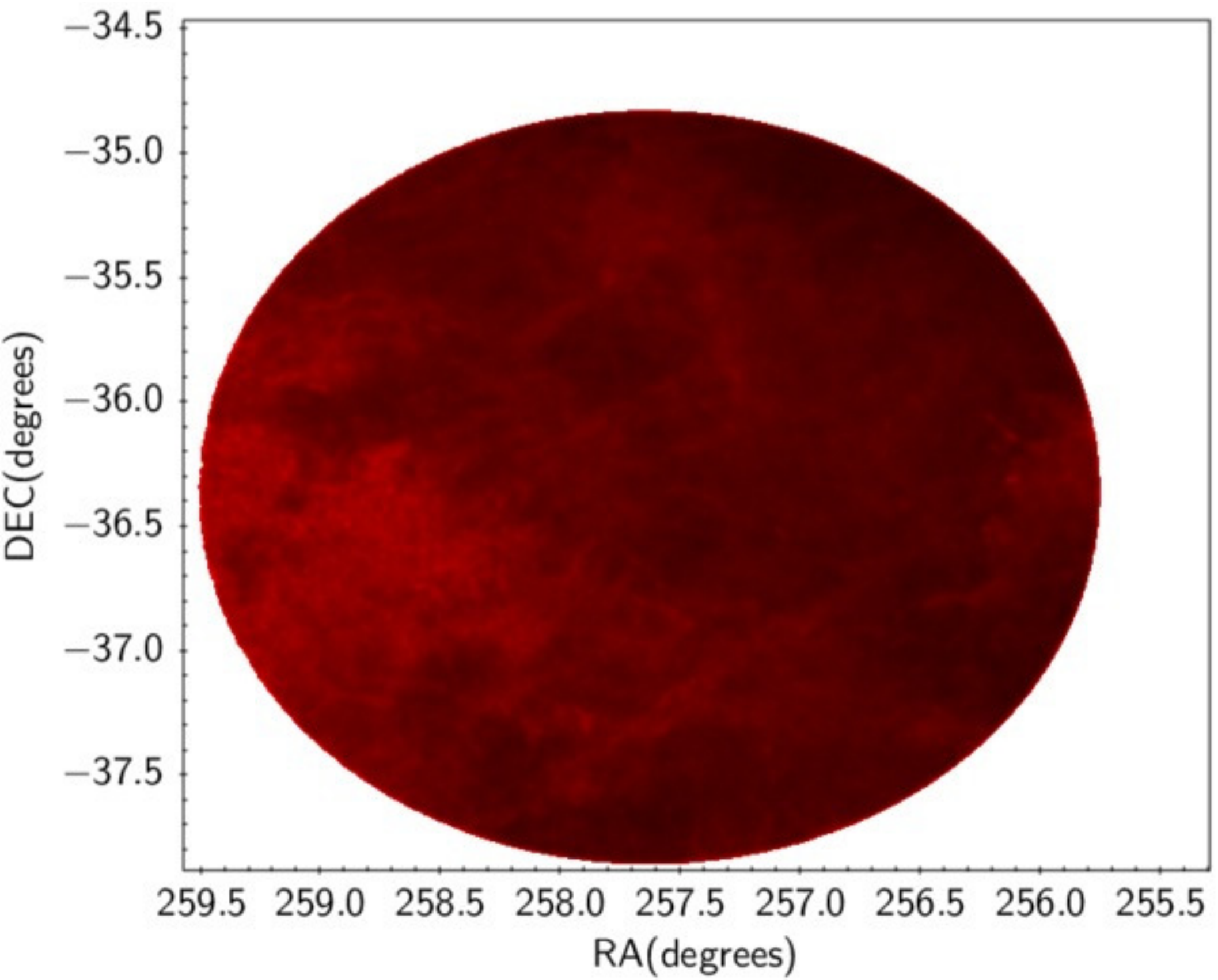} \\ \vspace{0.25cm}
\includegraphics[width=0.8\linewidth]{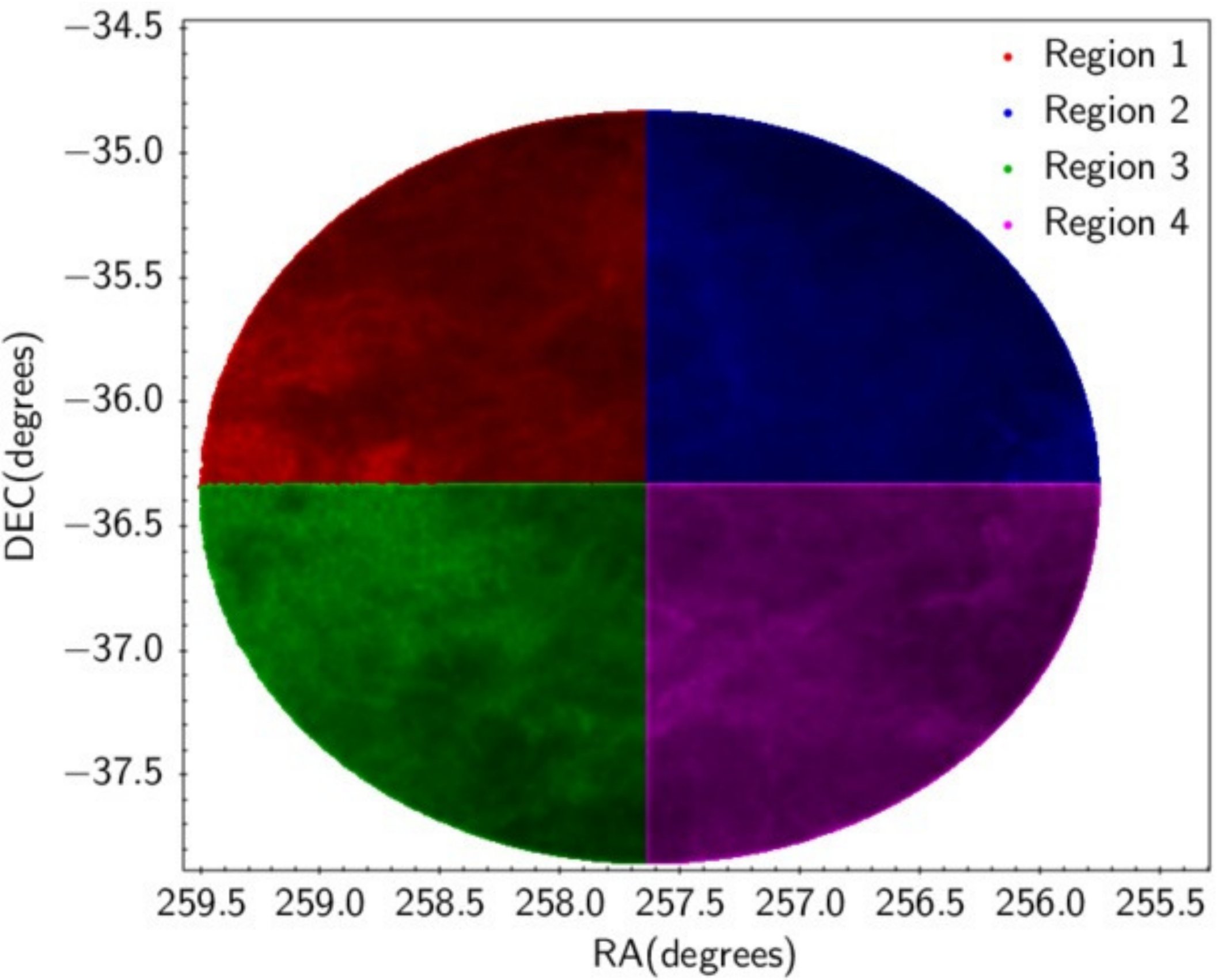}
\caption{Top: Star positions in a region of 1.5 degrees radius centred at coordinates of the field 90, RA=257.632$^{\circ}$ and DEC=-36.3409$^{\circ}$, the entire region comprises $\sim$2 million sources. Bottom: The same region, but divided by 4 sectors of same area.}
\label{fig:space_cut}
\end{figure}

Dividing the region in 4 sectors of same area diminishes the amount of stars in the vector point diagram (VPD) and colour-magnitude diagram (CMD), making it easier to find overdensities, as it increases the relative number of cluster stars over the field stars (i.e. the contrast) in both parameters spaces. Fig ~\ref{fig:space_cut} shows the spatial filter  applied to one of the analysed fields by dividing the full region in 4 smaller slices. The field is centred in the coordinates RA=257.632$^{\circ}$ and DEC=-36.3409$^{\circ}$ and contains $\sim$ 2 million sources. Each slice contains $\sim$4.0$\times 10^{5}$ sources (red), $\sim$7.4.$\times 10^{5}$ sources (blue), $\sim$2.7$\times 10^{5}$ sources (green) and $\sim$6.8$\times 10^{5}$ sources (purple).

\subsection{The colour and magnitude filters}
\label{sect:colour_filter}

For each sector, we build a $G$ versus $(G_{BP}-G_{RP})$ CMD and applied colour and magnitude filters. The first filter is a simple magnitude cut that keeps in our sample sources brighter than G = 18\,mag (similar to the filter 
used in \citeauthor{cjv18}\,\,2018a). This procedure produces smaller working subsamples and makes the following procedures computationally faster. 

\begin{figure}
\centering
\includegraphics[width=0.82\linewidth]{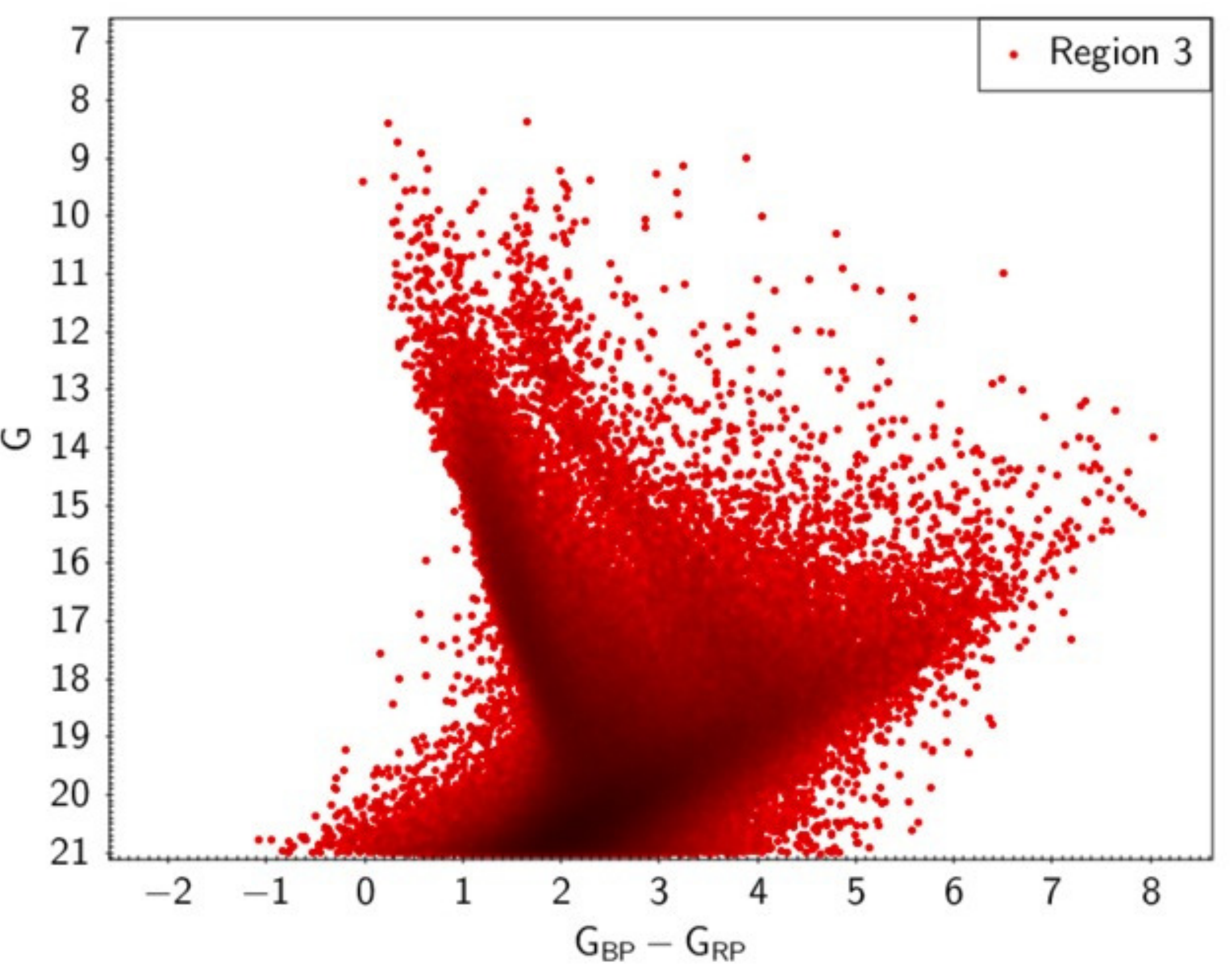} \\ \vspace{0.25cm}
\includegraphics[width=0.82\linewidth]{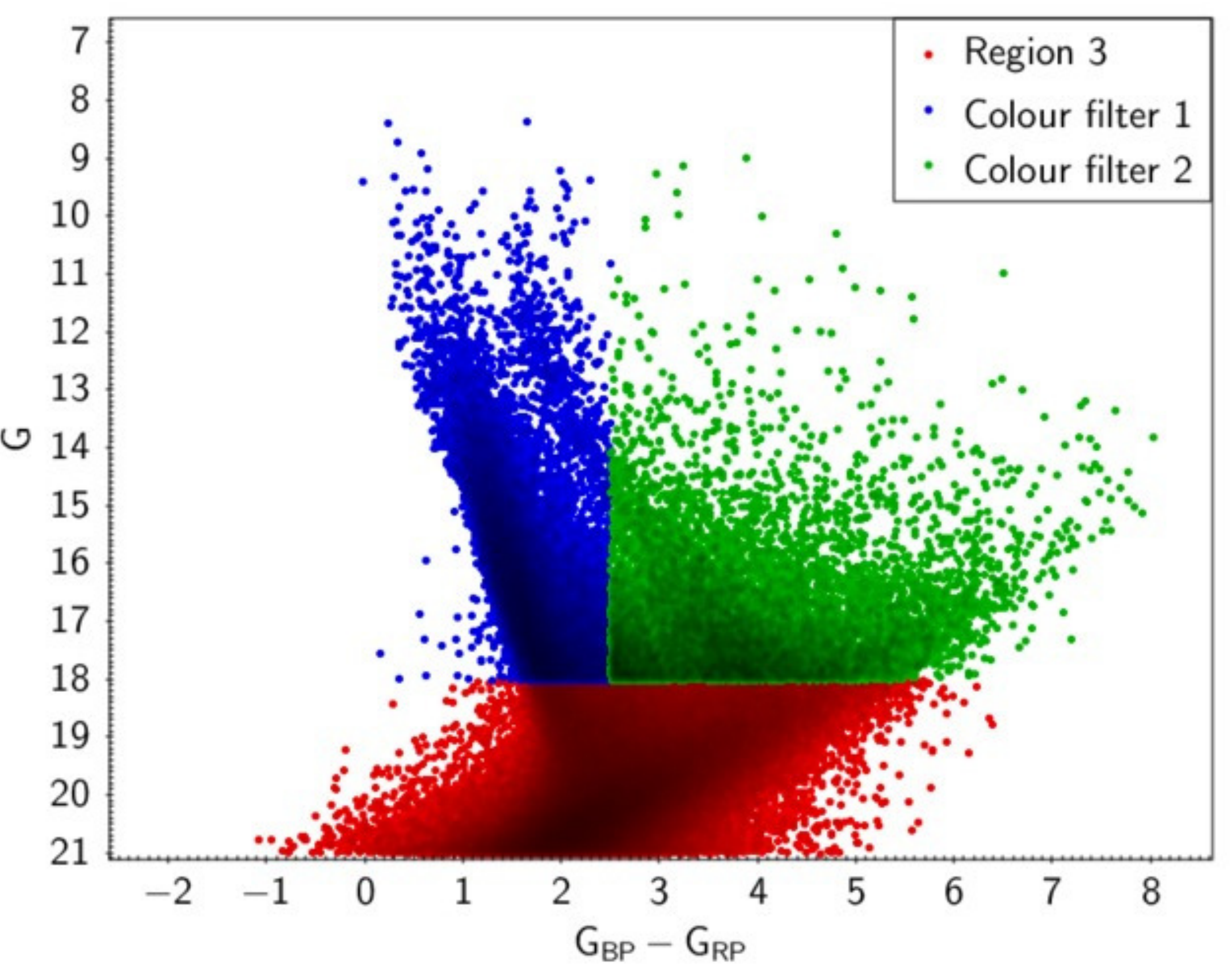}
\caption{Top: CMD of the sample highlighted and referred as Region 3 in the Fig ~\ref{fig:space_cut}. Bottom: CMD with the colour filters applied, where the first subsample covers $G_{BP}-G_{RP}<2.5$ and the second one covers $G_{BP}-G_{RP}>2.5.$}
\label{fig:CMD_cut}
\end{figure}

The second filter divides the remaining sample ($G<18\,$mag) in two sub-samples by a colour value. This colour value is a free parameter and will depend in the local reddening as we visually divide the denser portions of the CMD in half. Usually, the first sub-sample keeps stars with $G_{BP}-G_{RP}$ smaller than a limiting value between 1.5 and 2.5\,mag. The second sub-sample will contain stars with $G_{BP}-G_{RP}$ larger than this limit. This first filter is important to separate highly reddened stars, thus enhancing the cluster main sequence. A lower colour cut value is suitable to probe younger clusters with low reddening, while higher values are appropriate to probe older and/or more reddened ones.

Using region 3 of Fig \ref{fig:space_cut} as an example, the colour and magnitude filters reduced the number of sources from $\sim$2.7$\times 10^{5}$ to $\sim$2.3$\times 10^{4}$ in the bluer side of the cut, and to $\sim$1.1$\times 10^{4}$ in the red side of the cut (see Fig \ref{fig:CMD_cut}).

\subsection{The search for overdensities}
\label{sect:over_search}

\begin{figure*}
\includegraphics[width=0.42\linewidth]{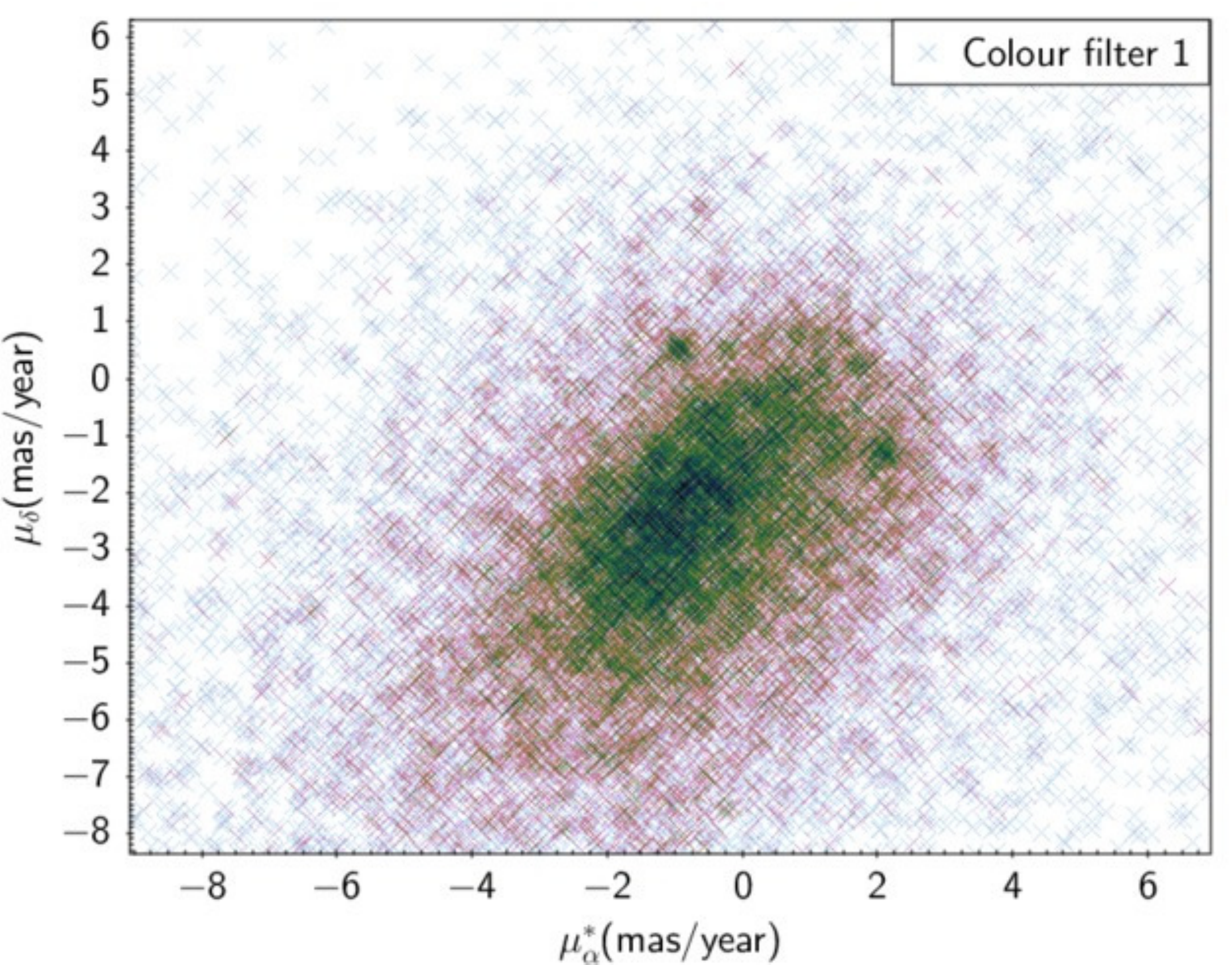} \hspace{1cm}
\includegraphics[width=0.42\linewidth]{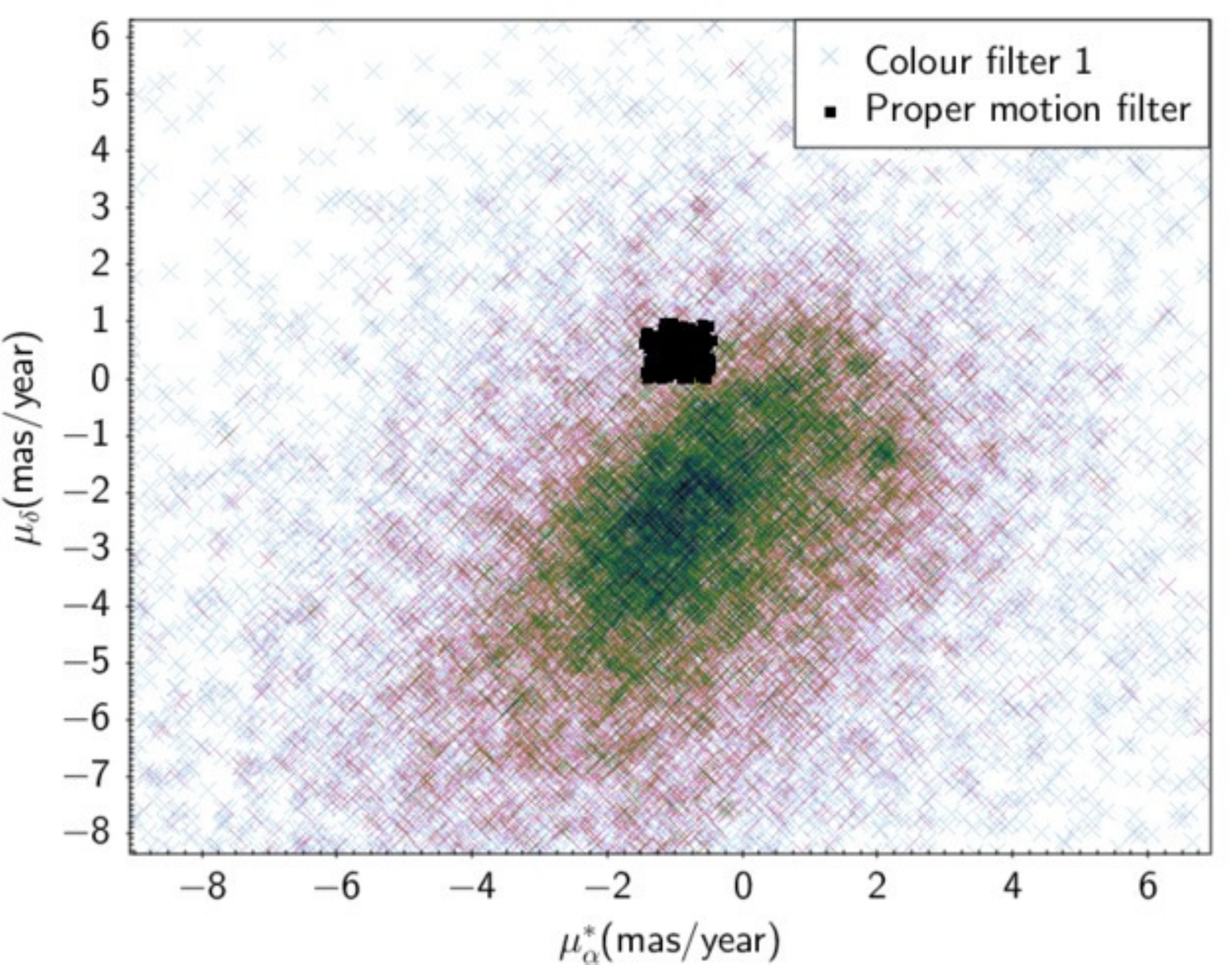}\\ \vspace{0.5cm}
\includegraphics[width=0.42\linewidth]{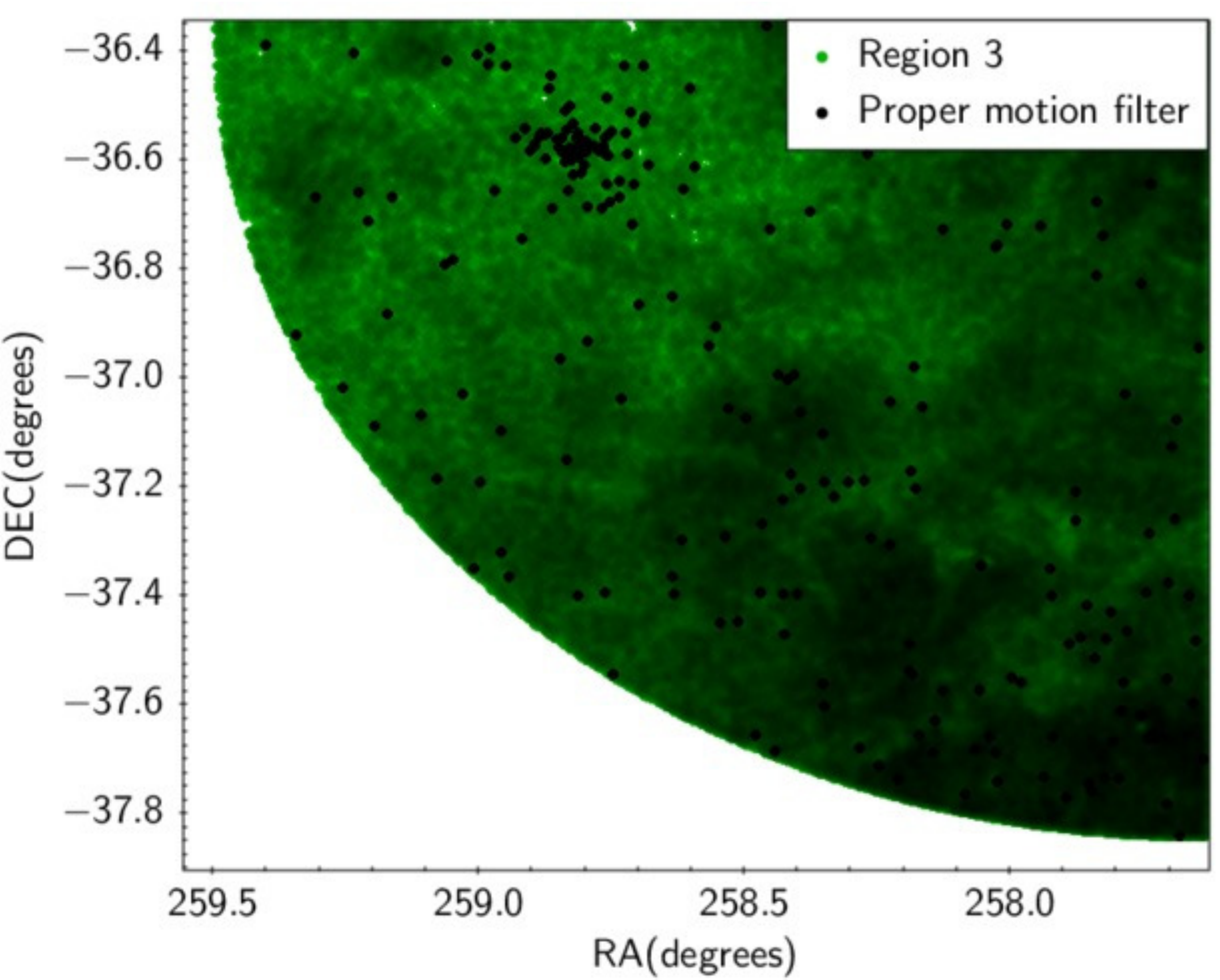} \hspace{1cm}
\includegraphics[width=0.42\linewidth]{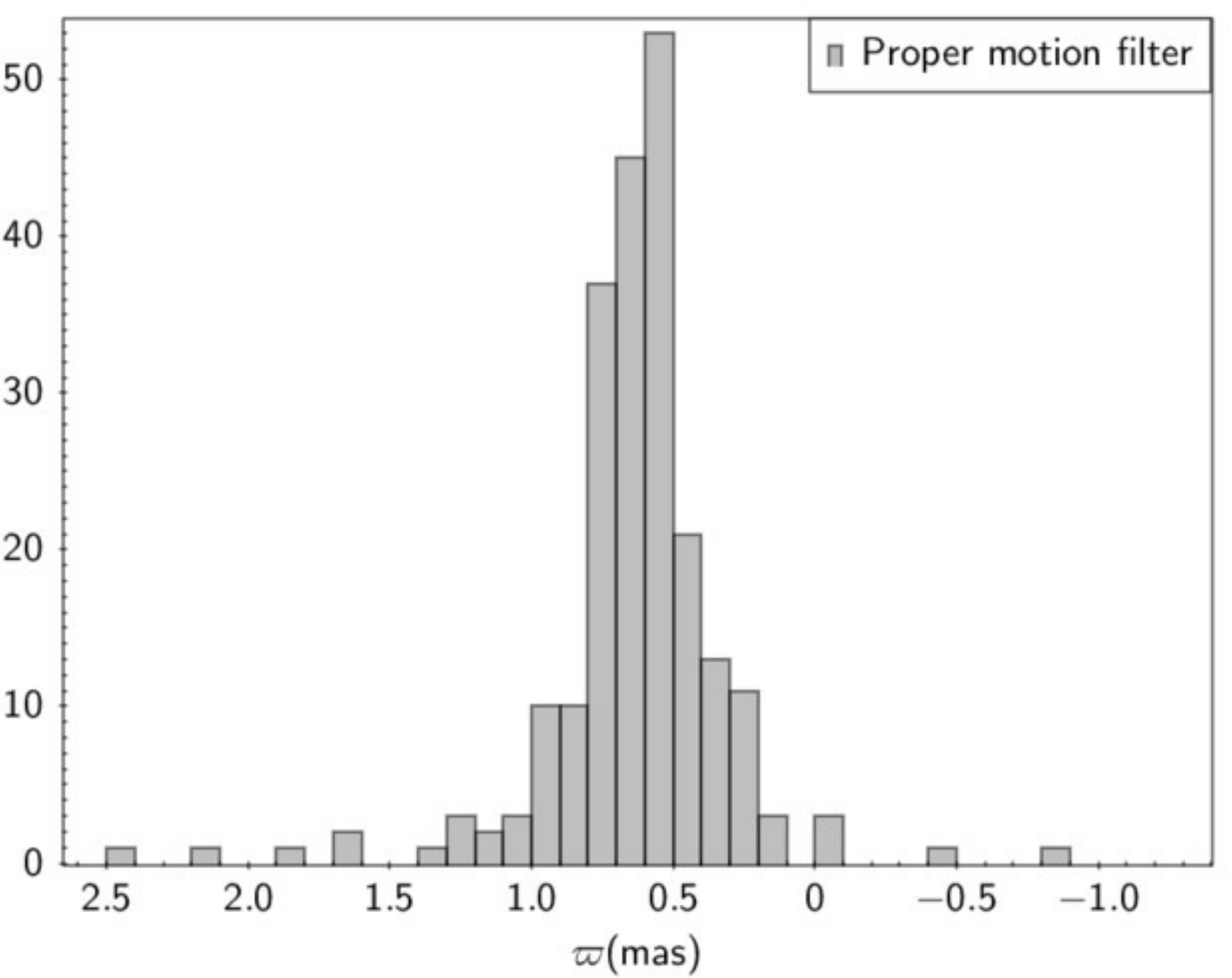}
\caption{Top left: VPD built from the Region 3 sample in the Fig~\ref{fig:CMD_cut} where an overdensity can be seen in the VPD. Top Right: A 1x1 mas\,yr$^{-1}$ extraction mask is applied over the visual centre of the detected overdensity. Bottom left: A Sky chart of all stars in the region (green) overploted by the subsample filtered by proper motion (black); we note the stars are spatially clustered. Bottom right: Parallax values of the subsample filtered by proper motion, showing that most of the stars also share a similar distance.}
\label{fig:pm_ra_dec}
\end{figure*}

With the subsamples settled, we analyze the stars distribution on sky charts and VPDs, visually searching for overdensities in both spaces. If any overdensity is detected in the VPD, we apply a box-shaped mask of size 1\,mas\,yr$^{-1}$ on it and analyze the spatial and parallax distribution of stars inside it. If this overdensity represents the typical proper motion of stars in a cluster, we expect a gaussian distribution of the parallax values and a spatial bound between the stars (usually with a dense core surrounded by a sparse halo), both with few outliers. This procedure is shown in Fig. ~\ref{fig:pm_ra_dec}. The VPD (top-left) contains $\sim$2.3 $\times 10^{4}$ sources and after applying the proper motion filter (top-right), this number is reduced to $\sim$ 200 stars, as showed by black dots (bottom-left) and histogram frequencies (bottom right). 

The majority of our new clusters were found by detecting a VPD overdensity, however there were some few cases where the colour cut revealed an overdensity more clearly in the sky chart, only becoming evident in the VPD after application of a spatial mask, as shown in the example below.

If an overdensity is detected on the sky charts, we apply an extraction mask of variable size, that depends on the visual concentration of the cluster. Usually, we adopt values between $\sim$5\arcmin\ $\times$ 5\arcmin\ and $\sim$15\arcmin\ $\times$ 15\arcmin\  around its visual centre and analyze the subsample proper motion and parallax distributions. If the overdensity detected on the sky chart represents a real cluster, 
the subsample spatially restricted must form a clump on the VPD space and a gaussian distribution of the parallax value, both with few outliers.

\begin{figure*}
\includegraphics[width=0.40\linewidth]{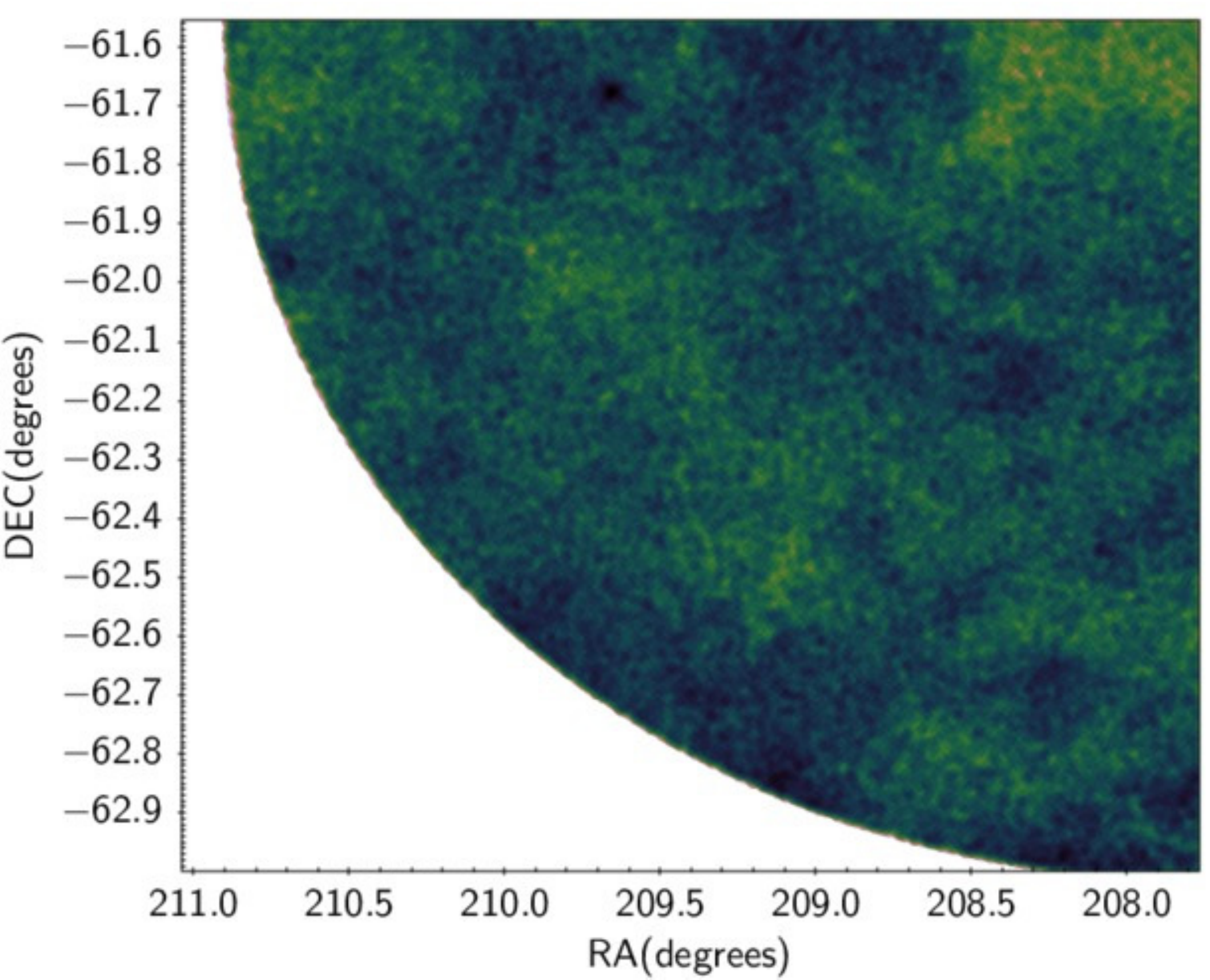} \hspace{0.5cm}
\includegraphics[width=0.4\linewidth]{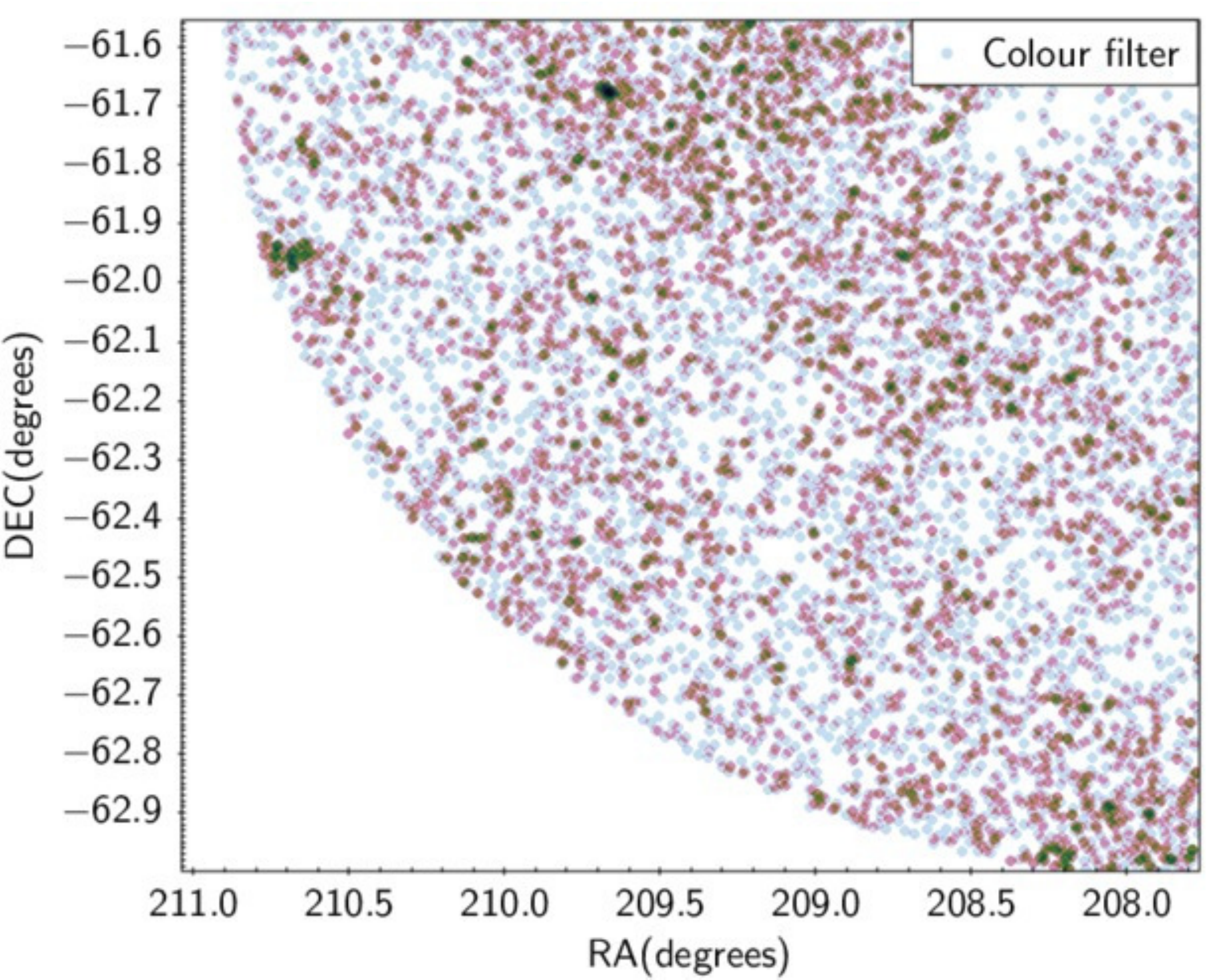}\\ \hspace{0.5cm}
\includegraphics[width=0.4\linewidth]{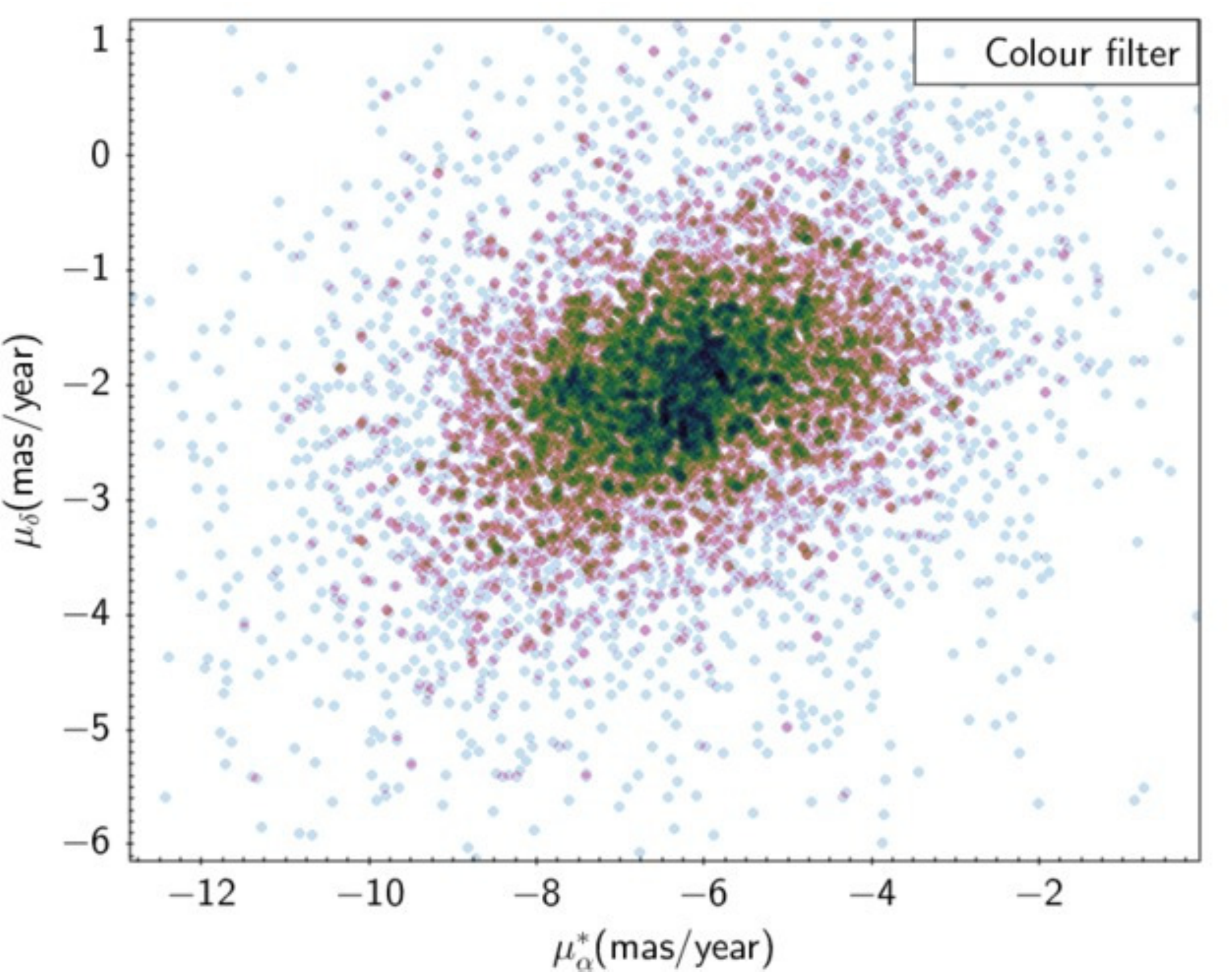} \vspace{0.5cm}
\includegraphics[width=0.4\linewidth]{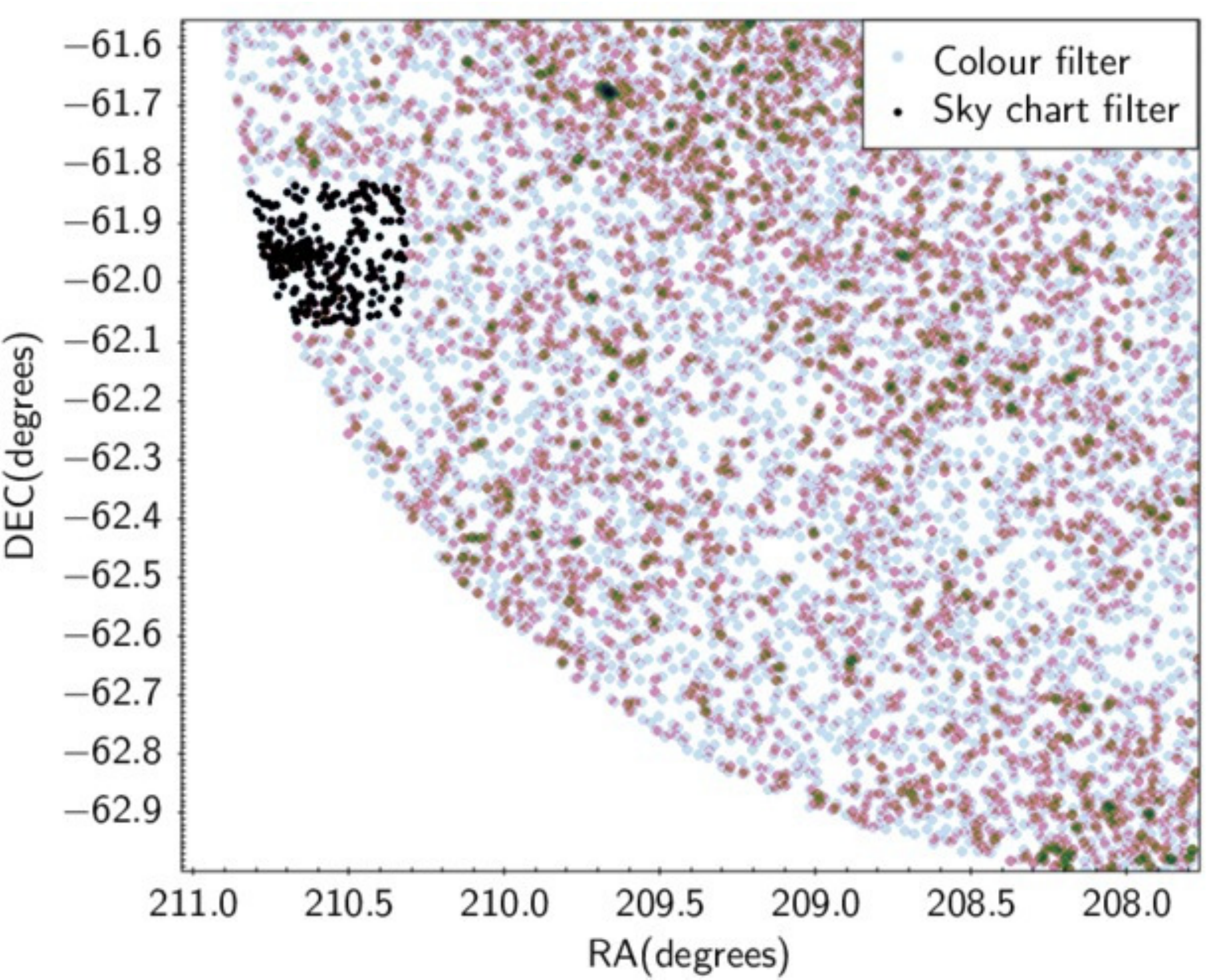} \\ \hspace{0.5cm}
\includegraphics[width=0.4\linewidth]{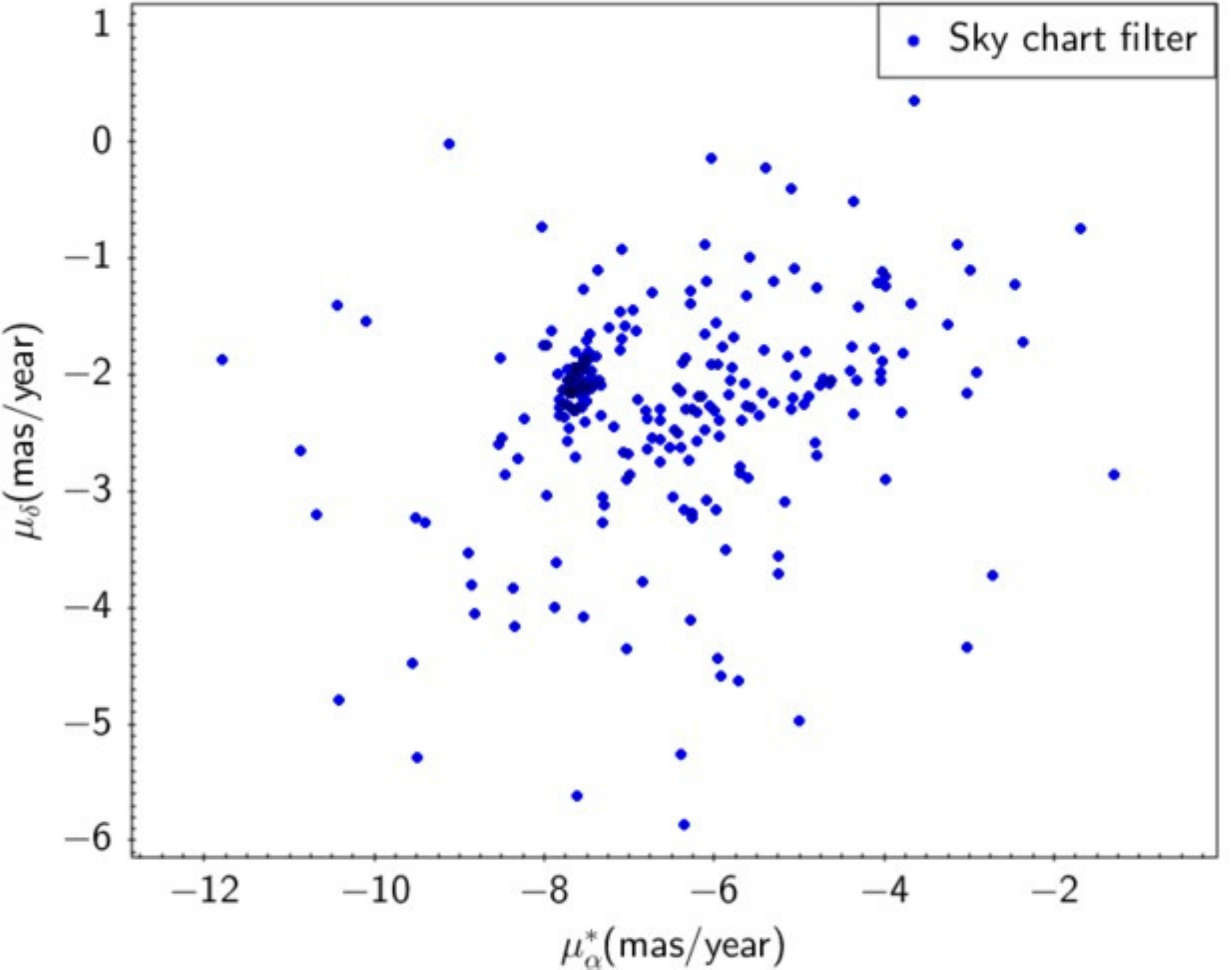} \hspace{0.5cm}
\includegraphics[width=0.4\linewidth]{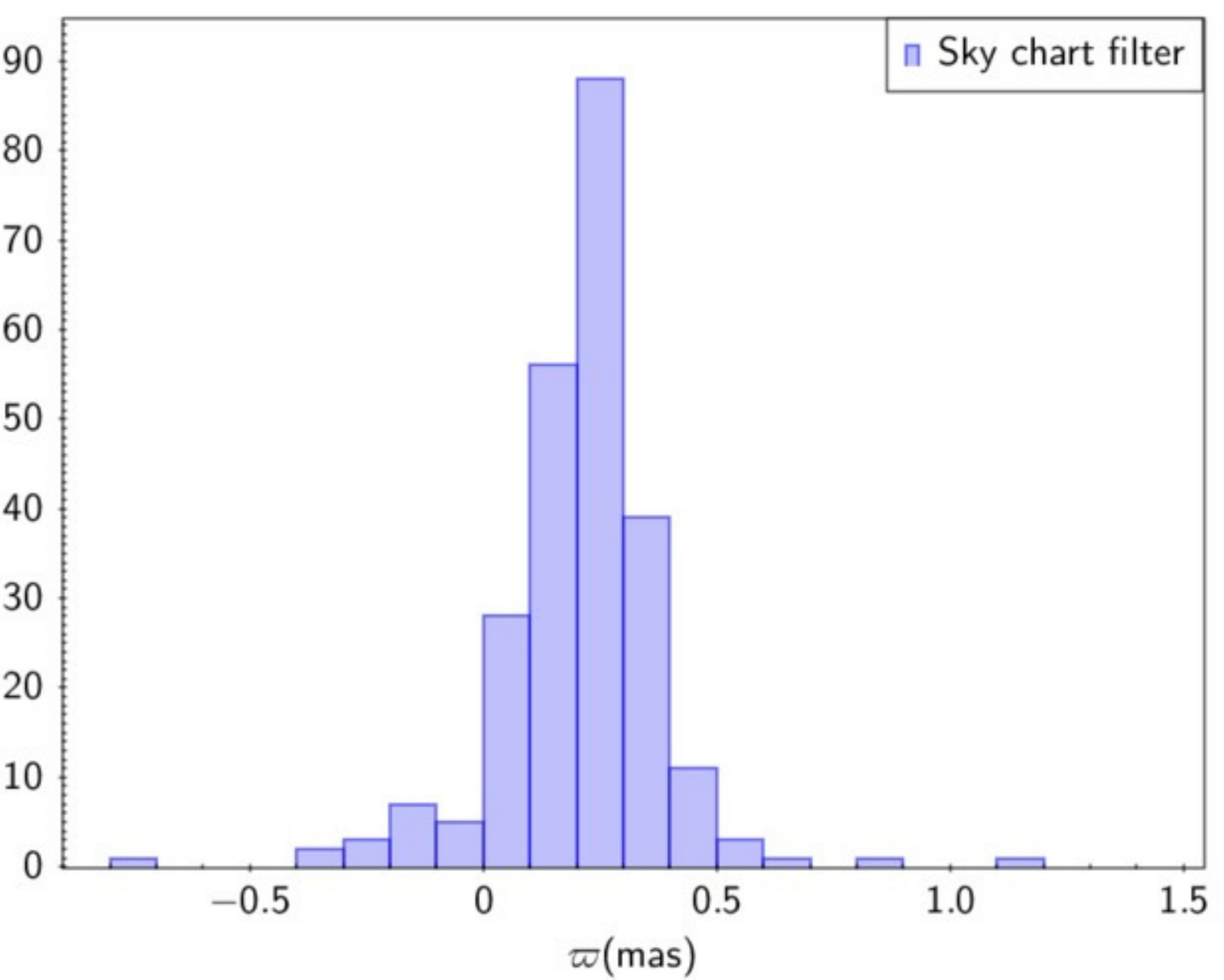}
\caption{Top left: A density plot of a sector of the field 32, centred at coordinates RA=207.893$^{\circ}$ and DEC=-61.5095$^{\circ}$, which the entire region of 1.5 $^{\circ}$ radii comprises $\sim$1.5 million sources}. Top middle: The same sky chart, but the sample is filtered by colour and magnitude. Top right: VPD of the sample filtered by colour. At this point, it is not possible to distinguish a typical cluster overdensity in VPD space. Bottom left: application of a filter over an overdensity found in sky chart to build a subsample. Bottom middle: VPD of a subsample filtered in sky chart where it is now possible to note an overdensity in VPD. Bottom right: Parallax values of the subsample filtered in sky chart.
\label{fig:ra_dec_clump}
\end{figure*}

The Fig.~\ref{fig:ra_dec_clump} exemplifies such procedure, where we analyse the field 32 (Tab.\ref{tab:survey_fields}) comprising $\sim$1.5 million sources. The upper left and upper right panels show the data, before and after a colour cut defined by $G_{BP}-G_{RP}>2.5$, respectively. After the sectioning and colour filtering, as explained before, the sample of $\sim$2.9$\times 10^{5}$ sources reduces to $\sim$7.7$\times 10^{3}$ sources. The middle left panel shows a VPD of the colour filtered sample. Even though, in this case, it is not possible to note a clear overdensity in the VPD, it can be seen an obvious concentration in the filtered sky chart (middle right panel).
After applying a spatial mask around an overdensity at edge of the region and analysing this subsample in a VPD (bottom left panel) and their parallax values (bottom right panel), we note the presence of a new cluster candidate UFMG 54.

The majority of the clusters studied in this work exhibit parallaxes values between $\sim$0.2 and 0.7 mas, where there is too much contamination by star fields. Therefore, diagrams of proper motions in right ascension or declination versus parallax were not very useful to find visual overdensities. The parallax is taken into account when we perform the membership assessment (Sect \ref{sect:sec4}).

In both cases, after the candidate detection, we used the subsamples to build histograms of the quantities RA, DEC, $\mu_{\alpha}^{*}$, $\mu_{\delta}$ and parallax and computed their mode values. 
We use these values in the cluster analysis by restricting our sample inside a smaller parameter space. It is important to emphasize that none of the filters applied to find the clusters (colour filters, proper motion filters or spatial filters) are kept in the subsequent analysis steps. Instead, the modal quantities computed are used as first guess and then refined. For example, the clusters centres are refined by building radial profiles, the proper motion final values are calculated after assessing membership, etc.

\section{Catalogues comparison and discoveries validation}

In the recent years new objects were discovered by many different authors, leading to a drastic increase in their numbers. Despite the efforts to compile all the information on open clusters, candidates and associations \citep[e.g.][]{1995ASSL..203..127M,Dias:2002,kps13},
there is not an absolute database available. To our knowledge the largest database of these objects so far was published by \cite{2019AJ....157...12B}. 

To validate our newly discovered open clusters, we matched our centres coordinates with those on the literature available, within 1$^{\circ}$. If the separation of the centres is larger than or equal the sum of our limiting radius and the literature radius, we assume it as a new cluster. Otherwise, we compare the other  parameters such as proper motion centre and parallax. Besides the base catalogue of \cite{2019AJ....157...12B}, we added the following available databases not incorporated in this reference catalogue:  \cite{Roser:2016}, \cite{cjl18},  \cite{2018MNRAS.481.3902B}, \cite{cjl19},  \cite{2019MNRAS.483.5508F}, \cite{cks19},  \cite{2019MNRAS.484.2181T},  \cite{2019arXiv190904612B}, Sim2019,  LP2019, CG2020 and \cite{2020PASP..132c4502H}. 
Besides the astrometric comparison, we also crossmatched our cluster memberlist with the members available in the above-mentioned works. This procedure shows if the detected cluster share the same stars or if it is a sub-structure of a known one.


At first, we found 59 targets not reported in the literature, however, during the elaboration of this paper, three works reporting new discovered open clusters have been published: 
Sim2019,
LP2019 and more recently CG2020. Sim2019 found 207 new open clusters within 1 kpc by visually inspecting \textit{Gaia} DR2 proper motion diagrams searching for overdensities. The authors reported the cluster that we identified as UFMG35; the proper motions difference are within 1 $\sigma$, our distance value of $955\pm 132$\,pc is compatible with the authors value of $887\pm 40$\,pc and our age $log(t)$ value of $7.50\pm 0.30$ is compatible with the authors value of $7.75$, so we assume they are identical objects. 

\begin{figure}
\centering
\includegraphics[width=\linewidth]{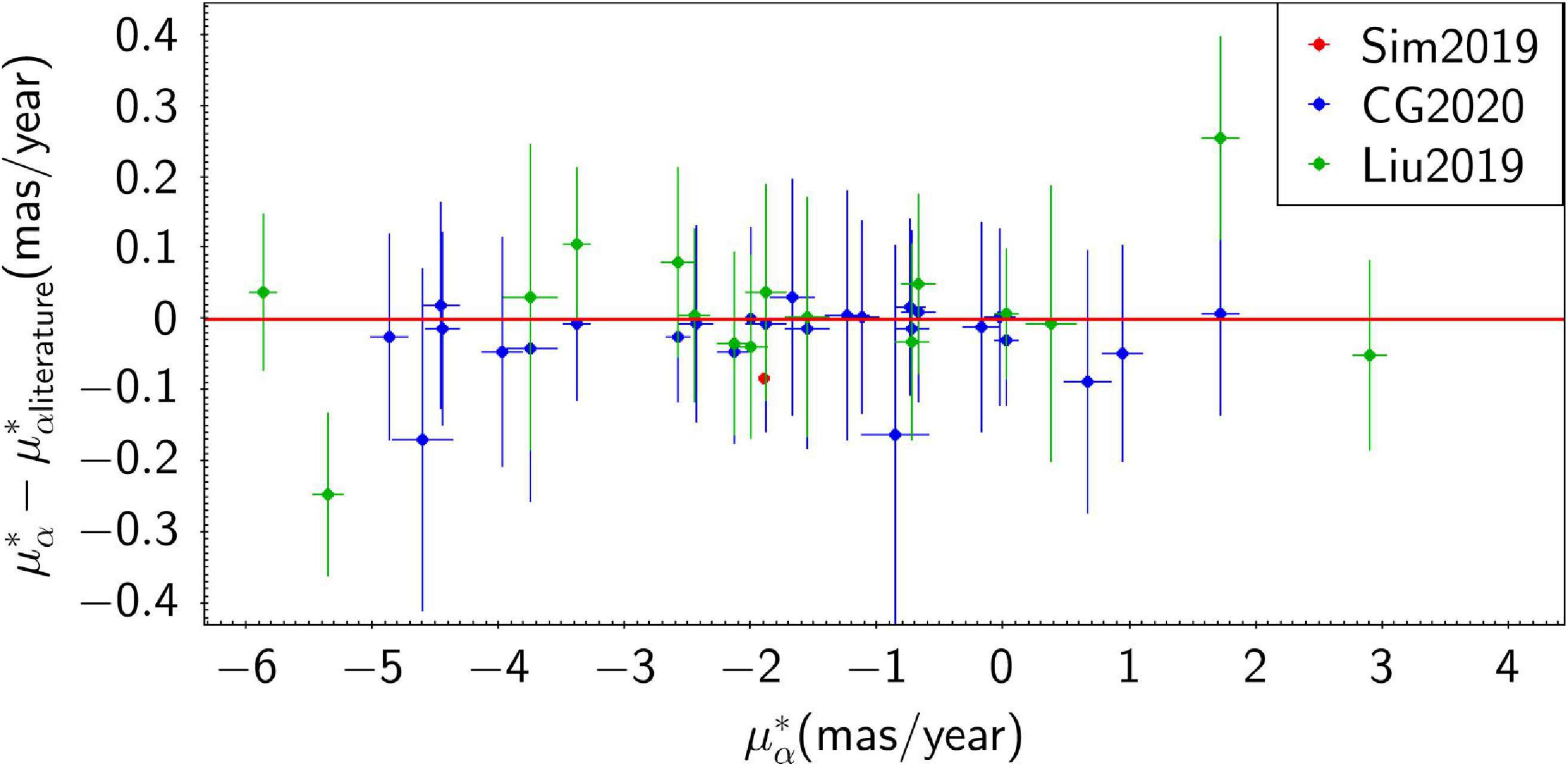} \\ \vspace{0.25cm}
\includegraphics[width=\linewidth]{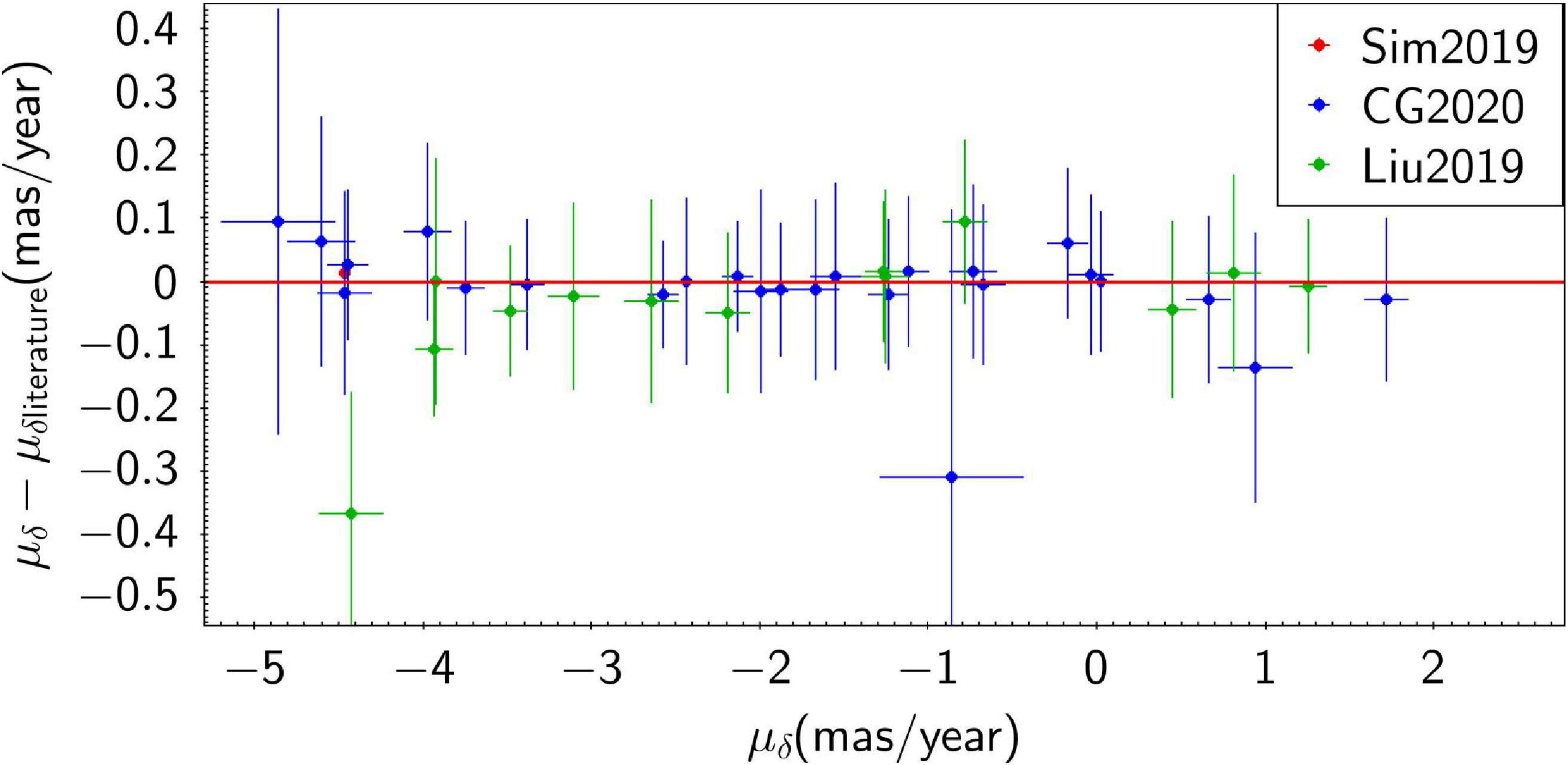} \\ \vspace{0.25cm}
\includegraphics[width=\linewidth]{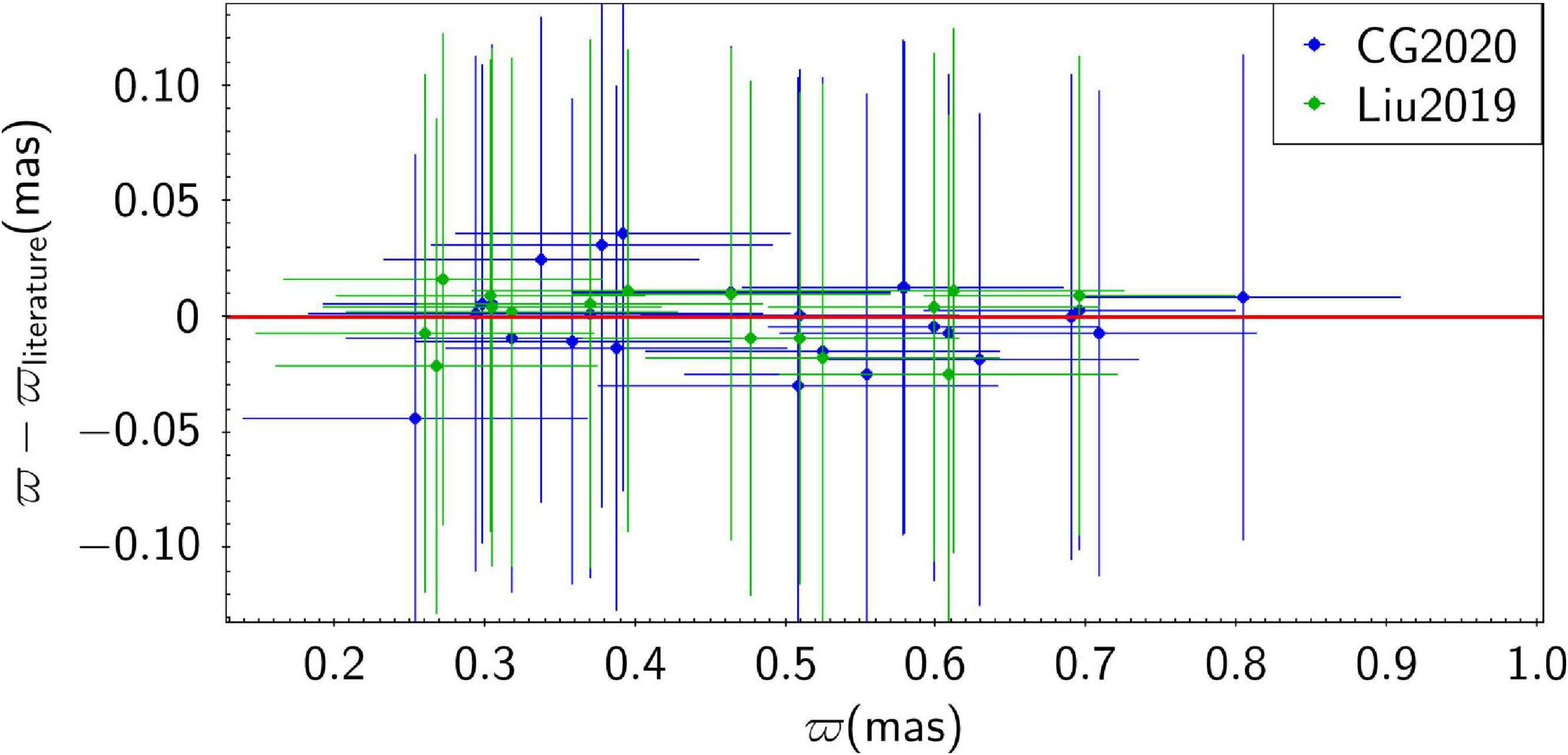}
\caption{Difference between our mean values of proper motion in right ascension ($\mu_{\alpha}^{*}$) (upper panel), declination ($\mu_\delta$) (middle panel) and parallax ($\varpi$) (bottom panel) and those found in Sim2019, LP2019 and CG2020. The bars represent 1$\sigma$ uncertainties of our mean values.}
\label{fig:validation}
\end{figure}

Table \ref{Tab:validation} compiles cross-identifications between the mentioned catalogues (id$_{UFMG}$, id$_{L2019}$, id$_{CG2020}$) and provides our 1$\sigma$ dispersion of the cluster astrometric parameters ($\sigma_{\mu_{\alpha}^{*}}$, $\sigma_{\mu_{\delta}}$, $\sigma_{\varpi}$) as well as the difference of their mean values respect to the ones on LP2019 ($\Delta\mu^{*}_{\alpha L2019}$, $\Delta\mu_{\delta L2019}$, $\Delta\varpi_{L2019}$) and CG2020 ($\Delta\mu^{*}_{\alpha CG2020}$, $\Delta\mu_{\delta CG2020}$, $\Delta\varpi_{CG2020}$). The separation between the centre coordinates derived in this work and the literature ones are also reported ($\Delta d_{L2019}$, $\Delta d_{CG2020}$). Fig.~\ref{fig:validation} depicts this information, showing the difference between the mean of our sample astrometric parameters $\mu_\alpha^*$, $\mu_\delta$ and $\varpi$ and those derived in the literature for the clusters in common.

LP2019 detected 2443 star clusters using a clustering algorithm in the 5-D astrometric space (l, b, ,$\varpi$, $\mu_{\alpha}^{*}$ , $\mu_{\delta}$ ). 76 of these objects were flagged as high confidence, new open clusters candidates. However, it is important to notice that their clusters 1180 and 1182 had been previously reported as the open clusters UFMG1 and UFMG3,  discovered in \cite{2019MNRAS.483.5508F}. Besides those, we found 7 more coincidences with the present work clusters (see Table~\ref{Tab:validation}).
In all the cases, the mean proper motions in right ascension  and declination and mean parallax are compatible within 1$\sigma$ of our value distributions.

We have also found 10 possible clusters, flagged as low confidence candidates by LP2019.
In general, the mean proper motions in right ascension  and declination and mean parallax are compatible within 1 $\sigma$ of our values, with two exceptions: UFMG13 and UFMG27. UFMG13 for which $\mu_{\alpha}^{*}$ is compatible within 2$\sigma$ and its centre differs by only 2.5 arcmin from the cluster candidate 140. The UFMG27, identified as cluster candidate 466 
for which $\mu_{\alpha}^{*}$ is compatible within 3$\sigma$, $\mu_{\delta}$ compatible within 2$\sigma$ and $\varpi$ compatible within 1$\sigma$. The centres differ by $\sim$470 \arcsec, which is a value greater than the limiting radii found for this object.  There exist an appreciable difference in proper motion in right ascension and, according to our analysis, the centre of the corresponding cluster is out of the found structure.
In spite of the centres separation be larger than our limiting radii, the targets share 11 stars in a total of 143 stars (88 from the cluster candidate 466 and 55 from UFMG27). We would like to to emphasize that the cluster candidate 466 presents high dispersion in the VPD and looks very sparse in sky chart, being reported as a class 3 object, that means a candidate of low confidence that need future confirmation. The two objects are not alike, but 
we assume that UFMG27 could actually be part of the cluster candidate 466 and decided to withdraw UFMG27 from our sample of new findings and give the discovery credit to LP2019, reducing our discovered cluster sample from 26 to 25 objects.

More recently, CG2020 have discovered 582 new open clusters by applying an unsupervised clustering algorithm to find overdensities in a five-dimensional parameter space ($l$,$b$,$\varpi$,$\mu_{\alpha}^{}$, $\mu_{\delta}$). The authors found 26 open clusters from the sample of 59 objects studied in our work, some of them being also reported in LP2019 (see Table~\ref{Tab:validation}).
These clusters astrometric parameters are compatible with our values within 1 $\sigma$, so we assume they are identical objects. 

However, neither LP2019 nor CG2020 derived structural parameters for these 33 coincident open clusters, making this work the first one to provide a detailed analysis of these targets. To the best of our knowledge, the remaining sample of 25 open clusters are newly discovered objects and are not listed in the literature.

\begin{table*}
\rotcaption{Astrometric properties of objects in common with previous works}
\begin{sideways}
\begin{minipage}{230mm}

\begin{tabular}{|l|r|l|r|r|r|r|r|r|r|r|r|r|r|}
\hline
\multicolumn{1}{|c|}{id} &
\multicolumn{1}{c|}{id} &
\multicolumn{1}{c|}{internal id} &
\multicolumn{1}{c|}{$\sigma_{\mu_{\alpha}^{*}}$} &
\multicolumn{1}{c|}{$\sigma_{\mu_{\delta}}$} &
\multicolumn{1}{c|}{$\sigma_{\varpi}$}  &
\multicolumn{1}{c|}{$\Delta\mu^{*}_{\alpha L2019}$} &
\multicolumn{1}{c|}{$\Delta\mu_{\delta L2019}$}&
\multicolumn{1}{c|}{$\Delta\varpi_{L2019}$}&
\multicolumn{1}{c|}{$\Delta d_{L2019}$} &
\multicolumn{1}{c|}{$\Delta\mu^{*}_{\alpha CG2020}$} &
\multicolumn{1}{c|}{$\Delta\mu_{\delta CG2020}$} &
\multicolumn{1}{c|}{$\Delta\varpi_{CG2020}$} &
\multicolumn{1}{c|}{$\Delta d_{CG2020}$} \\

\multicolumn{1}{|c|}{LP2019} &
\multicolumn{1}{c|}{CG2020} &
\multicolumn{1}{c|}{UFMG} &
\multicolumn{1}{c|}{mas.$yr^{-1}$} &
\multicolumn{1}{c|}{mas.$yr^{-1}$} &
\multicolumn{1}{c|}{mas}  &
\multicolumn{1}{c|}{mas.$yr^{-1}$} &
\multicolumn{1}{c|}{mas.$yr^{-1}$}&
\multicolumn{1}{c|}{mas}&
\multicolumn{1}{c|}{arcmin} &
\multicolumn{1}{c|}{mas.$yr^{-1}$} &
\multicolumn{1}{c|}{mas.$yr^{-1}$} &
\multicolumn{1}{c|}{mas} &
\multicolumn{1}{c|}{arcmin}  \\
\hline
867* &  &UFMG4 &   0.078 & 0.110 & 0.113 & -0.010 & -0.012 & 0.003 & 0.160 &  &  &  & \\
145 & UBC551 &UFMG5 &   0.129 & 0.087 & 0.106 & -0.034 & -0.003 & -0.009 & 1.138 & -0.045 & 0.010 & 0.001 & 0.696\\
860* & UBC310 &UFMG6 &   0.127 & 0.125 & 0.112 & 0.050 & -0.048 & 0.004 & 0.995 & 0.009 & -0.003 & 0.006 & 0.855\\
   & UBC537 &UFMG7 & 0.138 & 0.132 & 0.103 &  &  &  &  & -0.006 & 0.003 & 0.006 & 1.717\\
866 &  &UFMG8 &   0.132 & 0.138 & 0.104 & -0.050 & -0.044 & 0.011 & 0.645 &  &  &  & \\
 & UBC336 &UFMG9 &   0.185 & 0.130 & 0.105 &  &  &  &  & -0.087 & -0.028 & 0.025 & 0.966\\
2094 & UBC278 &UFMG10 &   0.214 & 0.104 & 0.118 & 0.031 & -0.007 & -0.018 & 1.511 & -0.042 & -0.010 & -0.015 & 1.129\\
2210 & UBC311 &UFMG12 &   0.092 & 0.111 & 0.103 & 0.007 & 0.018 & 0.009 & 2.416 & -0.029 & 0.002 & 0.003 & 1.196\\
140* &  &UFMG13 &   0.114 & 0.191 & 0.107 & -0.247 & -0.365 & -0.021 & 2.487 &  &  &  & \\
 436* &  &UFMG14 &  0.121 & 0.106 & 0.106 & 0.006 & -0.105 & 0.016 & 1.692 &  &  &  & \\
  & UBC322 &UFMG15 &  0.265 & 0.422 & 0.113 &  &  &  &  & -0.162 & -0.307 & 0.031 & 0.614\\
  & UBC303 &UFMG17 &  0.144 & 0.336 & 0.133 &  &  &  &  & -0.024 & 0.095 & -0.029 & 2.979\\
 861 &  &UFMG18 &  0.133 & 0.193 & 0.112 & 0.079 & 0.002 & -0.007 & 0.736 &  &  &  & \\
2100 & UBC305 &UFMG19 &   0.107 & 0.103 & 0.106 & 0.105 & -0.045 & 0.010 & 1.868 & -0.007 & -0.003 & 0.011 & 1.836\\
438* & UBC321 &UFMG20 &   0.168 & 0.147 & 0.114 & 0.004 & -0.023 & 0.006 & 0.550 & -0.013 & 0.010 & 0.001 & 0.468\\
  & UBC319 &UFMG21 &  0.124 & 0.130 & 0.105 &  &  &  &  & 0.018 & 0.017 & -0.007 & 0.685\\
  & UBC326 &UFMG23 &  0.136 & 0.118 & 0.106 &  &  &  &  & 0.003 & 0.018 & 0.013 & 0.545\\
  & UBC136 &UFMG25 &  0.123 & 0.126 & 0.105 &  &  &  &  & 0.003 & 0.011 & 0.000 & 1.042\\
  & UBC121 &UFMG26 &  0.174 & 0.119 & 0.107 &  &  &  &  & 0.006 & -0.021 & 0.013 & 0.882\\
 466 *& & UFMG27&0.192 & 0.164 & 0.116 &  0.550 & 0.195 & 0.003 & 474.34 & & & & \\
  & UBC550 & UFMG28 & 0.091 & 0.084 & 0.105 &  &  &  &  & -0.025 & -0.021 & -0.011 & 1.795\\
 1624 & UBC335 &UFMG31 &  0.142 & 0.128 & 0.111 & 0.255 & 0.096 & -0.009 & 2.566 & 0.009 & -0.027 & -0.009 & 2.207\\
  & UBC535 &UFMG32 &  0.161 & 0.139 & 0.114 &  &  &  &  & -0.046 & 0.081 & -0.044 & 1.702\\
 & UBC572 &UFMG33 &   0.152 & 0.213 & 0.106 &  &  &  &  & -0.048 & -0.136 & -0.018 & 1.597\\
 623$^a$ &  &UFMG35 &  0.149 & 0.146 & 0.108 & -0.083$^a$ & 0.014$^a$ &  & 3.932$^a$ &  &  &  & \\
  & UBC533 &UFMG36 &  0.166 & 0.141 & 0.105 &  &  &  &  & 0.032 & -0.012 & 0.009 & 0.958\\
 714* & UBC307& UFMG40  & 0.128 & 0.159 & 0.110 & -0.039 & -0.030 & 0.002 & 1.028 & 0.000 & -0.015 & -0.009 & 1.037\\
  & UBC295 &UFMG49 &  0.144 & 0.160 & 0.121 &  &  &  &  & 0.018 & -0.016 & -0.025 & 1.308\\
  & UBC293 &UFMG50 &  0.241 & 0.198 & 0.111 &  &  &  &  & -0.170 & 0.065 & 0.001 & 0.404\\
 2156* & UBC314 &UFMG51 &  0.152 & 0.106 & 0.110 & 0.039 & -0.062 & 0.004 & 0.529 & -0.006 & -0.013 & -0.004 & 2.308\\
 1204*& UBC334 &UFMG52 &  0.138 & 0.136 & 0.112 & -0.032 & 0.010 & -0.025 & 2.042 & -0.013 & 0.016 & -0.007 & 3.061\\
  & UBC522 &UFMG56 &  0.136 & 0.119 & 0.113 &  &  &  &  & -0.013 & 0.029 & -0.013 & 0.652\\
  & UBC552 &UFMG57 &  0.146 & 0.118 & 0.111 &  &  &  &  & -0.011 & 0.061 & 0.036 & 0.751\\
 676* &  &UFMG62 &  0.110 & 0.113 & 0.102 & 0.037 & -0.019 & 0.009 & 0.981 &  &  &  & \\
\hline\end{tabular}

$^a$ relative to the work of Sim2019 \\
$*$ relative to low confidence OC candidate of LP2019
\end{minipage}
\end{sideways}
\label{Tab:validation}
\end{table*}

\section{Analysis}
\label{sect:analysis}

\subsection{Centre and structural parameters}
\label{sect:4.2}

With the centres and proper motion mean values computed in the previous analysis, we followed a similar procedure adopted in \cite{2019MNRAS.483.5508F} and selected  stars inside 30\arcmin\ from the objects centres and with proper motions inside a square box around its mean values. This proper motion filter removes the excess of field stars, increasing the contrast of the cluster population over the field and allowing the construction of the radial density profiles (RDPs). 
Our objects exhibit parallax smaller than 1.2 mas.  In this case, a 1\,mas\,yr$^{-1}$ box around the mean proper motion value is expected to encompass more than 3$\sigma$ of the distribution, assuming an average proper motion dispersion of 0.13 mas\,yr$^{-1}$, as found by \citep{ca20}. Therefore, we adopted a 1\,mas\,yr$^{-1}$ box. When the field contamination was not critical, we allowed a 2\,mas\,yr$^{-1}$ proper motion box aiming at recovering more members, which was employed for seven clusters. On the other hand, for six critical cases, where the clusters candidates were less populous, we needed to adopt box size values of 0.8\,mas\,yr$^{-1}$, to enhance the contrast against the field.

At this point we have defined our working sample for the radial profile procedure: best quality \textit{Gaia} data within the square proper motion box and 30\arcmin\ from the objects centres. We did not find any systematic effect introduced by this procedure that could bias our determination of the structural parameters. 
Afterwards, we built RDPs on which we estimated the size of each object, the local background density and refined the centres previously computed. We built the RDP by counting the number of stars in concentric rings around the initial centre and dividing it by the rings area. Independent RDPs were made by using different ring widths and then merged into a single, final RDP. 

\begin{figure*}
\centering
\includegraphics[width=0.47\linewidth]{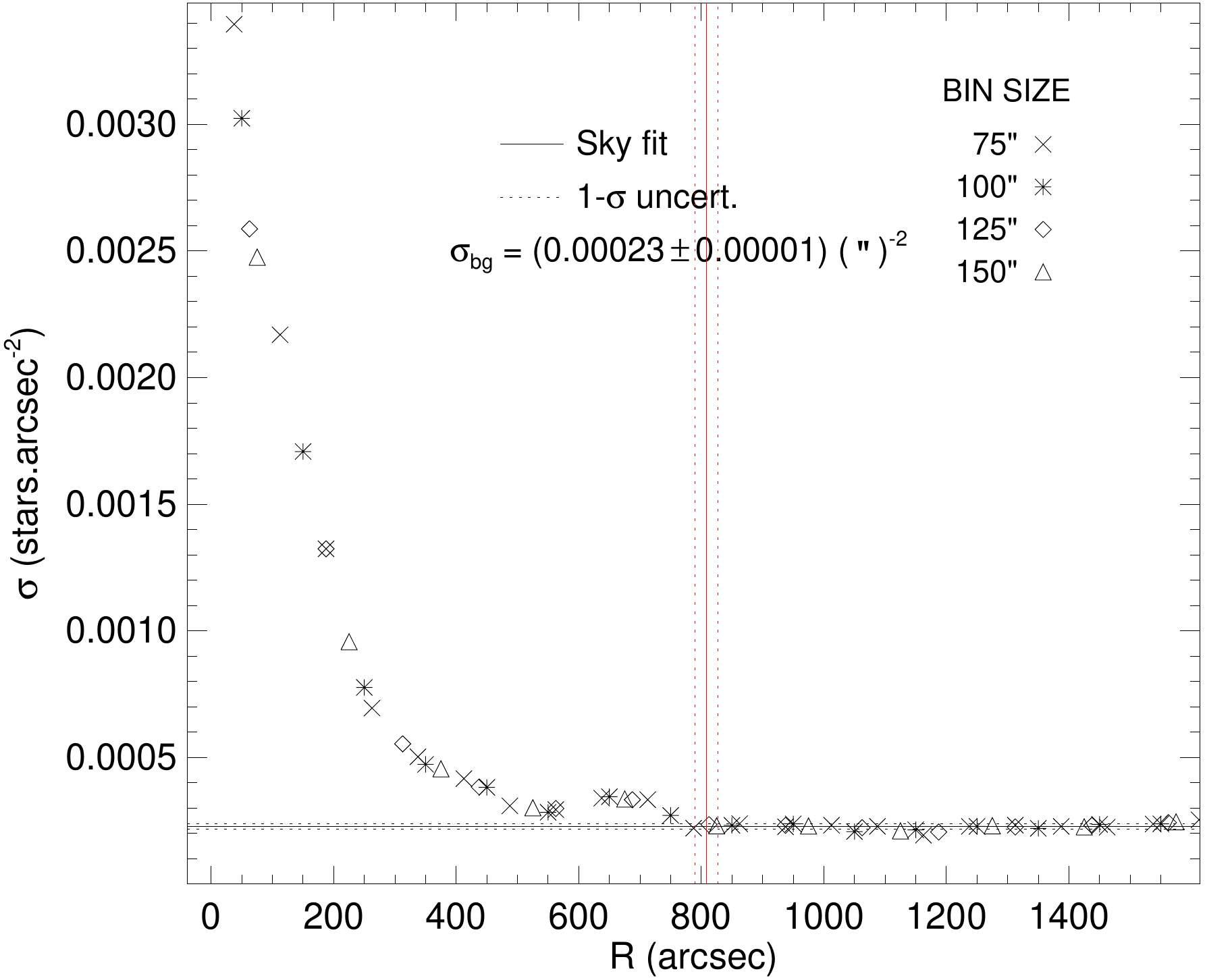} \hspace{0.0cm}  
\includegraphics[width=0.47\linewidth]{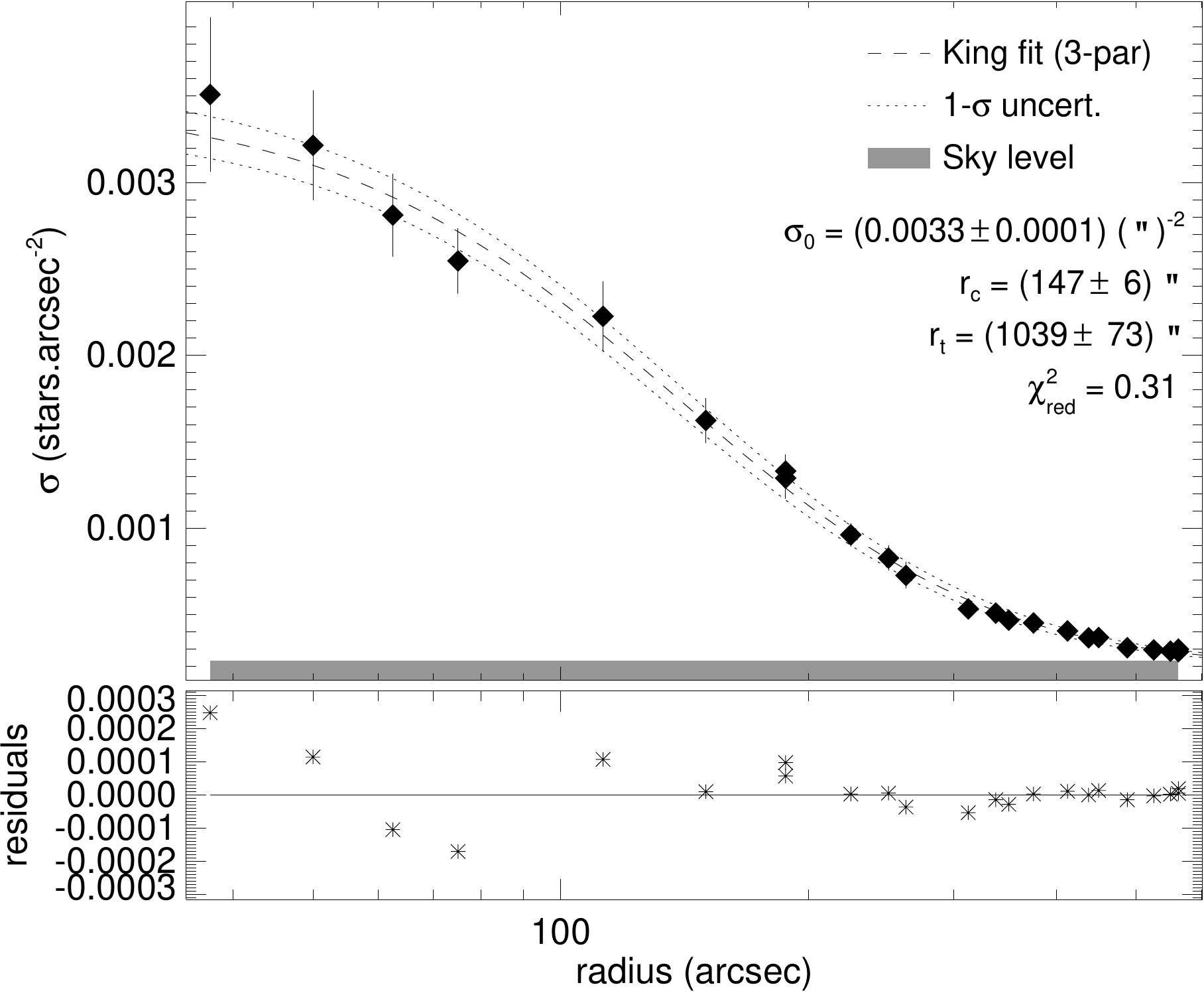} \\ \hspace{0.0cm}  
\includegraphics[width=0.47\linewidth]{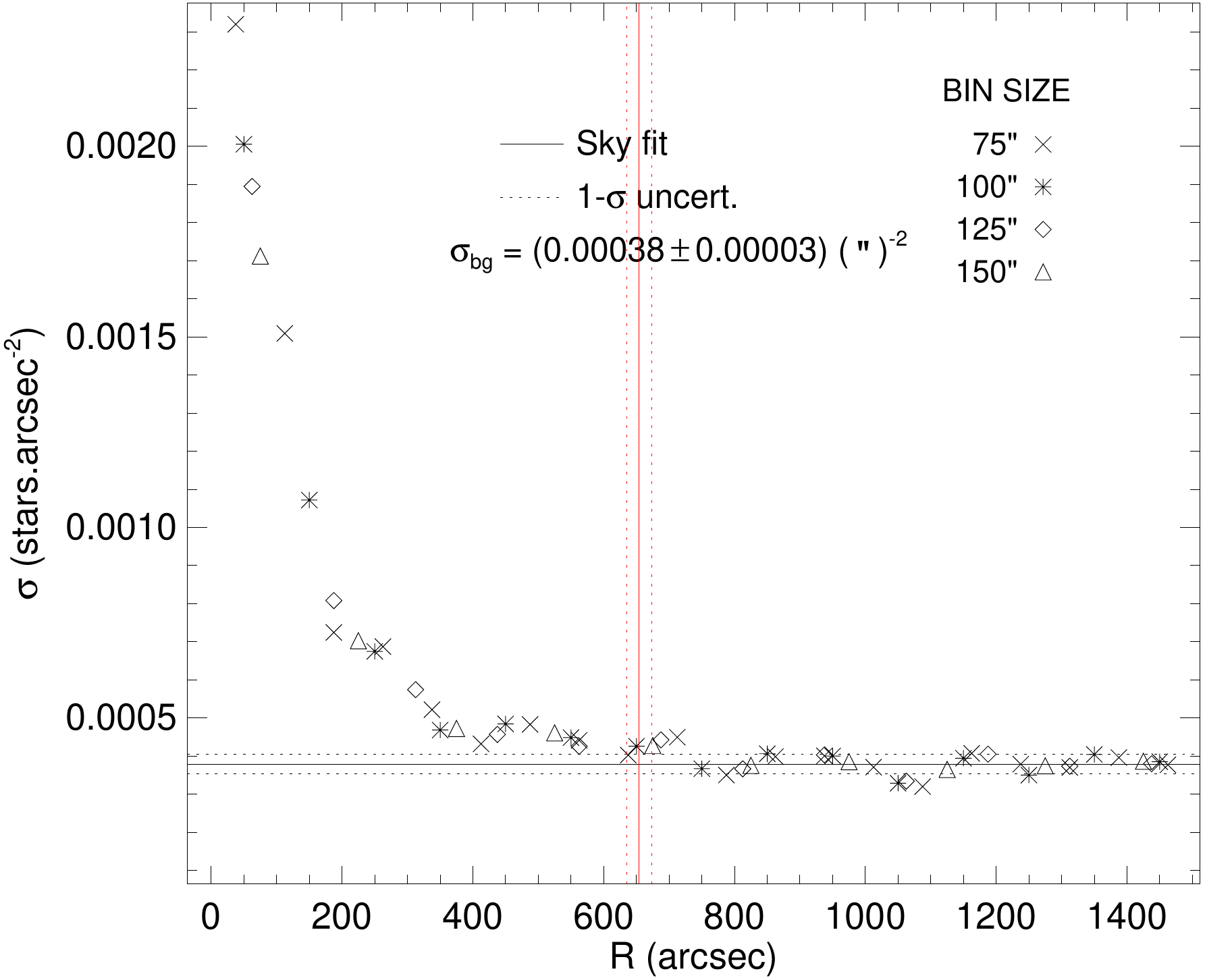}   \hspace{0.0cm}  
\includegraphics[width=0.47\linewidth]{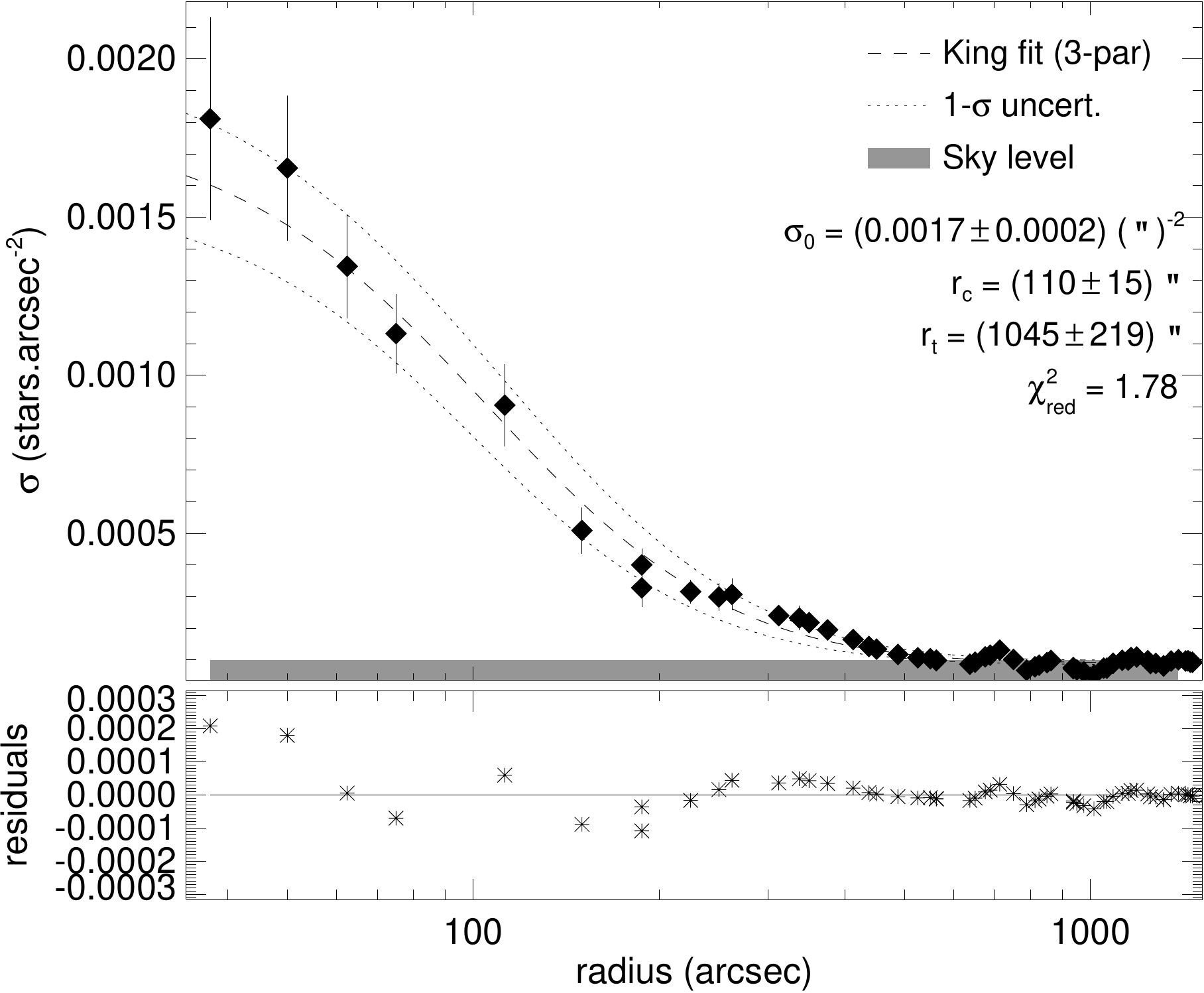} \\ \hspace{0.0cm}  
\includegraphics[width=0.47\linewidth]{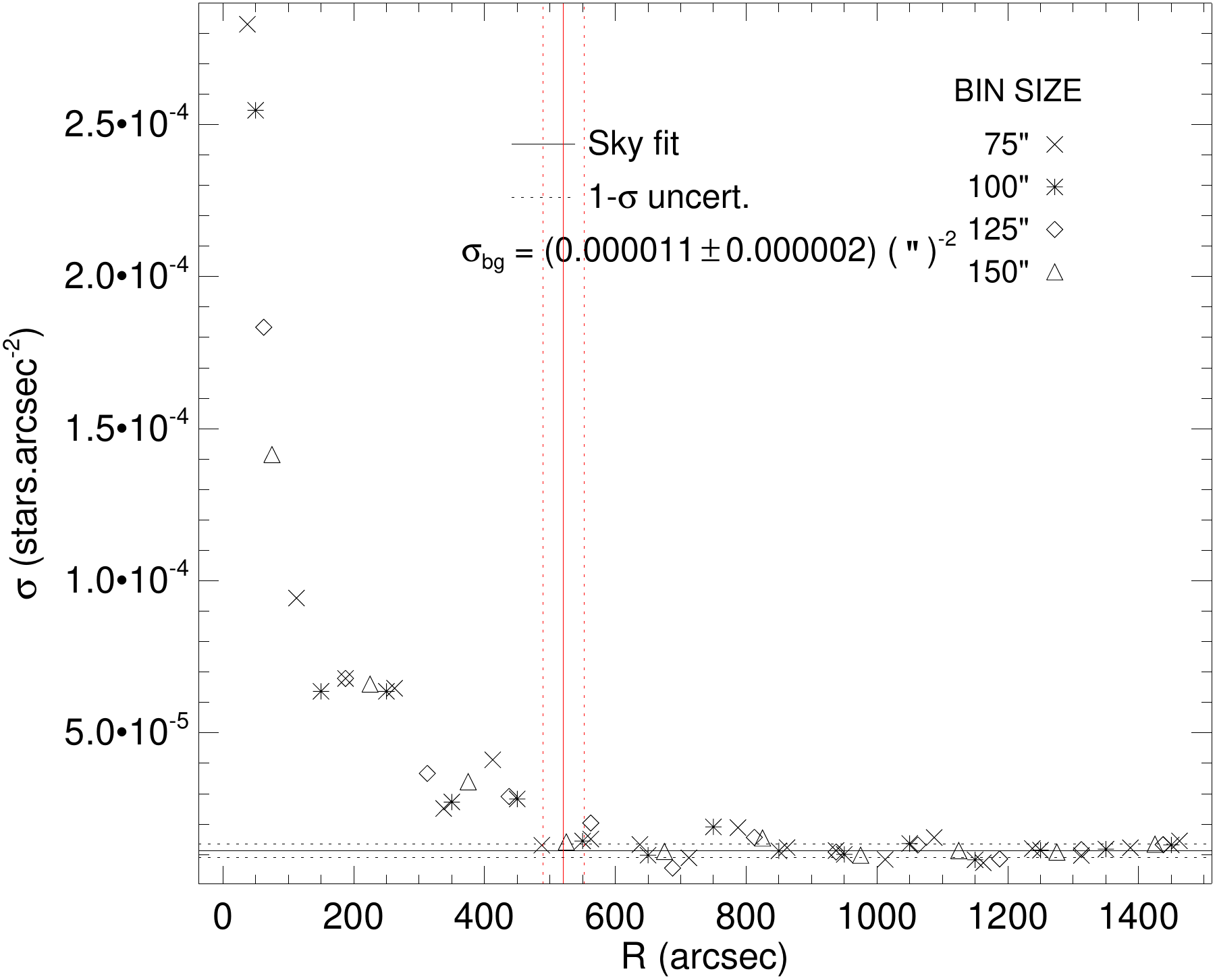}  \vspace{0.3cm}  
\includegraphics[width=0.47\linewidth]{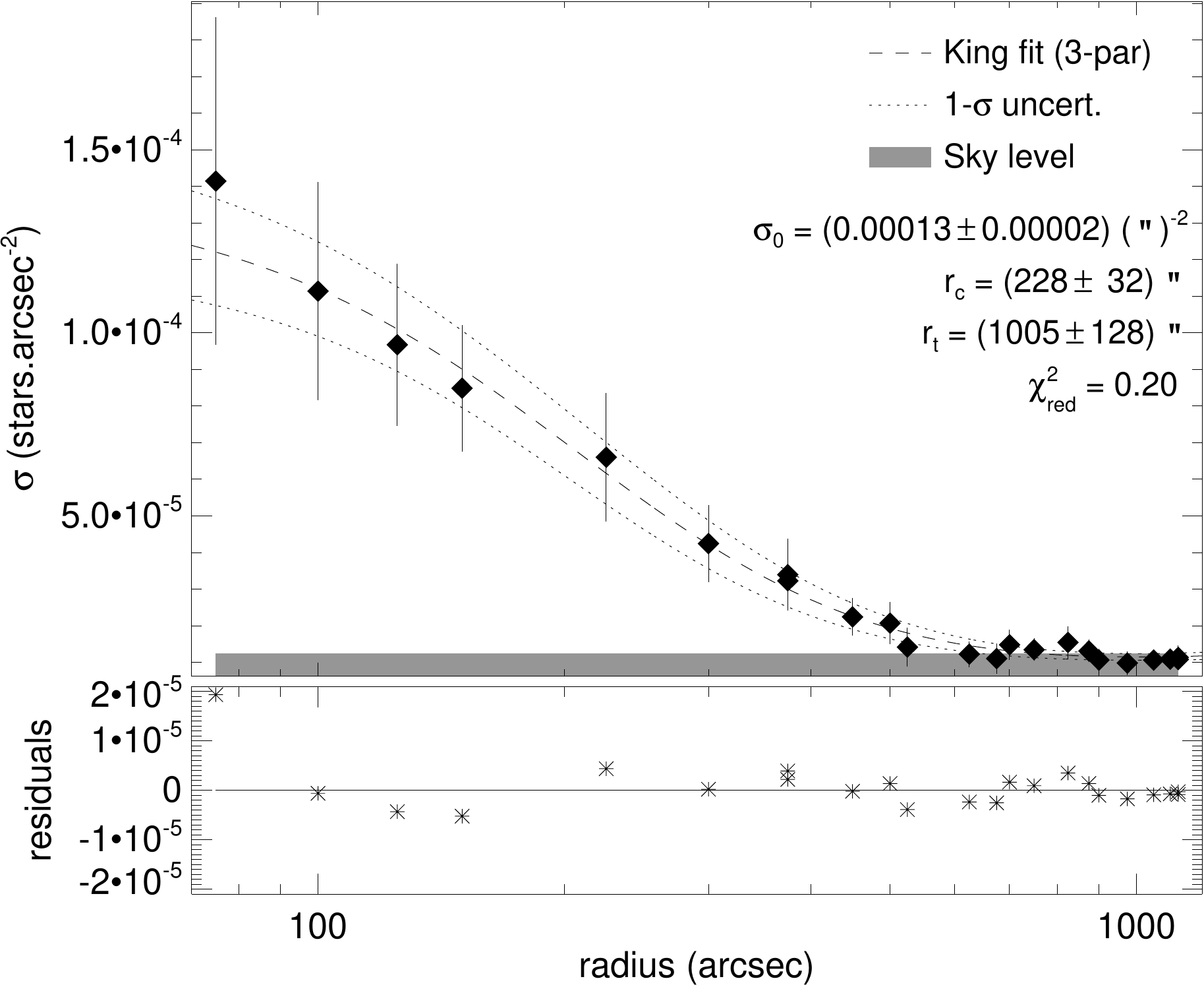}                  
\caption{Left column: radial density profiles of clusters UFMG04 (top panel), UFMG34 (middle panel) and UFMG59 (bottom panel) with their limiting radius (vertical line) and mean background density level (horizontal line) indicated. Right column: best-fitting King models (dashed line) of the same clusters with envelope of 1-$\sigma$ uncertainties (dotted lines). Error bars correspond to poissonian noise. The sky level and its fluctuation are indicated by the grey bar. The fitting residuals are also presented below the fittings. }
\label{fig:rdp_king}
\end{figure*}

We refined the centre values by making small adjustments to the coordinates previously set, accepting the solution with maximum central density as the new centre values. 
We fitted a straight line to the stellar density for rings well beyond the objects core, typically  between $\sim 15$\arcmin\ and $\sim 25$\arcmin, to compute the sky background level ($\sigma_{bkg}$) and its uncertainty.
The limiting radius ($r_{lim}$), defined as the radius where the stellar density reaches the sky level, resulted between 4\arcmin\ and 20\arcmin\  for all clusters. Fig.~\ref{fig:rdp_king} shows examples of the RDPs and the derived sky density levels used to determine the limiting radius for some cluster candidates.

Finally, the clusters' structural parameters were obtained by weighted fittings of the \cite{King:1962} analytical model over their RDP. 
We present some of the profiles fitted in Fig \ref{fig:rdp_king}. The error of each RDP datum correspond to its Poisson uncertainty and was used as weight in the fitting process. The structural parameters: central density ($\sigma_0$), core radius ($r_c$) and tidal radius ($r_t$) are presented in the Table \ref{Tab:clusters_prop2}. The number $N$ of cluster members is indicated in the last column of this table. 
This table is also available electronically through
Vizier\footnote{http://cdsarc.u-strasbg.fr/vizier/cat/J/MNRAS/{\bf vol/page}}.

\begin{table*}
\caption{Astrometric and structural properties of the investigated clusters}
\label{Tab:clusters_prop2}
\begin{tabular}{ll r@{$\,\pm\,$}l r@{$\,\pm\,$}l r@{$\,\pm\,$}l r@{$\,\pm\,$}l r@{$\,\pm\,$}l r@{$\,\pm\,$}l r@{$\,\pm\,$}l r}
\hline
\multicolumn{1}{|c|}{Name$^\dagger$} &
\multicolumn{1}{|c|}{internal id$^\dagger$} &
\multicolumn{2}{c|}{$r_{lim}$} &
\multicolumn{2}{c|}{$r_c$} &
\multicolumn{2}{c|}{$r_t$} &
\multicolumn{2}{c|}{$r_t$} &
\multicolumn{2}{c|}{$\mu_{\alpha}^{*}$} &
\multicolumn{2}{c|}{$\mu_{\delta}$} &
\multicolumn{2}{c|}{$\varpi$} &
\multicolumn{1}{c|}{$N$} \\

\multicolumn{1}{|c|}{} &
\multicolumn{1}{|c|}{} &
\multicolumn{2}{c|}{(arcsec)} &
\multicolumn{2}{c|}{(arcsec)} &
\multicolumn{2}{c|}{(arcsec)} &
\multicolumn{2}{c|}{(pc)} &
\multicolumn{2}{c|}{(mas$\cdot$yr$^{-1}$)} &
\multicolumn{2}{c|}{(mas$\cdot$yr$^{-1}$)} &
\multicolumn{2}{c|}{(mas)} &
\multicolumn{1}{c|}{$\#$} \\
\hline
UFMG11 && 546& 19 & 86& 18 & 673& 153 & 8.0&   2.1&      -2.083& 0.019      & -2.050& 0.039   & 0.300& 0.010       & 112\\
UFMG16 && 588& 43 & 286& 116 & 603& 139 &  3.4&0.9&       0.480&0.046         & -2.347&0.035 & 0.734&0.029 & 24\\
UFMG22 && 558& 7 & 64&       10 & 686& 135 &  5.0& 1.2&           -0.412&0.026    & -2.610&  0.020    & 0.358&0.011 & 101\\
UFMG24 && 304& 38 & 79&   25 & 847&261 & 5.2& 1.9  &          2.094&0.051       & -1.089&0.021  & 0.677&0.028 & 14\\
UFMG29 && 617& 47 & 156& 22 & 764&91 &  5.1 & 1.1 &            -0.986&0.023        & 0.524&0.022  & 0.585&0.020 & 29\\
UFMG30 && 917& 52 & 189& 53 & 579&  109 &  3.3 & 0.9 &        -0.422&0.025      & -0.820&0.021   & 0.817&0.019 & 29\\
UFMG34 & & 654& 19 & 110&   15 & 1045&219 &  7.0 & 2.0 &       -3.591&  0.012   & -5.729&0.013  & 0.380&0.012 & 87\\
UFMG37 & &333& 19 & 109& 37 & 367&  99 &  1.6 & 0.5 &           -5.466&0.032        & -5.308&0.030    & 0.720&0.023 & 21\\
UFMG38 && 808&   19 & 199& 26 & 1105&176 & 10.9 & 2.3 &            -1.641&0.010      & -2.594&0.010   & 0.255&0.009 & 137\\
UFMG39 && 408& 31 & 141& 43 & 482&118 & 2.9 & 0.9 &         -3.421&0.022      & -1.616&0.023 & 0.626&0.024 & 25\\
UFMG41 && 453&   30 & 96&   17 & 974&165  & 6.8 & 1.7 &           -1.026&0.028      & -1.897&0.033 & 0.503&0.020 & 27\\
UFMG42 && 433& 19 & 57&   8 & 498&70    &   12.1 & 3.3 &           -6.289&0.036         & -4.235&0.032 & 0.320&0.018 & 43\\
UFMG43 && 717& 31 & 351&  51 & 1020&117  &  8.4 & 1.8 &         -3.834&0.016    & -3.911&0.015 & 0.419&0.014 & 59\\
UFMG44 && 546& 19 & 291& 113 & 503&73   &  2.3 & 0.5 &         -3.718&0.023     & -5.372&0.022 & 0.561&0.021 & 25\\
UFMG45 && 488&38 & 451&  136 & 843&148 &   6.5 & 1.9 &       -2.548&0.009    & -2.908&0.013 & 0.314&0.020 & 28\\
UFMG46 && 408&31 & 77&    14 & 496&82    &  4.8 & 1.0 &           -1.718&0.038      & 0.526&0.027 & 0.350&0.025 & 24\\
UFMG47 && 333&19 & 59&      10 & 610&  105   & 7.8 & 2.0 &         -0.071&0.028    & -2.073&0.015 & 0.323&0.015 & 60\\
UFMG48 && 654&19 & 276&   64 & 735&100   &  10.8 & 2.5 &        -4.311&0.017     & -1.866&0.018 & 0.321&0.012 & 98\\
UFMG53 && 333& 19 & 183& 29 & 715&   87    & 9.6 & 1.8      &   -8.707&0.030      & -0.925&0.024   & 0.343&0.016 & 50\\
UFMG54 && 433& 19 & 75&   6 & 551&  38  &  10.2 & 1.6 &         -7.633&0.013    & -2.069& 0.017   & 0.274&0.013 & 71\\
UFMG55 && 883& 47 & 211&41 & 1385&353 & 19.4 & 6.1 &            -4.645&0.022    & -0.931&0.012   & 0.280&0.008 & 173\\
UFMG58 && 433& 19 & 47&   6 & 454&57 &   13.9 & 2.6 &       -2.953&0.029      & 3.356&0.027 & 0.180&0.013 & 75\\
UFMG59 && 521& 31 & 228&32 & 1005&128 & 6.5 & 1.5 &             -6.208&0.025     & 3.838&0.035 & 0.732&0.023 & 22\\
UFMG60 && 379& 31 & 164& 38 & 895&212 &  9.9 & 3.0 &            -3.852&0.016      & -2.519&0.023 & 0.428&0.018 & 37\\
UFMG61 && 340& 22 & 101& 25 & 798&256 &   9.5 & 3.8 &           -4.208&0.030        & -0.248&0.025 & 0.395&0.025 & 20\\
\hline
867* &UFMG4 & 808& 19 & 147& 6 & 1039& 73 & 10.1 & 1.0&               0.378&  0.011 & 0.779& 0.016     & 0.598& 0.015 & 47\\
145 &UFMG5 & 1067& 19 & 373& 34 & 999& 63 &  6.8 & 1.0 &            -2.137& 0.012 & -5.427& 0.008 & 0.509& 0.010 & 119\\
860* &UFMG6 & 671& 19 & 275& 18 & 806& 35 &  8.2 & 1.0 &           -0.679& 0.010      & -2.192& 0.010   & 0.304& 0.009 & 162\\
UBC537 &UFMG7 & 521& 31 & 122& 25 & 778& 179 &  7.5 & 2.1  &              -2.438& 0.016      & -2.019& 0.015 & 0.298& 0.012 & 77\\
866 &UFMG8 & 558& 7 & 224& 26 & 654& 44 &  7.6 & 1.2 &                2.897& 0.016            & 0.443& 0.016   & 0.395& 0.012 & 70\\
UBC336 &UFMG9 & 458& 26 & 122& 14 & 737& 88 &  6.8 & 1.3 &                0.659& 0.022         & 0.115& 0.015   & 0.337& 0.012 & 72\\
2094 &UFMG10 & 808& 19 & 223& 22 & 925& 83 &  7.4 & 1.1 &             -3.749& 0.024 & 1.250& 0.012     & 0.524& 0.013 & 81\\
2210 &UFMG12 & 558& 7 & 278& 38 & 1177& 165 &  7.5 & 1.5 &             0.018& 0.014     & -1.269& 0.017& 0.695& 0.016 & 44\\
140* &UFMG13 & 333& 19 & 103& 18 & 703& 137 &  7.5 & 1.7 &           -5.359&0.014       & -4.424&0.024  & 0.267&0.013 & 66\\
436* &UFMG14 & 458& 26 & 71&   11 & 654& 109 &  6.3 & 1.2 &             -2.446&0.017     & -3.940&0.015    & 0.272&0.015 & 51\\
UBC322 &UFMG15 & 654& 19 & 295& 33 & 942& 91 &    10.0 & 1.5 &            -0.867&0.025     & -2.036&0.040    & 0.377&0.011 & 112\\
UBC303 &UFMG17 & 546& 19 & 331& 49 & 876& 91 &   6.7 & 1.1 &            -4.866&0.016     & -3.778&0.036  & 0.508&0.014 & 86\\
861 &UFMG18 & 408& 31 & 59& 7 & 693& 86 &   11.7 & 2.2 &             -2.583&0.013       & -3.928&0.019  & 0.260&0.011 & 107\\
2100 &UFMG19 & 654& 19 & 251& 57 & 907& 193 &  6.4 & 1.6 &            -3.387&0.012     & -3.488&0.011 & 0.463&0.012 & 83\\
438* &UFMG20 & 521& 31 & 160& 19 & 736& 107 &  5.8 & 1.2 &          -1.558&0.012     & -3.106&0.011  & 0.370&0.008 & 185\\
UBC319 &UFMG21 & 1196&  38 & 180& 96 & 615& 269 &  3.0 & 1.4 &         -0.737&0.013    & -2.714&0.013 & 0.709&0.011 & 96\\
UBC326 &UFMG23 & 521& 31 & 257&  39 & 824& 95 & 6.1 & 1.3 &               -1.120&0.018        & -2.786&0.016  & 0.579&0.014 & 57\\
UBC136 &UFMG25 & 717& 31 & 535&93 & 835&56 &   3.8 & 0.7 &              -0.036&0.015         & -3.270&0.016    & 0.690&0.013 & 65\\
UBC121 &UFMG26 & 408& 31 & 183&36 & 884&186 &  4.1 & 1.0 &            -1.242&0.028       & -3.368&0.019 & 0.578&0.017 & 40\\
466* &UFMG27 & 408& 31 & 95&  13 & 515&62 &  4.1 & 0.8&            -1.030&0.026            & -3.697&0.022 & 0.236&0.016 & 55\\
UBC550 &UFMG28 & 750& 38 & 132&19 & 445&48 &  4.1 & 0.9 &            -2.583&0.012         & -4.144&0.011 & 0.358&0.014 & 57\\
1624 &UFMG31 & 783& 31 & 304& 19 & 1939&332 &  15.6 & 3.9 &               1.715&0.011     & -0.792&0.010   & 0.477&0.009 & 167\\
UBC535 &UFMG32 & 550& 66 & 136& 40 & 323&61 &  2.3 & 0.6 &               -3.980&    0.021       & -4.467&0.018 & 0.253&0.015 & 59\\
UBC572 &UFMG33 & 458& 26 & 121& 19 & 588&78 &  6.0 & 1.4 &            0.936&0.032          & -0.767&0.045 & 0.629&0.023 & 22\\
623 &UFMG35 & 885& 48 & 195&   79 & 851&372 &  3.9 & 1.8 &             -1.903&  0.029      & -4.466&0.028 &  1.109& 0.021 & 27\\
UBC533 &UFMG36 & 758& 63 & 294&   72 & 1323&400 & 7.03 & 2.5 &             -1.671&0.026   & -1.346&0.022  & 0.804&0.016 & 41\\
714* &UFMG40 & 654& 19 & 109& 9 & 643&58     &  7.0 & 0.9 &            -2.003&0.011       & -2.646&0.014 & 0.318&0.010 & 134\\
UBC295& UFMG49 & 262&32 & 130& 26 & 1038&268  &   8.8 & 2.8 &           -4.468&0.024   & -1.556&0.027 & 0.554&0.02 & 35\\
UBC293 &UFMG50 & 808&  19 & 168& 39 & 693&167   &  9.3 & 2.8 &             -4.610&0.015       & -1.645&0.013  & 0.294&0.007 & 244\\
2156* &UFMG51 & 717& 31 & 476& 46 & 971&  47    &  7.6 & 1.2 &          -1.884&0.014      & -5.407&0.010   & 0.598&0.010 & 112\\
1204* &UFMG52 & 1204& 43 & 619&   29 & 1692&50   &  10.3 & 2.4 &              -0.735&0.007    & -1.262&0.007  & 0.608&0.006 & 374\\
UBC522 &UFMG56 & 808& 19 & 471& 42 & 1498&114 &    15.2 & 3.0 &          -4.446&0.012    & -1.131&0.010   & 0.387&0.010 & 134\\
UBC552 &UFMG57 & 521& 31 & 165& 24 & 1164&195 &   14.2 & 3.5 &            -0.178&0.018    & -0.333&0.015 & 0.392&0.014 & 64\\
676* &UFMG62 & 494& 19 & 112& 9 & 533&41     & 5.9 & 0.7 &            -5.871&0.015       & 5.166&0.015 & 0.303&0.014 & 57\\
\hline
\end{tabular}
$\dagger$ the first part of the Table corresponds to OCs discovered in this work \\
$*$ relative to low confidence OC candidate of LP2019

\end{table*}

\subsection{Assessing membership}
\label{sect:sec4}
Most of our cluster candidates are projected against dense stellar fields. For this reason, in order to derive membership likelihoods to the stars, we applied a routine described in \cite{2019A&A...624A...8A} that evaluates statistically the overdensity of cluster 
stars in comparison to those in a nearby field in the 3D astrometric space ($\mu_{\alpha}^{*}$, $\mu_{\delta}$, $\varpi$).   
We call the region centred in the cluster's equatorial coordinates and restricted by its $r_{lim}$ as the cluster region. The comparison field is restricted by a ring-like region with inner radius  $r_{lim}$+3\arcmin. The outer radius is such that the field area is equal to 3 times the cluster region.
 
In order to improve the decontamination algorithm performance and minimize the presence of outliers (i.e., small groups of field stars with non-zero memberships), we build a box-like filter in the astrometric space centred in the peak values of $\mu_{\alpha}^{*}$, $\mu_{\delta}$ and $\varpi$ computed in the Sec 2.
We restricted the sample in box with width of 3\,mas\,yr$^{-1}$ for $\mu_{\alpha}^{*}$ and $\mu_{\delta}$ and 1 mas for $\varpi$.
 
The 3D astrometric space is divided in cells with widths of 1.0\,mas\,yr$^{-1}$, 1.0\,mas\,yr$^{-1}$ and 0.1\,mas in $\mu_{\alpha}^{*}$,$\mu_{\delta}$ and $\varpi$, respectively. These cell dimensions are large enough to accommodate a significant number of stars, but small enough to detect local fluctuations in the stellar density through the whole data domain.

A membership likelihood ($l_{\textrm{star}}$) is computed for cluster stars within each cell by using normalized multivariate Gaussian distributions, which formulation \citep[see eq. 1,][]{2019A&A...624A...8A} incorporates the uncertainties and correlations between the astrometric parameters. The same procedure is performed for stars in the comparison field. The degree of similarity between the two groups (cluster and field stars) within a given 3D cell is established by using an \textit{entropy-like} function $S$ \citep[see eq. 3,][]{2019A&A...624A...8A}. Stars within cells for which $S_{\textrm{cluster}}<S_{\textrm{field}}$ are considered possible members.

For cluster stars within those cells, we additionally computed an exponential factor ($L_{\textrm{star}}$) which evaluates the overdensity of cluster stars in the 3D space relatively to the whole grid configuration \citep[see eq. 4,][]{2019A&A...624A...8A}. This procedure was adopted to ensure that only significant local overdensities, statistically distinguishable from the distribution of field stars, would receive apreciable membership likelihoods.

\begin{figure*}
\centering
\includegraphics[width=0.34\linewidth]{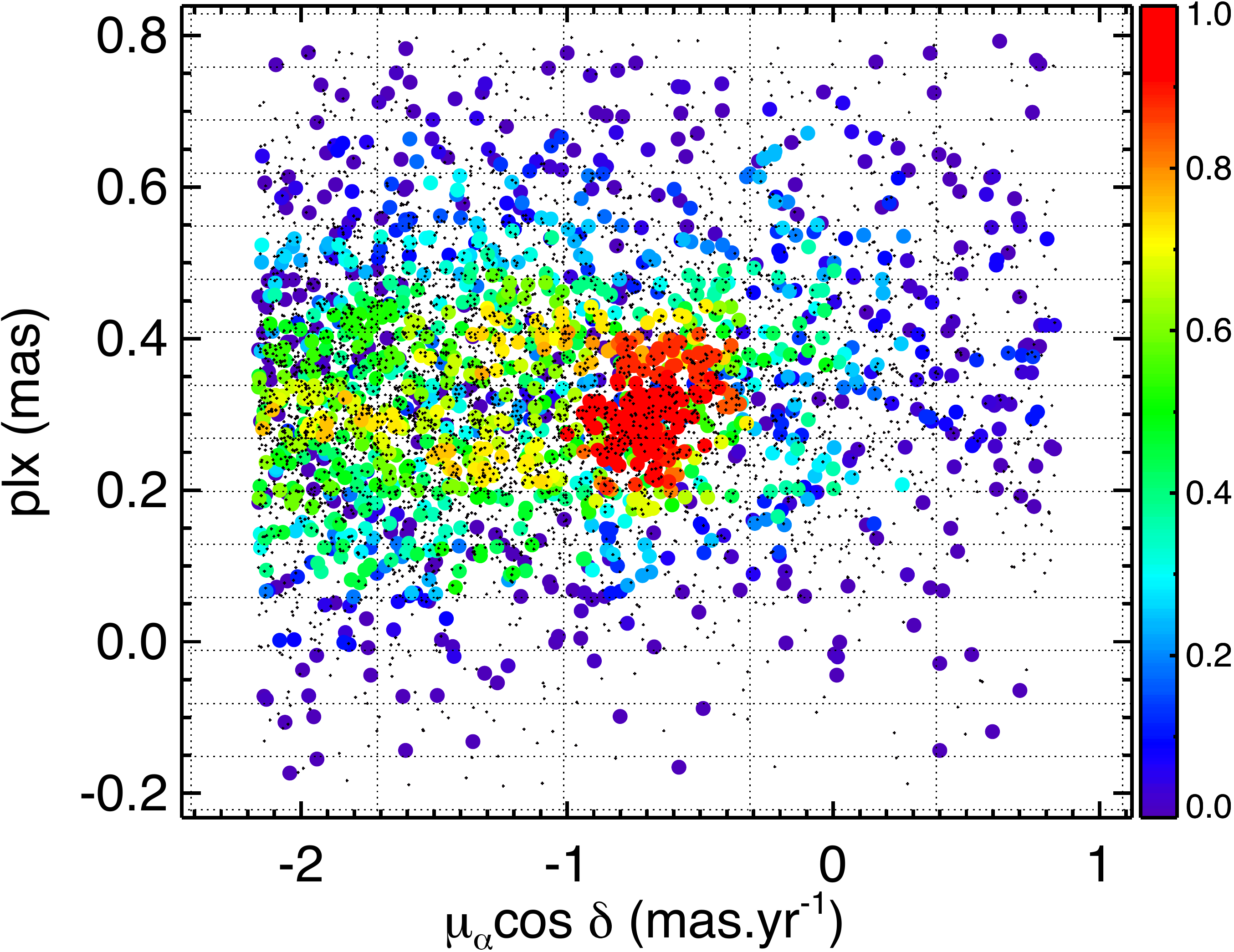} \hspace{0.1cm}
\includegraphics[width=0.34\linewidth]{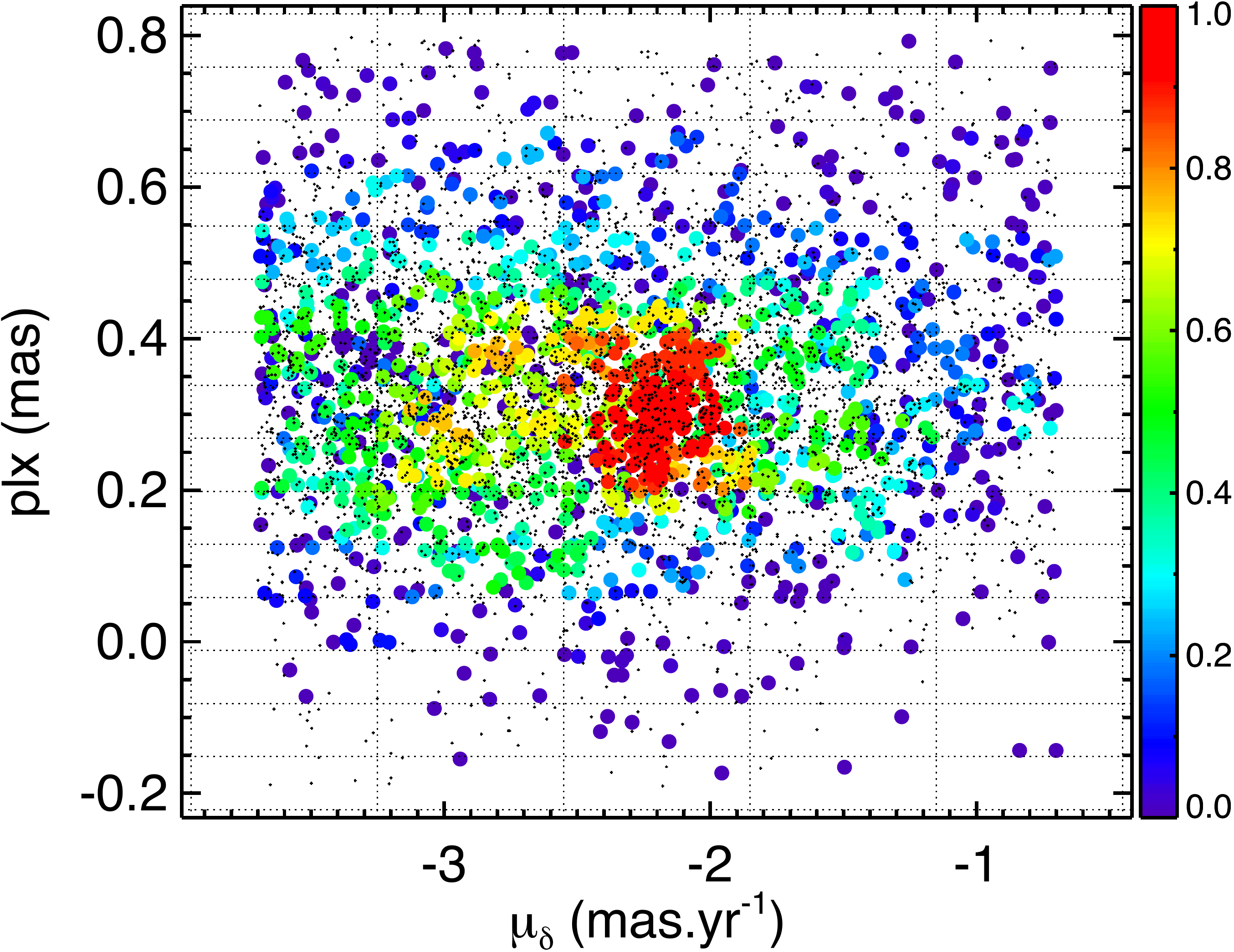} \hspace{0.1cm}
\includegraphics[width=0.272\linewidth]{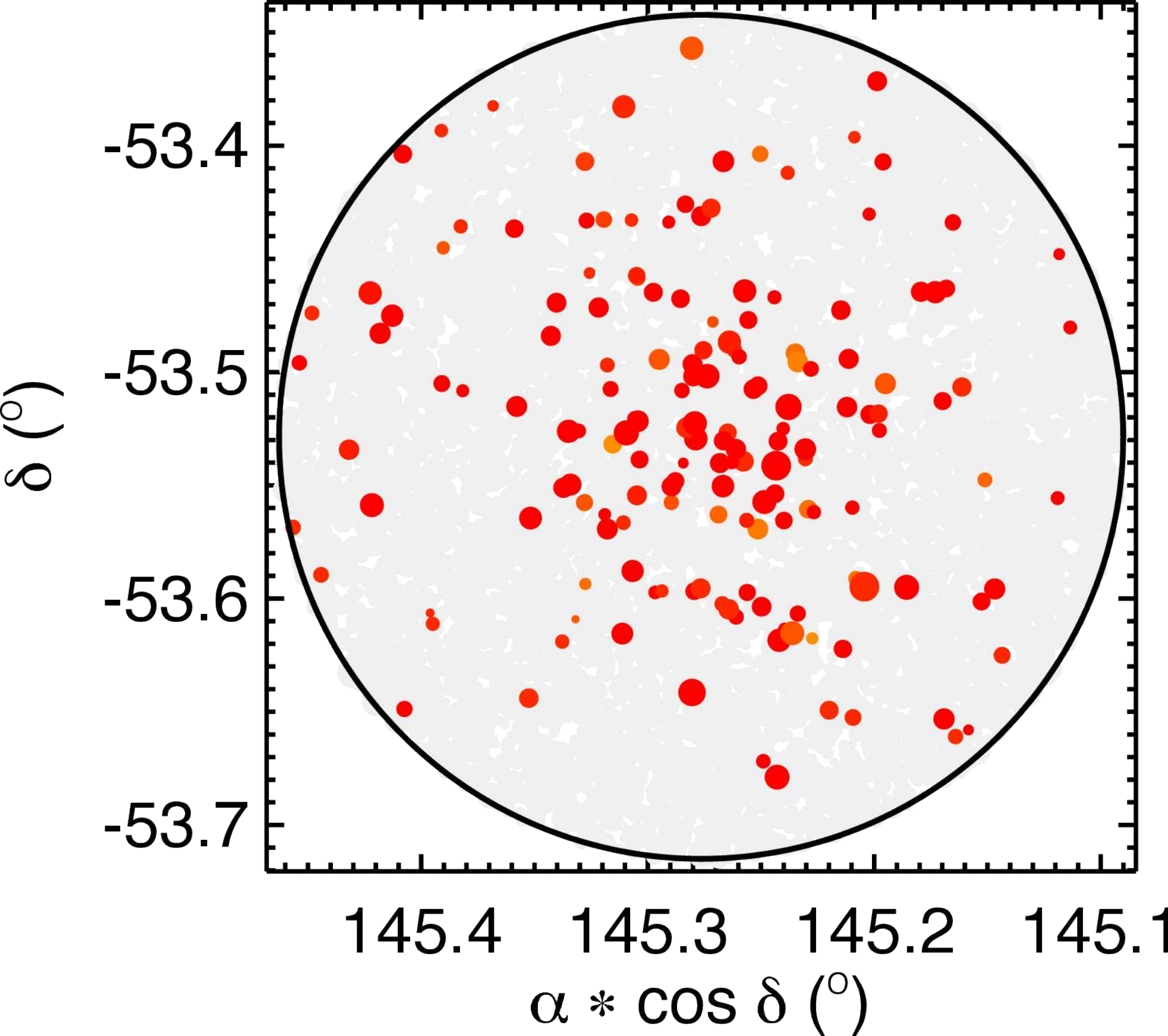}
\caption{A sequence of panels showing the results of the decontamination procedure applied to UFMG 6. It is possible to note concentrations in all astrometric space. Left: proper motion in right ascension versus parallax. Middle: proper motion in declination versus parallax Right: the spatial distribution of the most probable cluster members.} 
\label{fig:decont}
\end{figure*}

Finally, cell sizes are increased and decreased by 1/3 of their original sizes in each dimension and the procedure is repeated. The final likelihood for a given star corresponds to the median of the set of $L_{\textrm{star}}$ values taken through the whole grid configurations. Fig.~\ref{fig:decont} exemplifies the application of such procedure to the cluster UFMG6.

To build the final members sample, we adopted a membership probability cut  by keeping only stars with membership probabilities higher than 70$\%$. This is important to clean the sample of
less probable members. It makes evident the presence of the clusters stars loci in the CMDs and the existence of a clump in the 3D astrometric space  ($\mu_{\alpha}^{*}$,$\mu_{\delta}$ and $\varpi$), revealing 
the cluster nature. 

\subsection{Isochrone fittings}

We employed solar metallicity PARSEC-COLIBRI models \citep{Marigo:2017} 
to perform isochrone fittings on the decontaminated samples (i.e., stars with likelihoods > 70\%) to determine age, distance and reddening. We adopted a reddening law \citep{Cardelli:1989,Odonnell:1994} to convert $E(G_{BP}-G_{RP})$ to $E(B-V)$. 
Using a set of several solar metallicity isochrones covering a range of ages, we visually inspected the match between the model and the cluster stars loci, ensuring a good fitting in specific evolutionary regions
such as the main sequence, the turnoff, and the giant clump. To associate the uncertainties, we changed the reddening and distance modulus $(m-M)$ simultaneously searching for the maximum acceptable deviations from the best values fitted.

\begin{figure*}
\centering
\includegraphics[width=0.29\linewidth]{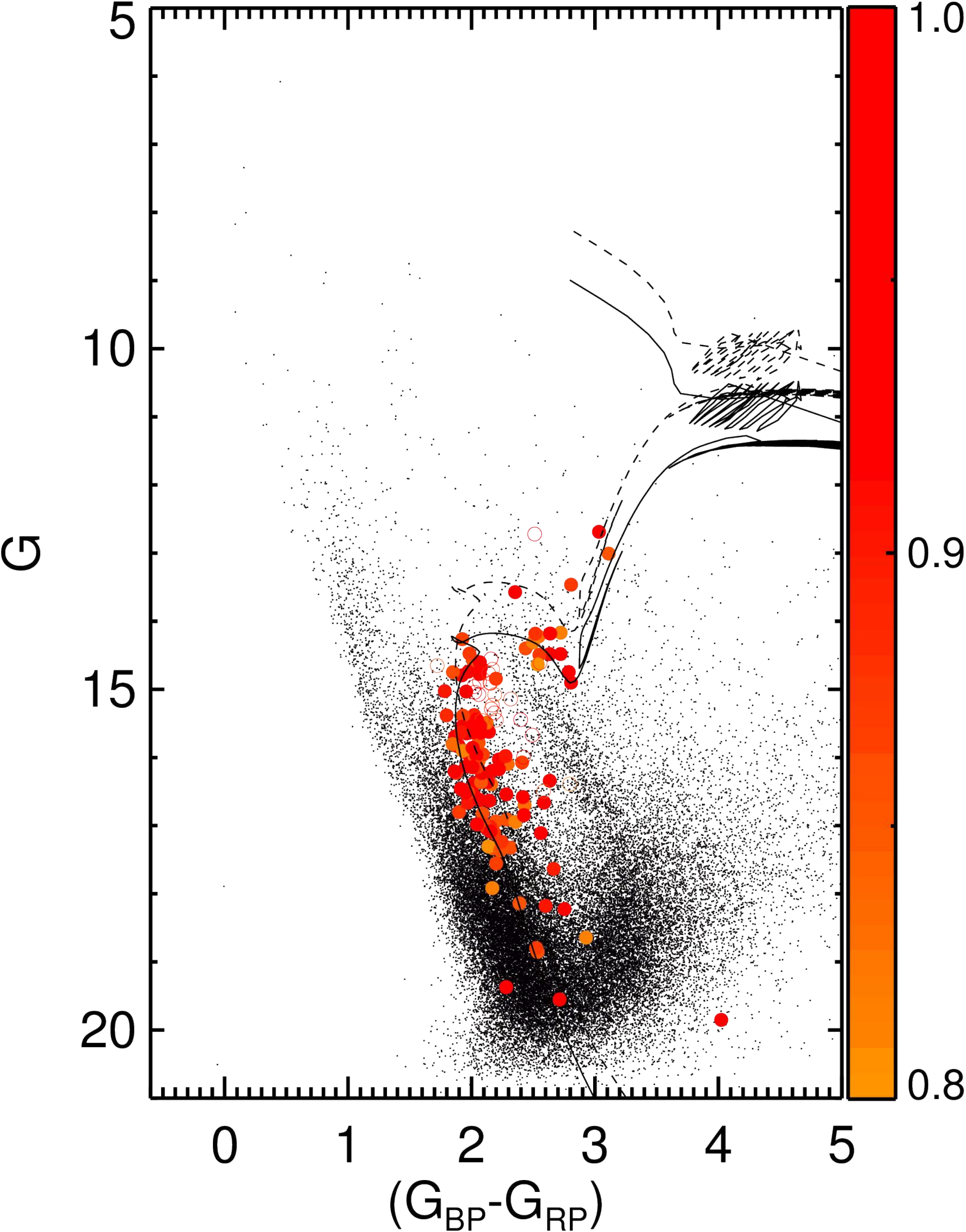} \hspace{0.25cm}
\includegraphics[width=0.29\linewidth]{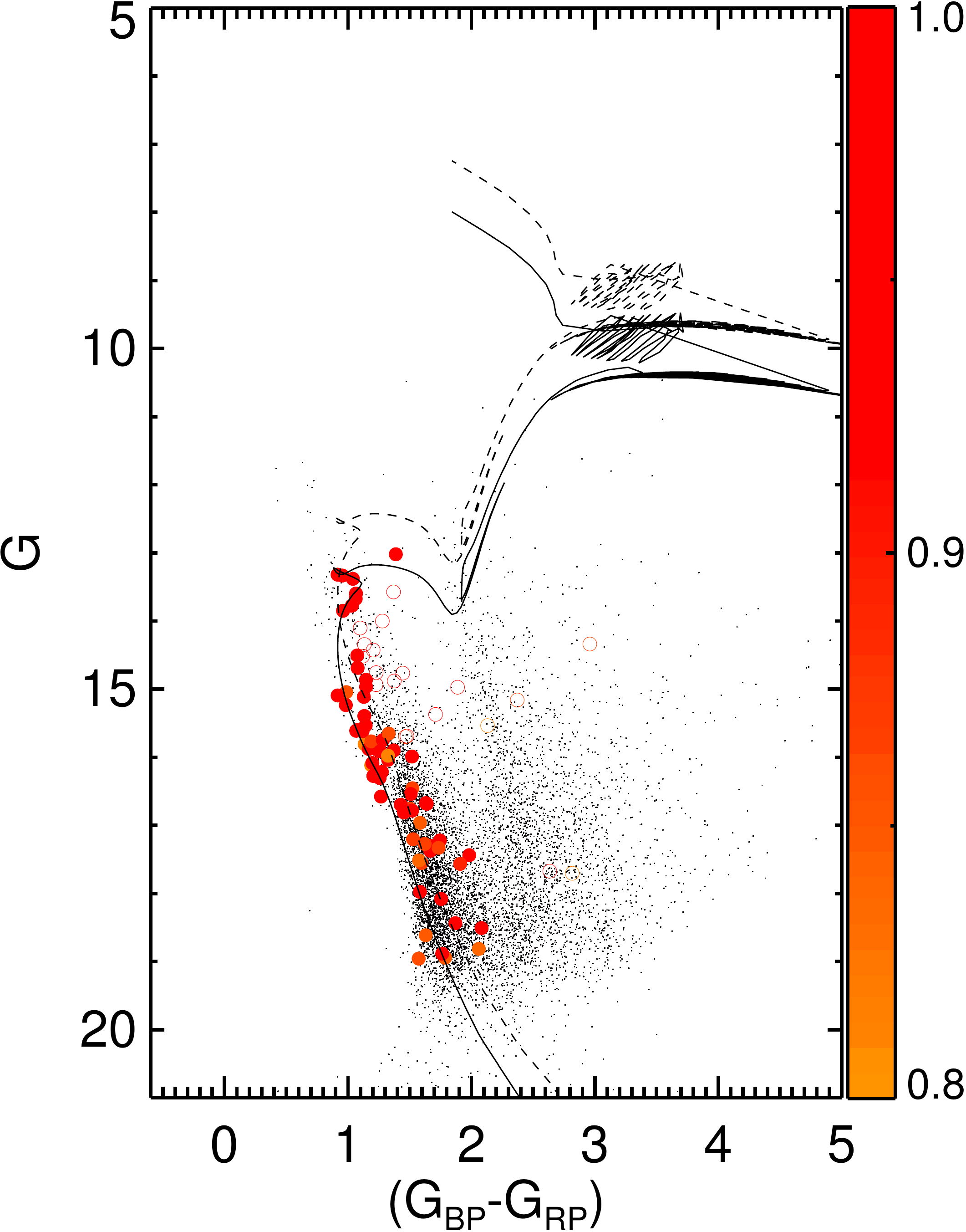}  \hspace{0.25cm}
\includegraphics[width=0.29\linewidth]{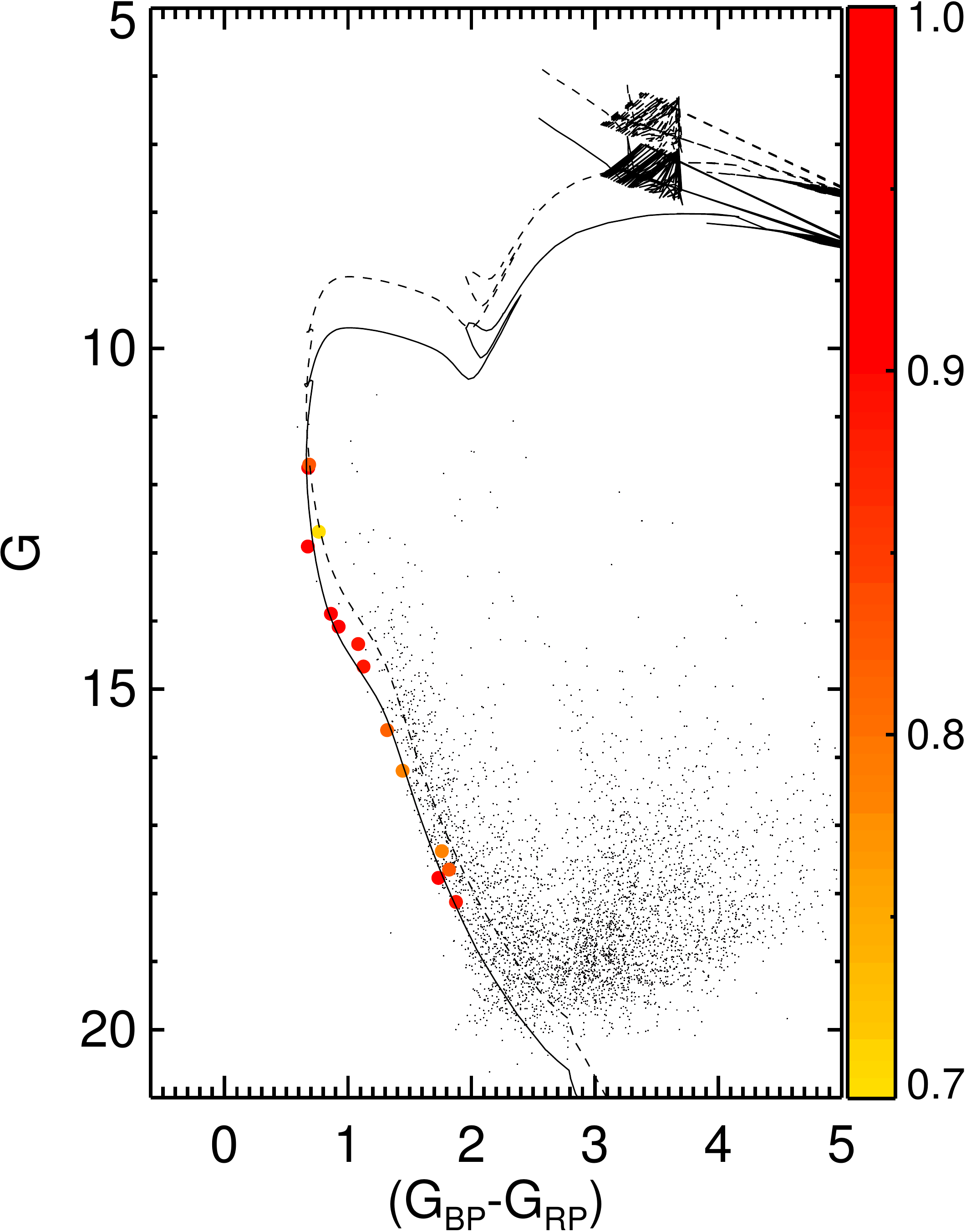}\\ \vspace{0.25cm}
\includegraphics[width=0.29\linewidth]{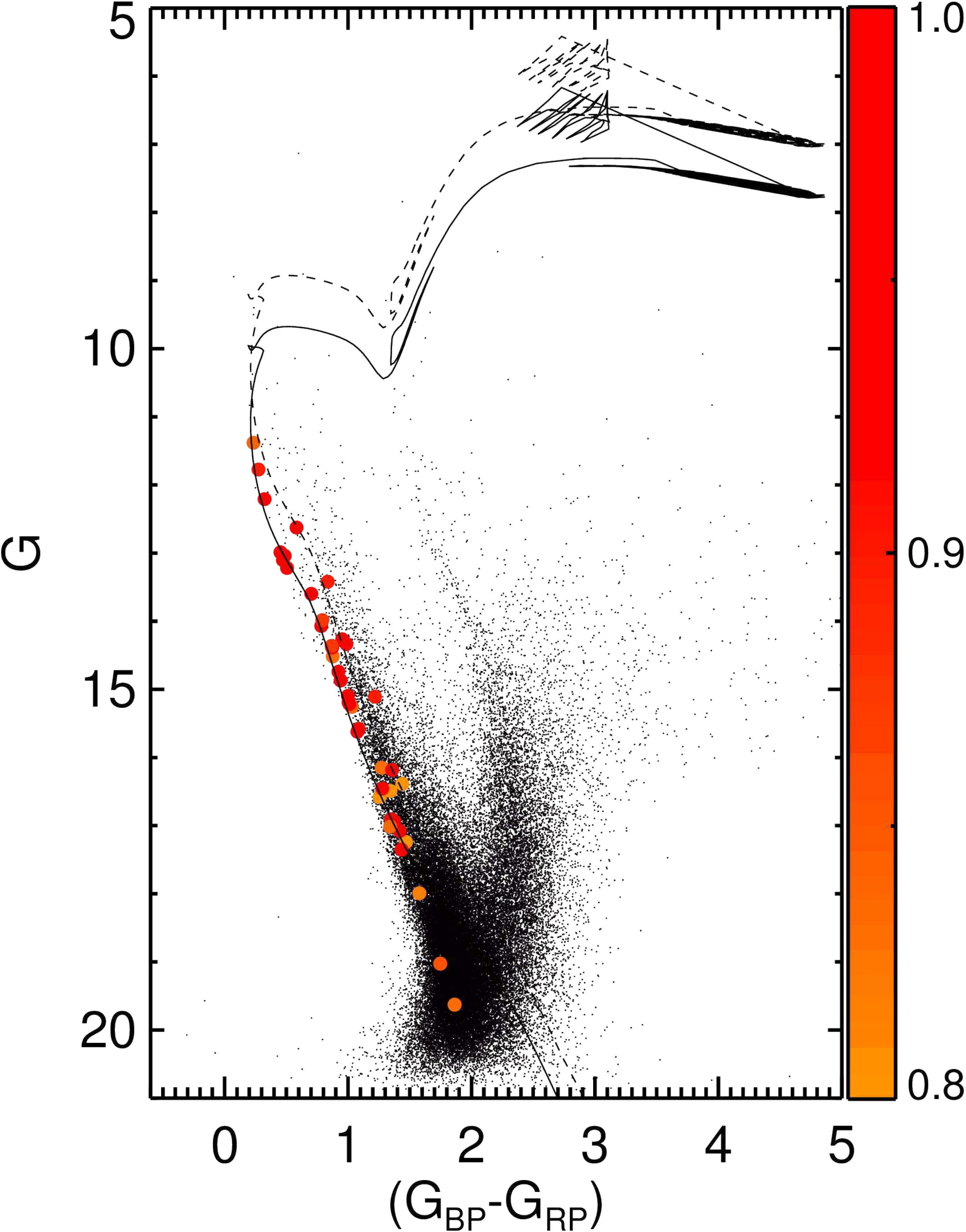} \hspace{0.25cm}
\includegraphics[width=0.29\linewidth]{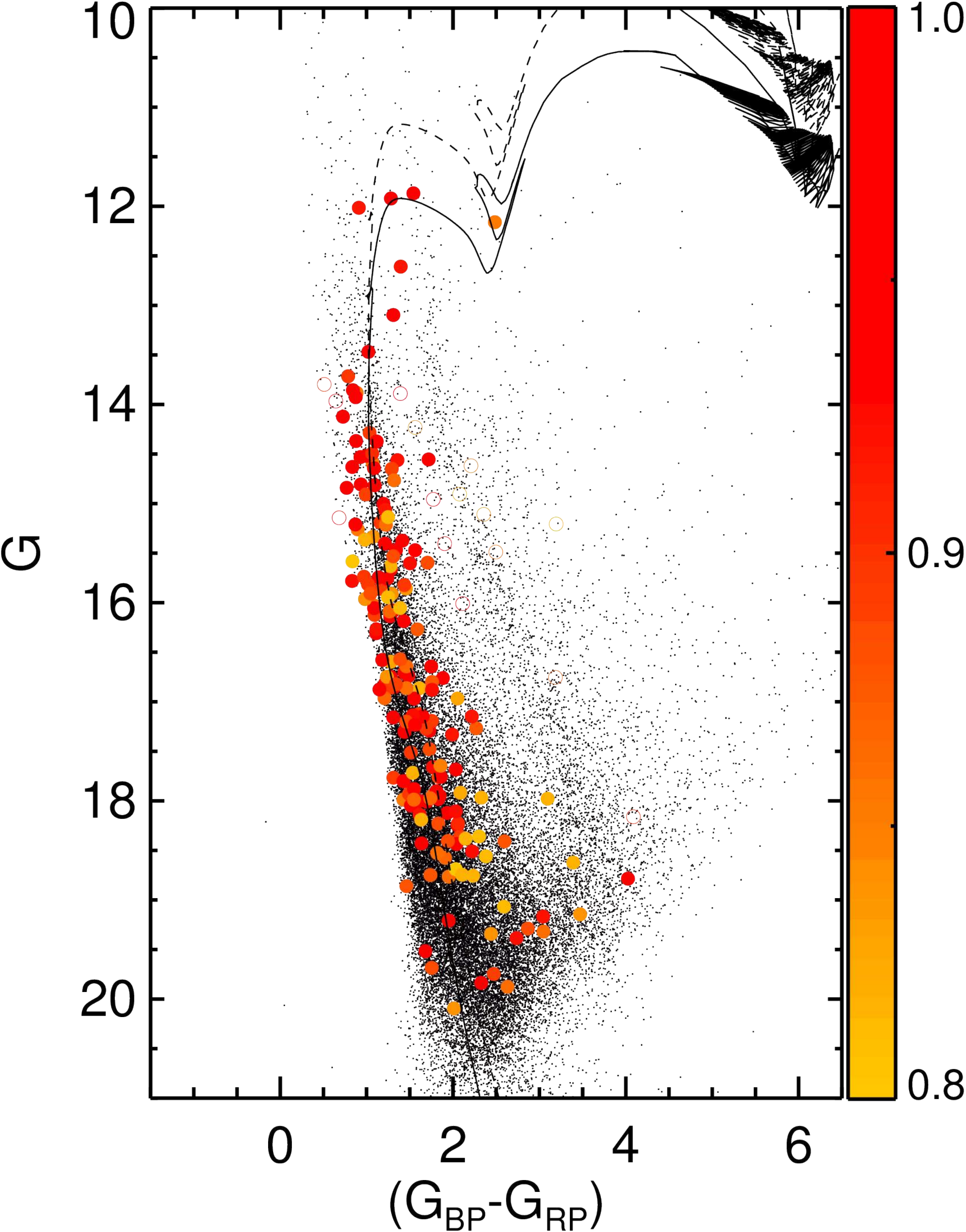} \hspace{0.25cm}
\includegraphics[width=0.29\linewidth]{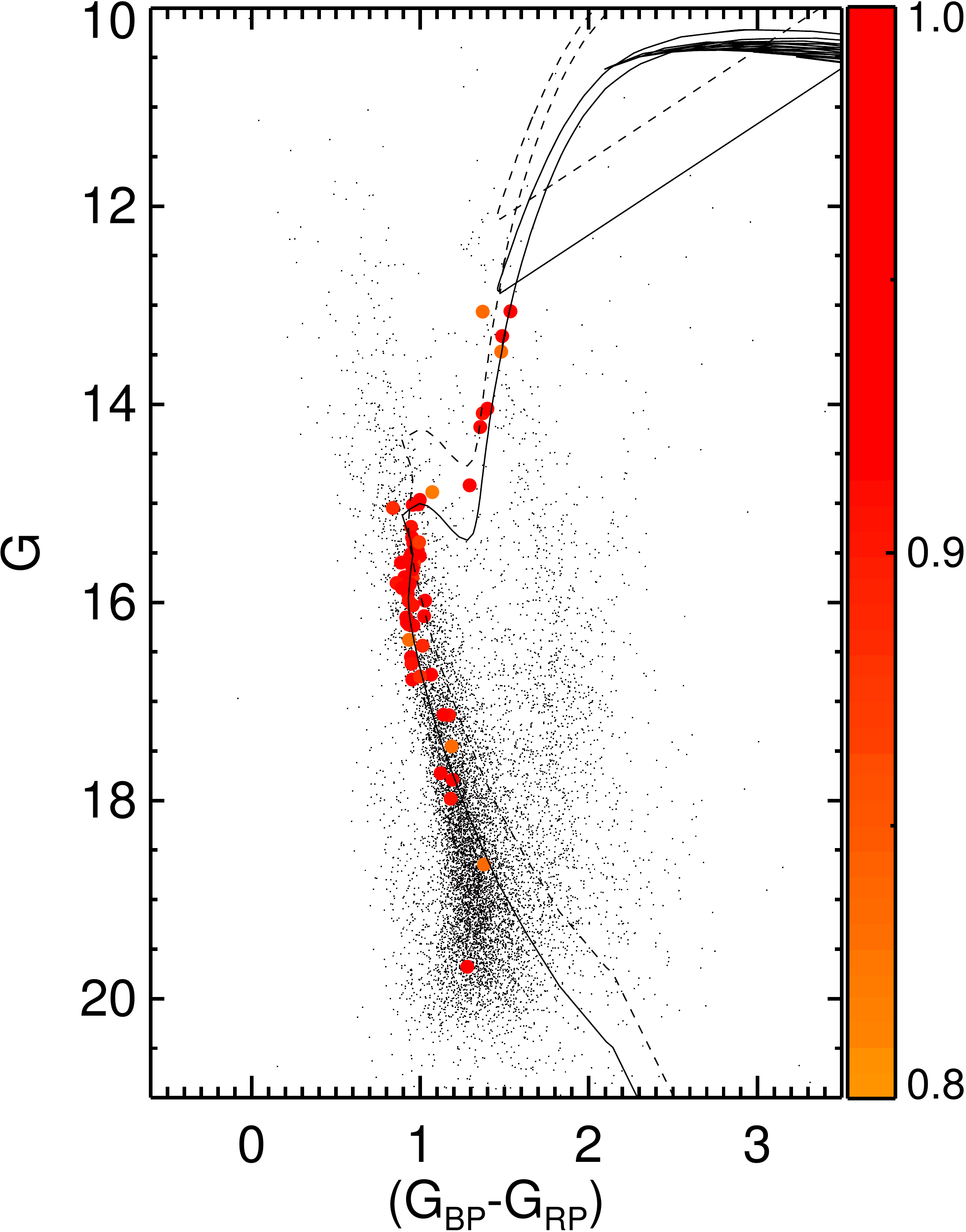} 
\caption{Examples of PARSEC-COLIBRI isochrone fitted (solid line) over
the cleaned CMD for UFMG5 (top left), UFMG13 (top middle), UFMG24 (top right), UFMG36 (bottom left), UFMG55 (bottom middle) and UFMG62 (bottom right). We overplotted the corresponding binary sequence (dashed lines) by deducing 0.75 mag from the G magnitude values.} 
\label{fig:isoc1}
\end{figure*}

The resulting values of the colour excess, distance module and age for all the clusters are present in the Table \ref{Tab:clusters_prop1}. The full table containing the astrophysical parameters and member lists derived for the clusters stars in this work are available as electronic tables through Vizier. Fig.~\ref{fig:isoc1} shows the best-fitting isochrones to the field-cleaned CMDs of some of the clusters investigated.

 \citet{2020arXiv200407274C} have obtained astrophysical parameters for 1867 open clusters previously identified with Gaia DR2 astrometry, including many  discovered open clusters present in LP2019 and CG2020. Their reported parameters (private communication) for the 29 clusters in common with our sample tend to agree with our values for close (d < 2 kpc), relatively low reddening (E(B-V) < 0.9) and relatively old (log(t) > 8.8) or young (log(t) < 8.0) clusters. There are a few discrepancies which we attribute to the very reddened, and scarce nature of some clusters, often lacking a sizeable population of giant stars needed to properly constrain their astrophysical parameters.

\begin{table*}
\caption{Astrophysical parameters of the investigated clusters}
\label{Tab:clusters_prop1}
\begin{tabular}{ll l l r r r@{$\,\pm\,$}l r@{$\,\pm\,$}l r@{$\,\pm\,$}l  r@{$\,\pm\,$}l}
\hline
\multicolumn{1}{|c|}{Name$^\dagger$} &
\multicolumn{1}{|c|}{Internal id$^\dagger$} &
\multicolumn{1}{c|}{$RA$} &
\multicolumn{1}{c|}{$DEC$} &
\multicolumn{1}{c|}{$l$} &
\multicolumn{1}{c|}{$b$} &
\multicolumn{2}{c|}{$Dist$} &
\multicolumn{2}{c|}{$(m-M)$} &
\multicolumn{2}{c|}{$log(t)$} &
\multicolumn{2}{c|}{$E(B-V)$} \\

\multicolumn{1}{|c|}{} &
\multicolumn{1}{|c|}{} &
\multicolumn{1}{c|}{(hh:mm:ss)} &
\multicolumn{1}{c|}{(dd:mm:ss)} &
\multicolumn{1}{c|}{$(deg)$} &
\multicolumn{1}{c|}{$(deg)$} &
\multicolumn{2}{c|}{$(pc)$} &
\multicolumn{2}{c|}{(mag)} &
\multicolumn{2}{c|}{} &
\multicolumn{2}{c|}{(mag)} \\
\hline
UFMG11 && 16:13:58.80 & -52:57:48.1 & 330.694 & -1.375 & 2455&340 & 11.95&0.30 & 8.40&0.15 & 0.67 &0.05\\
UFMG16 && 16:20:16.66 & -44:52:24.5 & 337.071 & 3.713 & 1148&132 & 10.30&0.25 & 8.60&0.50 & 0.50&  0.10\\
UFMG22 && 17:32:47.82 & -37:36:42.6 & 351.230 & -2.355 & 1514&210 & 10.90&0.30 & 9.00&0.10 & 0.92&0.10\\
UFMG24 && 17:12:02.31 & -35:56:03.4 & 350.254 & 2.019 & 1259&233 & 10.50&0.40 & 8.20&0.40 & 0.65  &0.10\\
UFMG29 && 17:15:15.21 & -36:34:42.7 & 350.112 & 1.117 & 1380&256 & 10.70&0.40 & 8.70&0.15 & 0.77&  0.10\\
UFMG30 && 17:11:14.71 & -48:45:11.0 & 339.800 & -5.418 & 1175&218 & 10.35&0.40 & 7.90&0.40 & 0.55&0.15\\
UFMG34 && 15:35:21.36 & -57:39:00.0 & 323.482 & -1.472 & 1380&256 & 10.70&0.40 & 9.25&0.10 & 0.95  &0.05\\
UFMG37 && 15:20:36.04 & -57:44:20.9 & 321.801 & -0.439 & 912&169 & 9.80&0.40 & 8.85&0.15 & 0.86  &0.10\\
UFMG38 && 16:05:59.49 & -52:39:57.1 & 330.009 & -0.338 & 2042&283 & 11.55&0.30 & 8.90&0.20 & 1.35  &0.20\\
UFMG39 && 14:36:02.40 & -57:51:54.0 & 316.500 & 2.225 & 1259&233 & 10.50&0.40 & 8.75&0.20 & 0.50  &0.15\\
UFMG41 && 15:40:30.83 & -55:46:12.3 & 325.164 & -0.375 & 1445&268 & 10.80&0.40 & 8.20&0.40 & 1.26&0.10\\
UFMG42 && 15:44:08.55 & -55:15:22.9 & 325.884 & -0.275 & 5012&1164 & 13.50&0.50 & 7.95&0.15 & 2.20&  0.15\\
UFMG43 && 15:42:52.39 & -55:25:14.9 & 325.641 & -0.296 & 1698&315 & 11.15&0.40 & 8.50&0.20 & 1.35&0.10\\
UFMG44 && 15:42:19.39 & -55:27:50.9 & 325.552 & -0.284 & 933&173 & 9.85&0.40 & 8.95&0.10 & 0.22&  0.05\\
UFMG45 && 15:59:09.94 & -55:47:56.6 & 327.193 & -2.036 & 1585&368 & 11.00&0.50 & 8.80&0.20 & 0.50  &0.15\\
UFMG46 && 18:18:11.93 & -25:04:13.5 & 6.949 & -4.399 & 1995&277 & 11.50&0.30 & 8.90&0.15 & 0.61&0.15\\
UFMG47 && 16:52:08.81 & -47:24:19.9 & 338.916 & -2.070 & 2630&487 & 12.10&0.40 & 8.25&0.10 & 1.45  &0.10\\
UFMG48 && 14:09:53.31 & -59:18:59.6 & 312.801 & 2.044 & 3020&559 & 12.40&0.40 & 8.35&0.35 & 1.10  &0.15\\
UFMG53 && 13:01:35.28 & -64:01:28.2 & 304.043 & -1.175 & 2754&382 & 12.20&0.30 & 8.40&0.15 & 1.35  &0.15\\
UFMG54 && 14:02:46.63 & -61:57:13.3 & 311.212 & -0.234 & 3802&527 & 12.90&0.30 & 7.60&0.15 & 2.32&0.20\\
UFMG55 && 13:08:07.60 & -61:14:24.8 & 304.939 & 1.566 & 2884&534 & 12.30&0.40 & 8.10&0.20 & 0.95  &0.15\\
UFMG58 && 08:19:33.99 & -38:16:14.8 & 256.220 & -1.164 & 6310&874 & 14.00&0.30 & 7.95&0.10 & 1.25  &0.10\\
UFMG59 && 08:26:01.20 & -38:20:00.2 & 256.996 & -0.161 & 1337&248 & 10.63&0.40 & 7.65&0.30 & 0.50  &0.15\\
UFMG60 && 14:18:42.52 & -62:40:16.2 & 312.759 & -1.476 & 2291&424 & 11.80&0.40 & 7.65&0.40 & 0.90&0.15\\
UFMG61 && 12:21:58.58 & -63:36:33.3 & 299.665 & -0.925 & 2455&570 & 11.95&0.50 & 7.85&0.30 & 0.55  &0.10\\
\hline
867* &UFMG4 & 18:11:58.32 & -22:58:51.6 & 8.116 & -2.157 & 1995&138 & 11.50&0.15 & 8.60&0.05 & 1.20  &0.10\\
145 &UFMG5 & 17:08:00.70 & -44:07:43.1 & 343.203 & -2.229 & 1413&196 & 10.75&0.30 & 8.70&0.10 & 1.40  &0.15\\
860* &UFMG6 & 16:17:54.90 & -53:31:06.2 & 330.734 & -2.183 & 2089&241 & 11.60&0.25 & 9.00&0.20 & 0.45  &0.10\\
UBC537& UFMG7 & 16:11:33.60 & -53:02:01.3 & 330.381 & -1.176 & 1995&323 & 11.50&0.35 & 8.90&0.15 & 0.85  &0.15\\
866 &UFMG8 & 17:27:02.63 & -39:24:12.4 & 349.113 & -2.401 & 2399&332 & 11.90&0.30 & 9.30&0.10 & 0.73  &0.15\\
UBC336& UFMG9 & 17:51:57.39 & -27:51:08.2 & 1.647 & -0.636 & 1905&264 & 11.40&0.30 & 8.90&0.15 & 0.80  &0.08\\
2094 &UFMG10 & 11:41:40.42 & -61:56:45.3 & 294.829 & -0.177 & 1660&191 & 11.10&0.25 & 8.15&0.15 & 0.34  &0.05\\
2210 &UFMG12 & 16:28:49.74 & -54:57:31.6 & 330.829 & -4.330 & 1318&183 & 10.60&0.30 & 8.60&0.25 & 0.40  &0.07\\
140*& UFMG13 & 15:07:38.77 & -60:24:08.3 & 318.996 & -1.850 & 2208&255 & 11.72&0.25 & 8.80&0.10 & 0.66  &0.07\\
436* &UFMG14 & 16:44:09.11 & -44:06:33.2 & 340.537 & 1.121 & 1995&184 & 11.50&0.20 & 8.95&0.10 & 1.12&0.10\\
UBC322& UFMG15 & 17:00:39.22 & -44:10:25.1 & 342.366 & -1.207 & 2399&221 & 11.90&0.20 & 8.10&0.05 & 1.17&0.08\\
UBC303 &UFMG17 & 14:47:26.92 & -56:14:52.3 & 318.582 & 3.058 & 1585&183 & 11.00&0.25 & 8.90&0.10 & 0.50  &0.05\\
861 &UFMG18 & 16:18:48.61 & -52:53:13.7 & 331.272 & -1.825 & 3467&481 & 12.70&0.30 & 9.05&0.05 & 0.59  &0.05\\
2100 &UFMG19 & 15:10:06.76 & -60:29:38.3 & 319.213 & -2.082 & 1445&200 & 10.80&0.30 & 8.90&0.15 & 0.46&  0.10\\
438*& UFMG20 & 16:47:36.00 & -44:32:02.0 & 340.615 & 0.376 & 1622&225 & 11.05&0.30 & 9.05&0.10 & 0.84  &0.05\\
UBC319 &UFMG21 & 16:38:48.11 & -44:45:17.2 & 339.421 & 1.412 & 1000&139 & 10.00&0.30 & 8.50&0.15 & 0.50&0.10\\
UBC326 &UFMG23 & 17:12:38.47 & -36:02:57.0 & 350.233 & 1.853 & 1514&280 & 10.90&0.40 & 8.35&0.25 & 1.20&  0.10\\
UBC136& UFMG25 & 20:18:02.56 & +32:35:51.9 & 71.358 & -1.776 & 933&173 & 9.85&0.40 & 8.80&0.10 & 0.70&0.10\\
UBC121& UFMG26 & 19:15:33.59 & +12:50:25.5 & 47.085 & 0.583 & 955&132 & 9.90&0.30 & 9.00&0.10 & 0.75  &0.10\\
466* &UFMG27 & 19:08:23.98 & +10:10:38.8 & 43.912 & 0.905 & 1660&230 & 11.10&0.30 & 8.95&0.10 & 1.15&0.05\\
UBC550& UFMG28 & 16:50:24.62 & -45:04:20.0 & 340.523 & -0.352 & 1905&353 & 11.40&0.40 & 8.50&0.15 & 0.93  &0.10\\
1624 &UFMG31 & 17:46:18.23 & -29:09:03.2 & 359.894 & -0.239 & 1660&307 & 11.10&0.40 & 8.30&0.10 & 0.80&0.10\\
UBC535& UFMG32 & 15:50:14.52 & -56:44:35.8 & 325.636 & -1.971 & 1445&268 & 10.80&0.40 & 8.95&0.20 & 0.45&  0.10\\
UBC572 &UFMG33 & 18:41:40.83 & -21:58:49.6 & 12.176 & -7.790 & 2089&387 & 11.60&0.40 & 7.75&0.15 & 1.05&0.10\\
623 &UFMG35 & 15:24:36.35 & -53:58:23.2 & 324.327 & 2.411 & 955&132 & 9.90&0.30 & 7.50&0.30 & 0.27  &0.10\\
UBC533& UFMG36 & 15:21:09.36 & -53:11:24.0 & 324.326 & 3.347 & 1096&203 & 10.20&0.40 & 8.60&0.30 & 0.20&0.10\\
714* &UFMG40 & 15:57:40.50 & -57:16:35.0 & 326.077 & -3.026 & 2239&207 & 11.75&0.20 & 9.00&0.05 & 0.57&0.05\\
UBC295& UFMG49 & 13:50:18.79 & -60:25:03.1 & 310.140 & 1.627 & 1738&322 & 11.20&0.40 & 8.35&0.15 & 0.50&0.15\\
UBC293 &UFMG50 & 13:30:56.62 & -61:46:12.6 & 307.586 & 0.744 & 2754&510 & 12.20 &0.40 & 7.70&0.20 & 0.85&0.15\\
2156* &UFMG51 & 16:13:29.45 & -50:08:56.5 & 332.574 & 0.721 & 1622&246 & 11.05&0.30 & 8.85&0.10 & 0.80&0.10\\
1204* &UFMG52 & 17:42:16.89 & -30:41:22.7 & 358.127 & -0.302 & 1259&292 & 10.50&0.50 & 7.55&0.30 & 1.47&0.10\\
UBC522 &UFMG56 & 12:43:09.42 & -61:37:11.2 & 301.949 & 1.240 & 2089&387 & 11.60&0.40 & 7.10&0.30 & 0.97  &0.10\\
UBC552 &UFMG57 & 16:58:51.31 & -42:02:36.9 & 343.839 & 0.367 & 2512&465 & 12.00&0.40 & 8.00&0.15 & 1.45  &0.10\\
676* &UFMG62 & 09:56:01.78 & -60:11:44.8 & 282.728 & -4.397 & 2291&211 & 11.80&0.20 & 9.55&0.10 & 0.22  &0.05\\
\hline
\end{tabular}
$\dagger$ the first part of the Table corresponds to OCs discovered in this work \\
$*$relative to low confidence OC candidate of LP2019
\end{table*}

\section{Discussion}
\label{sect:sec6}
In this Section, we present evidence for the 25 new findings to be real physical systems and place the OCs' derived properties in the Milky Way context. As the cluster 466 has central coordinates that differ from those of UFMG27 by more than the limiting radius determined by us (see Sect. 3), we grouped UFMG27 with the 25 newly discovered clusters in the following graphical analyses.
 
 \subsection{Comparing the new discoveries with confirmed open clusters}

\begin{figure}
\includegraphics[width=0.99\linewidth]{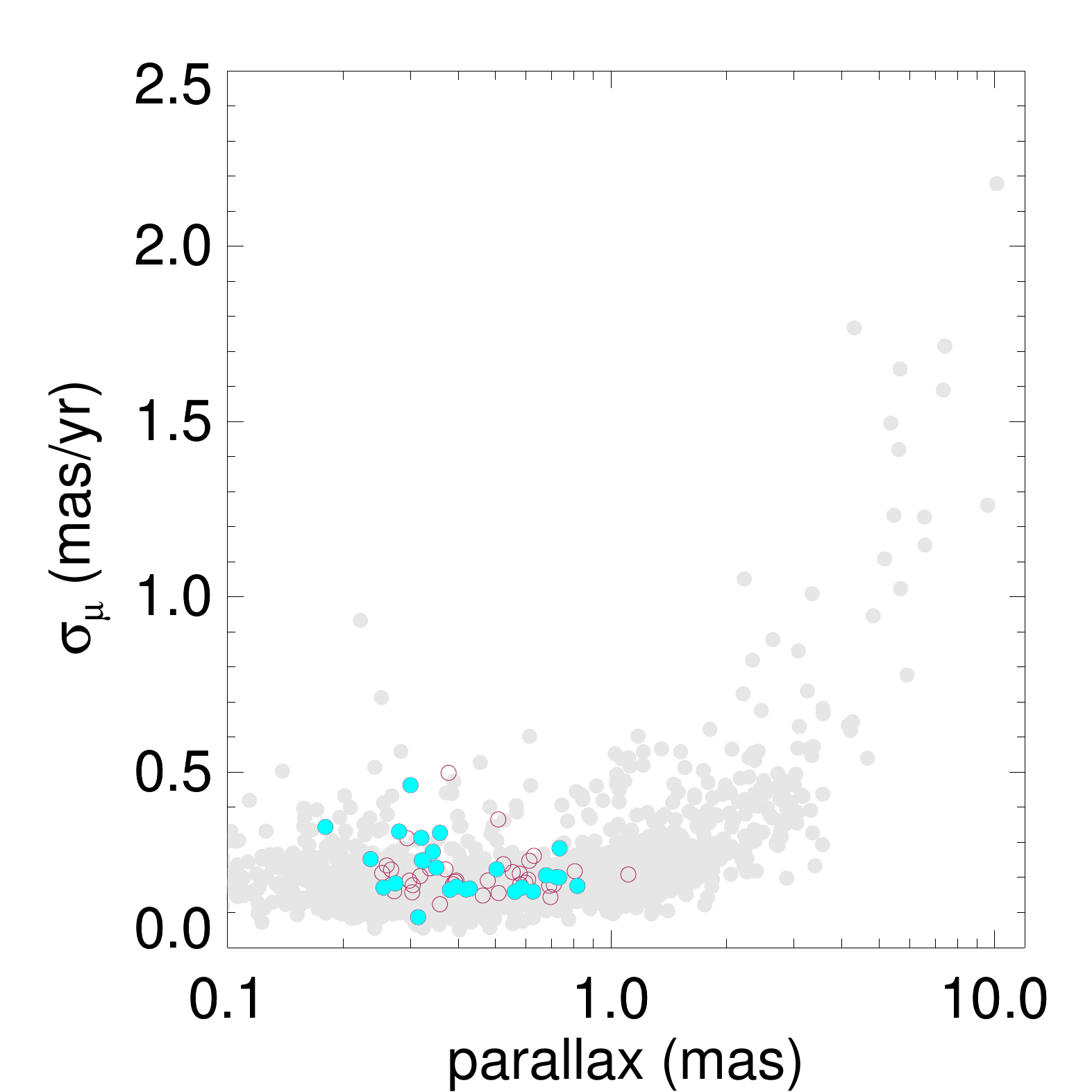}
\caption{Dispersion in proper motion versus parallax for confirmed OCs  (gray circles) compared with our sample of recently discovered OCs using \textit{Gaia} DR2 (red circles) and those reported in the present work (filled cyan circles).}
\label{fig:pmplx}
\end{figure}

The dispersion in proper motion is correlated with parallax for coherent stellar systems. Such a relationship is presented in Fig.~\ref{fig:pmplx} for the \textit{Gaia} DR2 confirmed OCs (gray circles) according with \cite{ca20}, our sample of recently discovered OCs (open red circles) and the newly discovered OCs (filled cyan circles). All OCs studied here have CMDs and structural properties compatible with real physical systems. This is corroborated by the fact that they lie in the diagram region where the bulk of confirmed OCs is.

\begin{figure}
\centering
\includegraphics[width=\linewidth]{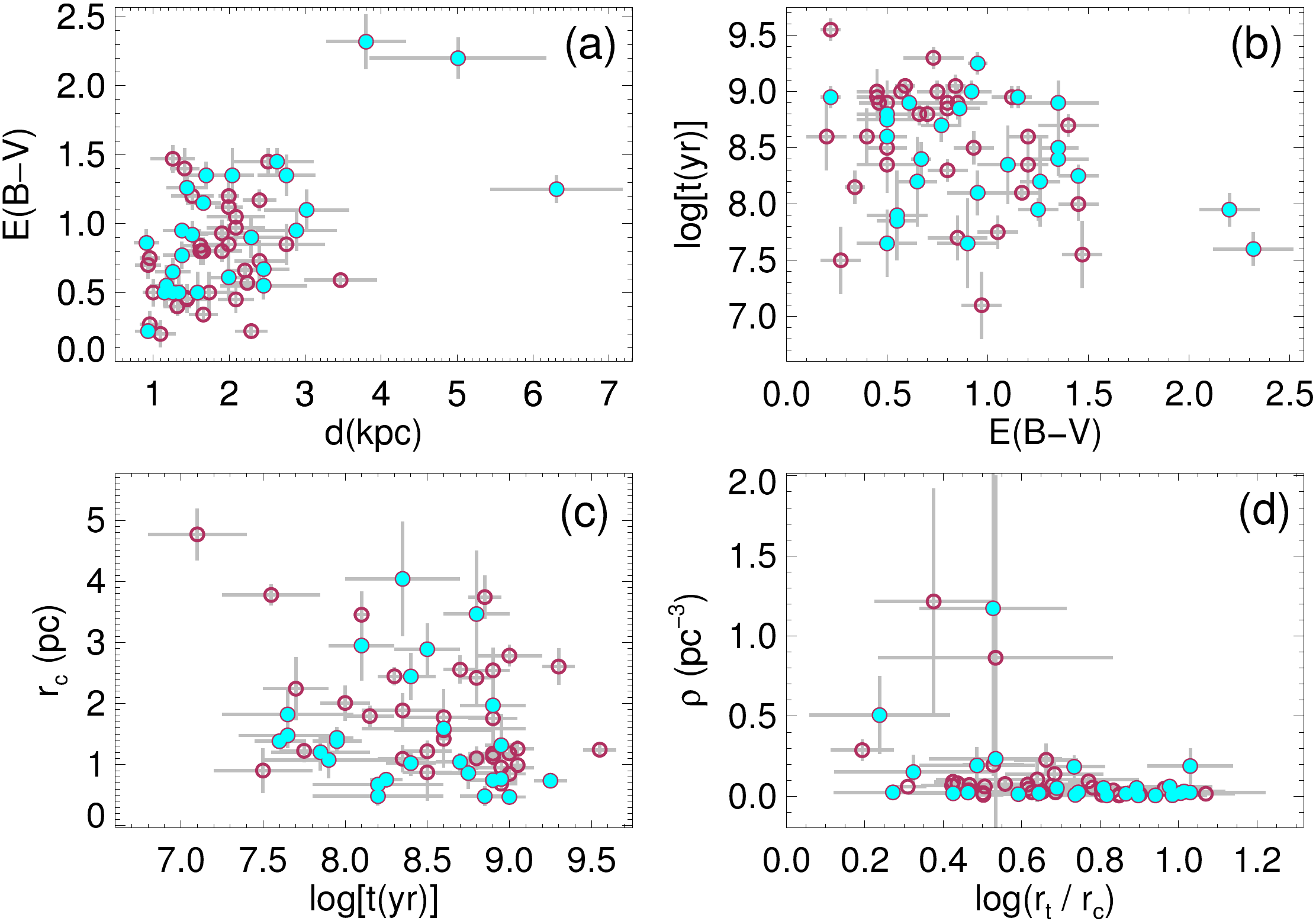}
\caption{Relations between parameters derived for the new OCs (filled cyan circles) and recently discovered OCs (open red circles). (a)~Reddening versus distance; (b) logarithmic age versus reddening; (c)~core radius versus logarithmic age; (d) average stellar density versus concentration parameter. }
\label{fig:new}
\end{figure}

Fig.~\ref{fig:new} displays the distribution of the derived parameters for the 25 newly discovered OCs (filled cyan circles) compared with the recently discovered ones (open red circles). Panel (a) shows the reddening as a function of the heliocentric distance; panel (b) is the distribution of logarithmic age with the reddening; panel (c) presents the core radius versus the logarithmic age; and panel (d) shows the average stellar density of members within the tidal radius as a function of the concentration parameter.  The properties of these two OC groups are similar, corresponding to low concentrated OCs, the majority of them located within 3\,kpc from the Sun, with ages ranging from 30\,Myr to 3.2\,Gyr and reddening limited to $E(B-V)=2.5$.
  
\subsection{The cluster sample from the Milky Way perspective}

\begin{figure}
\includegraphics[width=0.99\linewidth]{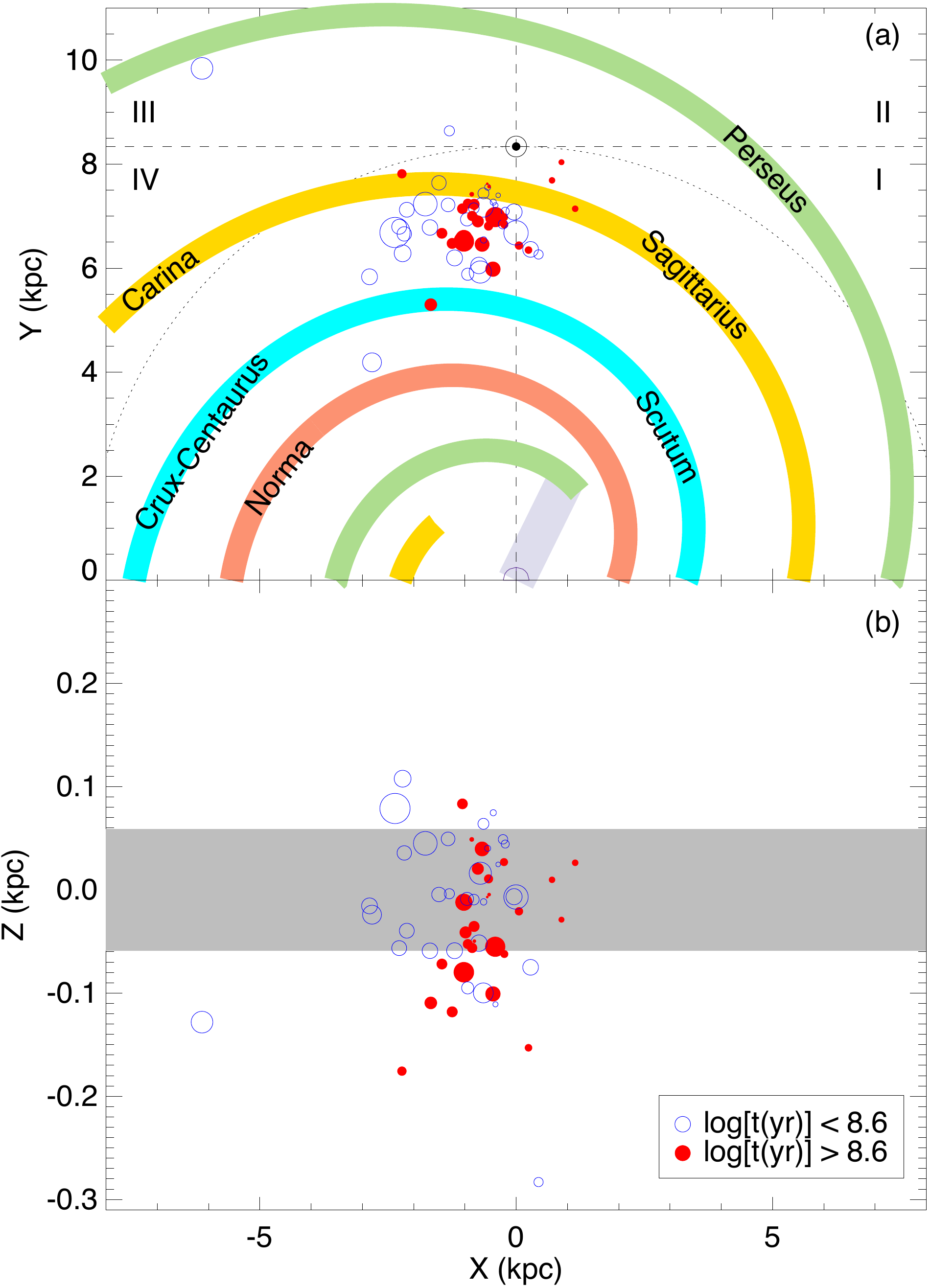}
\caption{(a) Location of the studied OCs in the Galactic plane where a scheme of the spiral arms is represented. The sun at 8.34\,kpc, the Galactic centre and bar are also sketched. The four quadrants are separated by dashed lines. Open blue circles correspond to OCs younger than 400\,Myr and filled red circles to OCs older than 400\,Myr. Symbol sizes represent the OCs' tidal radii. (b) Vertical distribution of the OCs. The gray band represents the disc vertical scale-height.} 
\label{fig:loc}
\end{figure}

Most of the OCs analysed in this work are located in the IV quadrant of the Galactic disc at low latitudes, therefore they are immersed in dense fields characteristic of the inner Milky Way. Fig.~\ref{fig:loc} shows the position of our sample projected onto the Galactic disc (panel a) and perpendicularly to it (panel b). The OCs are depicted in two age groups, below and above 400\,Myr, indicated by open blue circles and filled red circles, respectively. The OCs tidal radii are represented by the symbol size. A schematic drawing of the spiral arms \citep{v17}, the four quadrant limits, the Galactic centre, the Sun's location \citep[at 8.34\,kpc;][]{rmb14} and the central bar are shown in Fig.~\ref{fig:loc}(a), while the OCs' vertical distribution  and scale-height of 60\,pc \citep[gray band;][]{bkb06} are indicated in Fig.~\ref{fig:loc}(b).

The challenging task of separating fiducial members from field stars in a high density region of the Galactic disc was only possible due to the high precision of \textit{Gaia} DR2 astrometric data and to the developed method optimised for cluster detection. In this method, as explained in Sect.~\ref{sect:search}, we employed a sequence of steps in the search of stars within delimited fields whose astrometric and photometric coherence are compatible with a star cluster. It is worth noticing that specialized automated searches with different algorithms using \textit{Gaia} DR2 (\citealp{cjl18,cjl19,cjv18,cks19};LP2020;\citealp{ca20};CG2020) are useful and necessary tools to detect such objects. Nevertheless, improvements are still needed for the detection of low concentration systems immersed in crowded fields. More akin to our technique, Sim2019 performed visual inspection of the \textit{Gaia} DR2 astrometric and photometric data, resulting in the  discovery of 207 new OCs. The 25 new entries reported in the present analysis amounts to $\sim$\,45 percent of our sample. Based on this, (we speculate that) a non-negligible fraction of OCs may have been missed by automatic detection algorithms.

\begin{figure*}
\includegraphics[width=0.99\linewidth]{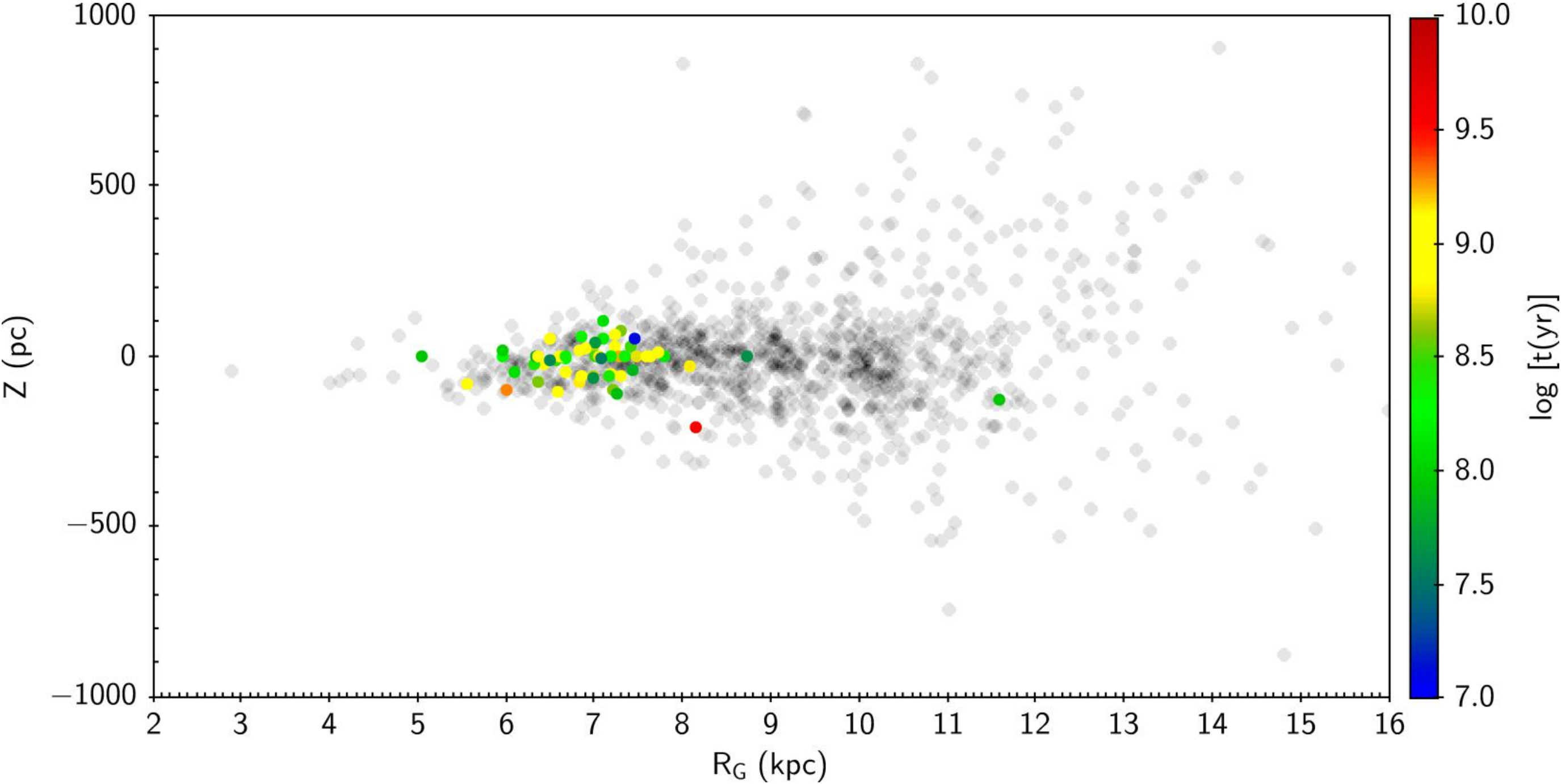}
\caption{Vertical distance from the Galactic plane versus Galactocentric distance for our OC sample (coloured circles) compared with Kharchenko's catalogue (gray circles) for confirmed clusters. The OC ages are indicated by the colourbar. }
\label{fig:ZRg}
\end{figure*}

Most of the studied OCs in the present work are inside the solar circle and have ages between 30\,Myr and 3.2\,Gyr (average and dispersion of $\log[t(yr)]=8.49\pm0.52$), as can be inferred from Fig.~\ref{fig:ZRg}, which presents the vertical distribution as a function of the Galactocentric distance for the \cite{kps13}'s OC catalogue matched to \textit{Gaia} DR2 confirmed OCs \citep{ca20}. In this Figure, our sample is indicated by filled circles, with colours attributed according to the colourbar. 

Among the 59 OCs in our sample, 27 are older than 500\,Myr. Among the 25 new discoveries, 10 are older than 500\,Myr. This suggests that nearly equal numbers of young and old clusters contribute to the inner disc uncompleteness. The disc flaring towards larger Galactocentric distances (see Fig.~\ref{fig:ZRg}) is a well-known feature of the Milky Way morphology, detected not only with OCs \cite[e.g.][]{bkb06}, but also using field stars \citep[e.g.][]{arr17} and H\,I \citep[e.g.][]{kd08}.

\begin{figure*}
\includegraphics[width=0.99\linewidth]{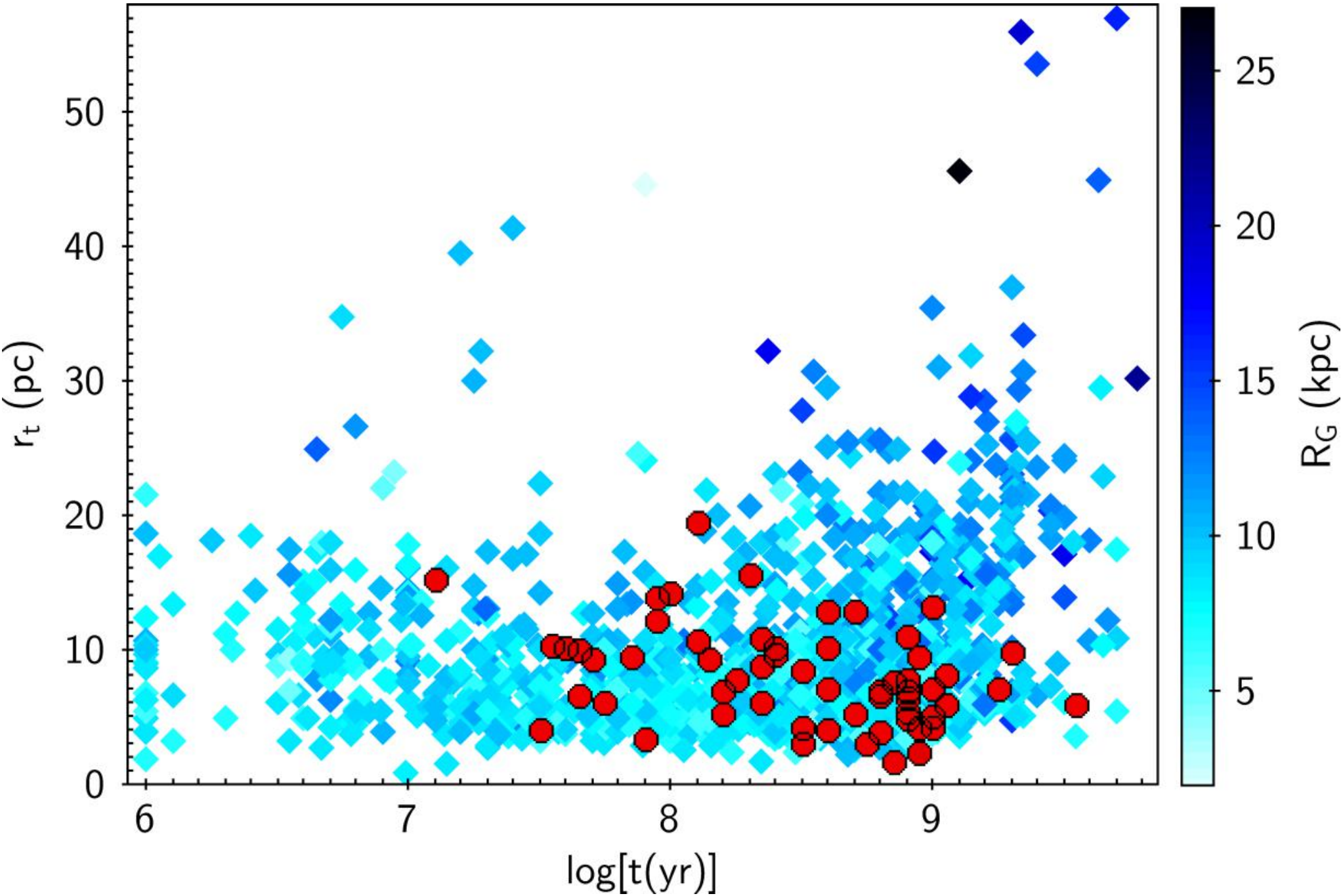}
\caption{Tidal radius versus age with Galactocentric distance indicated by the colourbar. OCs tidal radii and ages from \citep{kps13} for \textit{Gaia} DR2 confirmed clusters \citep{ca20} are represented by {\bf filled blue diamonds}, while the values determined in the present work are represented by filled {\bf red} circles. Galactocentric distances as in \citet{ca20}.}
\label{fig:rtlogt}
\end{figure*}

If surviving the early gas ejection phase, OCs change their structure with time in response to external dynamical effects combined with stellar evolution mass loss and internal gravitational interactions between the stars \citep[][and references therein]{pmg10,r18,kmb19}. These  processes lead the OCs to dissolution through  evaporation with loss of stars that in some cases produce tidal tails \citep{blg01}. External factors like the tidal Galactic field strength and OC encounters with the spiral arms \citep{mpm16}, molecular clouds \citep{s58,gpb06} and disc shocking \citep{osc72,mpp17} are more relevant in the denser environment of the inner Galactic regions. As an OC revolves around the Galaxy in $\sim$\,200\,Myr at the Galactocentric distance of Sun's orbit, its content is depleted by the aforementioned effects, which are more effective the older the OC is. For OCs orbiting closer to the Galactic centre, the external tidal field is stronger and the disrupting gravitational encounters proceed in a shorter time-scale. On the other hand, in the outer disc, OCs can expand without losing stars, surviving longer. Therefore, old clusters tend to be found in the outer disc \citep{vm80,l82,ktl88}.

Our sample follows the general distribution of tidal radius with age expected for OCs orbiting in the inner Galactic disc.  This can be seen in Fig.~\ref{fig:rtlogt}, where \textit{Gaia} DR2 confirmed OCs \citep{ca20} were matched to \cite{kps13}'s catalogue from which ages and tidal radii were extracted for 1140 OCs. The colourbar indicates the Galactocentric distance from \cite{ca20}, calculated with a heliocentric distance of $8.34$\,kpc \citep{rmb14}. Our results for the OC sample characterized in the present analysis are plotted in Fig.~\ref{fig:rtlogt} as filled circles. 

Being inserted in the inner Galaxy under relevant Galactic tidal field stresses and higher probability of encounters with molecular clouds and spiral arms, our sample contains mostly moderate size OCs, with $r_t=7.9\pm3.7$\,pc, compared to the following overall averages and dispersion for the confirmed sample of OCs: $r_t=8.3\pm4.9$\,pc for $R_G\leq 9.5$\,kpc (663 OCs) and $r_t=13.0\pm7.9$\,pc for $R_G>9.5$\,kpc (477 OCs). The interplay of tidal radius, age and galactocentric distance reveals the connection between OC structure evolution and the Galactic environment where they live. On the observational side, few works were dedicated so far to this aspect of the OC population in the Milky Way  \citep[e.g.][]{nsp02,ccs04,asc20}. 

\begin{figure}
\includegraphics[width=0.99\linewidth]{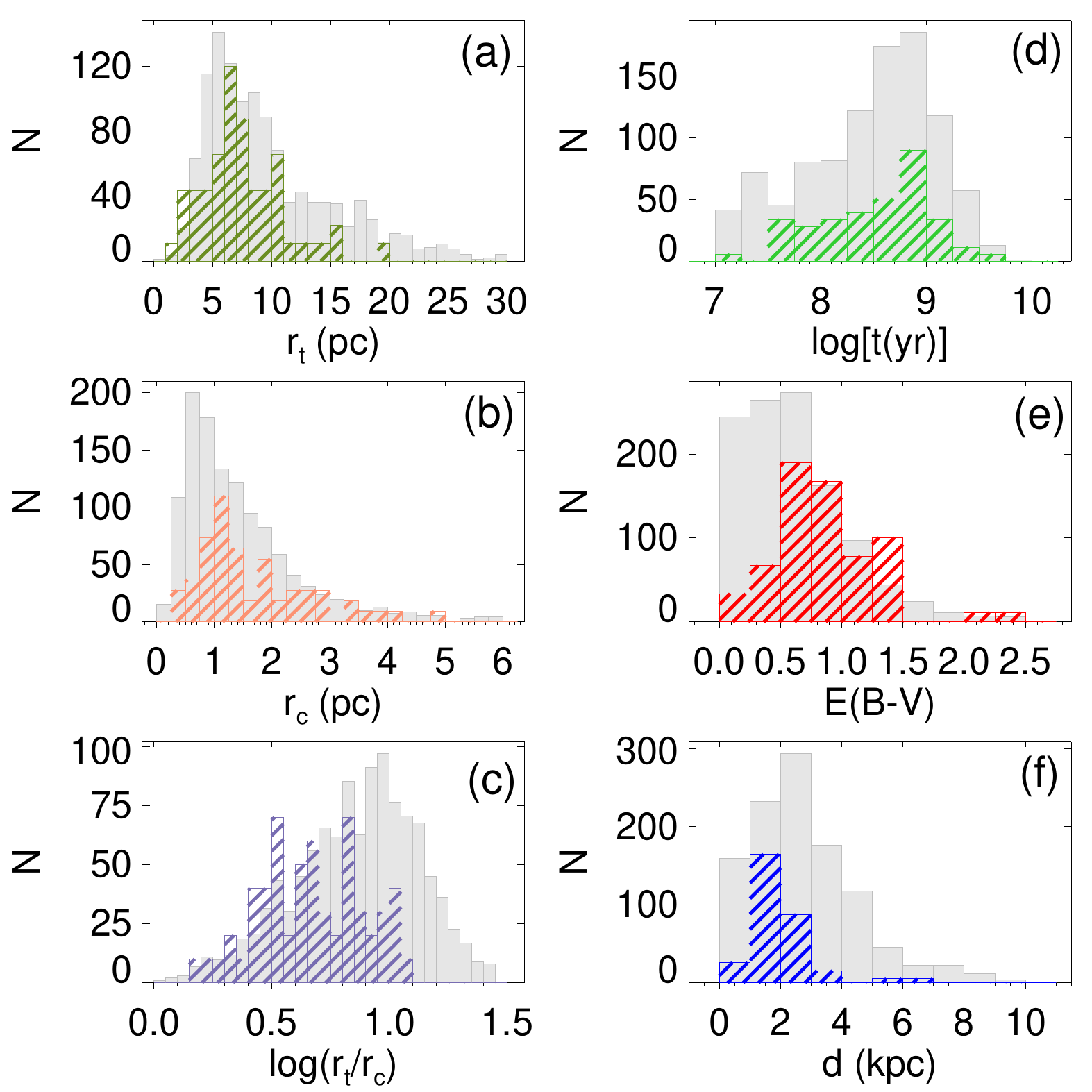}
\caption{Frequency distribution of OCs for parameters derived in the present study compared to those of \textit{Gaia} DR2 confirmed OCs in the Kharchenko's catalogue: (a) tidal radius, (b) core radius, (c) concentration parameter, (d) logarithm of age, (e) reddening and (f) heliocentric distance. For better visualization, the OC numbers per bin of our sample have been multiplied by 10, except for the age (panel d) and distance (f) histograms where the multiplying factor was 5.}
\label{fig:param}
\end{figure}

Fig.~\ref{fig:param} shows the distribution of derived parameters for our sample OCs compared with that for \textit{Gaia} DR2 confirmed OCs. The literature structural parameters, age and reddening were extracted from \cite{kps13}'s catalogue while the distance was obtained from \cite{ca20}. From a qualitative analysis of the structural parameter distributions, as shown in Fig.~\ref{fig:param}(a,b,c), our sample contains OCs with the mode of the core radii larger than that for the literature sample. Given that the distribution of tidal radii is not significantly different of that for the literature confirmed OCs, our sample OCs tend to be less concentrated. The ages present similar distributions, while most of our sample OCs have larger reddenings and smaller distances, as can be seen in Fig.~\ref{fig:param}(d,e,f), reflecting their low Galactic latitudes, where dust is concentrated and the stellar field is crowded. With many of the recently discovered clusters being closer than $\sim 2$\,kpc, it seems that their number is far from complete as discussed by \cite{cjv18} and \cite{ca20}.

\section{Summary and concluding remarks}

We derived astrophysical and structural parameters for 59 open clusters projected in high stellar density regions by using \textit{Gaia} DR2 data, where 25 objects were not reported in the literature yet. The high precision of \textit{Gaia} DR2 astrometric and photometric data allowed us to find the open clusters in the astrometric space by restricting samples by colour and magnitude. 

We were capable of disentangling fiducial members from field stars in high density regions of the Galactic disc. We analyzed the clusters'   stellar distributions and determined their centres, limiting radii and structural parameters. With the addition of \textit{Gaia} photometric information, we were able to discriminate cluster members and perform isochrone fitting to derive their astrophysical parameters. Our analysis showed that the sample of OCs investigated in this work lies in the same loci of the bulk of confirmed OCs in different parameters.

The investigated clusters in this work are mainly located within 3 kpc from the Sun in the IV quadrant of the Galactic disc at low latitudes. Their ages are comprised between $\sim 30$ Myr and $\sim 3$ Gyr, exhibiting a similar distribution presented by the \textit{Gaia} DR2 confirmed open clusters. The structural parameters derived by means of King-profile fittings revealed that the average core radii of the investigated sample are larger than those of the literature sample. In the same way, the average tidal radii are not significantly different than those of the literature, showing that our clusters have less concentrated structures.

Our optimised method for the detection of star clusters was effective to find less concentrated objects. This showed that improvements are still needed for the detection of low concentration systems immersed in crowded fields by specialized automated searches.

\section*{Acknowledgements}

The authors wish to thank the Brazilian financial agencies FAPEMIG, CNPq and CAPES (finance code 001). W. Corradi wishes to thank the LNA staff. 
We thank the referee for the comments that helped us to improve the paper.
This research has made use of the VizieR catalogue access tool, CDS, Strasbourg, France. This work has made use of data from the European Space Agency (ESA) mission \textit{Gaia} (\url{https://www.cosmos.esa.int/gaia}), processed by the \textit{Gaia} Data Processing and Analysis Consortium (DPAC, \url{https://www.cosmos.esa.int/web/gaia/dpac/consortium}). Funding for the DPAC has been provided by national institutions, in particular the institutions participating in the \textit{Gaia} Multilateral Agreement. This  research  has  made  use  of  TOPCAT \citep{Taylor:2005}.



\bibliographystyle{mnras}
\bibliography{references} 



%
%

\appendix

\section{Surveyed fields}

This appendix lists the fields surveyed in this work in the search for new cluster candidates. The 96 investigated fields are listed in Table \ref{tab:survey_fields}.

\begin{table}
\caption{Surveyed fields.}
\label{tab:survey_fields}
\begin{tabular}{|r|r|l|l|r|r|}
\hline
  \multicolumn{1}{|c|}{Field} &
  \multicolumn{1}{c|}{UFMGs} &
  \multicolumn{1}{c|}{RA} &
  \multicolumn{1}{c|}{DEC} &
  \multicolumn{1}{c|}{l} &
  \multicolumn{1}{c|}{b} \\
    \multicolumn{1}{|c|}{ } &
  \multicolumn{1}{c|}{ } &
  \multicolumn{1}{c|}{($^h$ : \arcmin : \arcsec)} &
  \multicolumn{1}{c|}{($^\circ$ : \arcmin : \arcsec)} &
  \multicolumn{1}{c|}{(deg)} &
  \multicolumn{1}{c|}{(deg)} \\
  \hline
  1 & 9,31 & 17:50:25 & -28:43:39 & 0.722 & -0.792\\
  2 &  & 17:57:11 & -24:55:40 & 4.762 & -0.168\\
  3 & 46 & 18:16:27 & -25:42:11 & 6.204 & -4.35\\
  4 & 4 & 18:15:52 & -22:11:50 & 9.236 & -2.575\\
  5 &  & 17:57:10 & -18:55:00 & 9.966 & 2.845\\
  6 & 33 & 18:38:31 & -22:17:14 & 11.572 & -7.27\\
  7 &  & 18:24:42 & -16:57:56 & 14.836 & -1.95\\
  8 & 27 & 19:01:48 & +09:12:27 & 42.304 & 1.905\\
  9 & 26 & 19:20:35 & +12:05:51 & 47.002 & -0.849\\
  10 & 25 & 20:18:02 & +32:35:51 & 71.341 & -1.736\\
  11 &  & 05:52:34 & +32:32:17 & 177.678 & 3.133\\
  12 &  & 07:44:31 & -23:53:27 & 240.07 & 0.121\\
  13 & 58, 59 & 08:21:38 & -38:25:59 & 256.585 & -0.921\\
  14 & 62 & 09:56:01 & -60:11:44 & 282.723 & -4.388\\
  15 &  & 11:42:07 & -58:45:24 & 294.034 & 2.912\\
  16 & 10 & 11:36:32 & -61:36:26 & 294.151 & -0.019\\
  17 &  & 12:03:13 & -60:31:09 & 297.031 & 1.799\\
  18 &  & 12:24:45 & -58:11:34 & 299.411 & 4.496\\
  19 &  & 12:25:00 & -60:25:36 & 299.673 & 2.278\\
  20 & 61, 56 & 12:29:33 & -63:02:52 & 300.455 & -0.281\\
  21 &  & 12:39:02 & -68:35:24 & 301.796 & -5.746\\
  22 &  & 12:41:03 & -64:42:59 & 301.824 & -1.867\\
  23 &  & 12:43:42 & -59:46:02 & 301.957 & 3.09\\
  24 & 53 & 12:58:19 & -63:16:30 & 303.707 & -0.414\\
  25 & 55 & 13:02:10 & -60:30:10 & 304.255 & 2.341\\
  26 &  & 13:16:20 & -66:10:56 & 305.446 & -3.432\\
  27 &  & 13:21:05 & -69:59:58 & 305.482 & -7.275\\
  28 & 50 & 13:24:57 & -62:40:56 & 306.768 & -0.061\\
  29 &  & 13:36:26 & -58:47:21 & 308.745 & 3.573\\
  30 &  & 14:00:01 & -68:20:17 & 309.218 & -6.308\\
  31 &  & 13:56:32 & -65:14:55 & 309.681 & -3.234\\
  32 & 49, 54 & 13:51:34 & -61:30:34 & 310.04 & 0.53\\
  33 &  & 13:54:45 & -59:09:04 & 310.982 & 2.727\\
  34 & 60 & 14:17:36 & -62:46:53 & 312.604 & -1.539\\
  35 &  & 14:05:18 & -57:19:39 & 312.803 & 4.121\\
  36 & 48 & 14:15:17 & -59:58:36 & 313.246 & 1.205\\
  37 &  & 14:36:24 & -60:34:31 & 315.479 & -0.286\\
  38 &  & 14:46:57 & -63:01:52 & 315.602 & -3.037\\
  39 & 17, 39 & 14:41:47 & -57:19:34 & 317.423 & 2.408\\
  40 &  & 15:03:00 & -61:44:43 & 317.846 & -2.745\\
  41 &  & 14:56:06 & -59:53:01 & 317.984 & -0.706\\
  42 &  & 15:27:29 & -62:04:44 & 320.131 & -4.537\\
  43 & 13, 19 & 15:13:24 & -59:22:54 & 320.136 & -1.338\\
  44 &  & 15:05:52 & -57:07:50 & 320.418 & 1.105\\
  45 &  & 15:04:35 & -54:22:58 & 321.61 & 3.588\\
  46 &  & 15:30:26 & -59:08:56 & 322.09 & -2.325\\
  47 &  & 15:54:48 & -62:28:43 & 322.434 & -6.771\\
    48 & 34, 37 & 15:27:47 & -56:46:42 & 323.141 & -0.171\\
   \hline\end{tabular}
\end{table}

\begin{table}
\contcaption{}
\label{tab:Continued}
\begin{tabular}{|r|r|l|l|r|r|}
\hline
  \multicolumn{1}{|c|}{Field} &
  \multicolumn{1}{c|}{UFMGs} &
  \multicolumn{1}{c|}{RA} &
  \multicolumn{1}{c|}{DEC} &
  \multicolumn{1}{c|}{l} &
  \multicolumn{1}{c|}{b} \\
      \multicolumn{1}{|c|}{ } &
  \multicolumn{1}{c|}{ } &
  \multicolumn{1}{c|}{($^h$ : \arcmin : \arcsec)} &
  \multicolumn{1}{c|}{($^\circ$ : \arcmin : \arcsec)} &
  \multicolumn{1}{c|}{(deg)} &
  \multicolumn{1}{c|}{(deg)} \\
\hline 
  49 & 35, 36 & 15:22:40 & -54:29:05 & 323.809 & 2.138\\
  50 & 40 & 15:54:39 & -58:41:47 & 324.855 & -3.856\\
  51 & 32, 41 & 15:42:15 & -56:32:24 & 324.896 & -1.136\\
  52 &  & 16:21:05 & -60:59:39 & 325.752 & -7.791\\
  53 & 44, 43, 42 & 15:40:44 & -54:49:51 & 325.753 & 0.358\\
  54 &  & 15:30:11 & -51:50:23 & 326.223 & 3.705\\
  55 &  & 15:39:17 & -53:25:03 & 326.426 & 1.619\\
  56 &  & 15:28:25 & -50:01:03 & 327.029 & 5.364\\
  57 & 45 & 16:08:57 & -56:11:41 & 327.956 & -3.241\\
  58 &  & 15:41:21 & -51:17:10 & 327.956 & 3.137\\
  59 & 38 & 15:57:49 & -53:32:37 & 328.509 & -0.193\\
  60 &  & 15:40:04 & -47:54:10 & 329.829 & 5.966\\
  61 &  & 15:28:25 & -45:00:29 & 329.901 & 9.486\\
  62 & 6 & 16:14:49 & -53:49:14 & 330.195 & -2.083\\
  63 &  & 15:53:52 & -49:49:21 & 330.428 & 3.053\\
  64 & 12 & 16:27:38 & -53:55:20 & 331.462 & -3.49\\
  65 & 7,11,18,51 & 16:15:20 & -51:15:38 & 332.02 & -0.286\\
  66 &  & 15:54:41 & -46:11:14 & 332.862 & 5.763\\
  67 &  & 16:36:44 & -51:59:08 & 333.816 & -3.172\\
  68 &  & 16:13:27 & -47:53:34 & 334.121 & 2.364\\
  69 &  & 15:51:33 & -41:38:49 & 335.357 & 9.613\\
  70 &  & 16:59:41 & -52:43:41 & 335.501 & -6.336\\
  71 &  & 16:11:26 & -45:23:34 & 335.587 & 4.421\\
  72 &  & 16:31:40 & -48:48:01 & 335.613 & -0.42\\
  73 &  & 16:24:45 & -46:44:16 & 336.298 & 1.845\\
  74 &  & 16:09:11 & -43:25:23 & 336.639 & 6.136\\
  75 &  & 16:49:53 & -49:36:52 & 336.974 & -3.19\\
  76 & 16 & 16:26:12 & -44:33:57 & 338.032 & 3.179\\
  77 & 30 & 17:10:50 & -50:17:49 & 338.505 & -6.271\\
  78 &  & 17:04:29 & -48:10:42 & 339.594 & -4.175\\
  79 & 20,21,28,47 & 16:48:35 & -45:51:39 & 339.714 & -0.614\\
  80 &  & 16:15:43 & -39:51:21 & 340.009 & 7.877\\
  81 & 14 & 16:43:09 & -42:57:40 & 341.286 & 2.009\\
  82 &  & 16:34:30 & -41:06:58 & 341.595 & 4.435\\
  83 & 5, 15 & 17:03:46 & -43:41:03 & 343.097 & -1.351\\
  84 &  & 17:26:57 & -46:32:17 & 343.144 & -6.337\\
  85 &  & 16:26:30 & -36:57:41 & 343.589 & 8.404\\
  86 & 57 & 16:56:52 & -40:33:45 & 344.767 & 1.58\\
  87 &  & 17:20:19 & -43:04:01 & 345.37 & -3.421\\
  88 &  & 17:13:34 & -39:39:17 & 347.422 & -0.413\\
  89 & 8 & 17:27:43 & -39:28:14 & 349.13 & -2.548\\
  90 & 23, 24, 29 & 17:10:31 & -36:20:27 & 349.746 & 2.025\\
  91 &  & 17:23:33 & -36:04:20 & 351.486 & 0.036\\
  92 &  & 17:13:13 & -33:16:11 & 352.557 & 3.385\\
  93 & 22 & 17:39:34 & -36:47:27 & 352.649 & -3.053\\
  94 &  & 17:54:22 & -34:40:55 & 356.012 & -4.539\\
  95 & 52 & 17:40:33 & -32:13:48 & 356.626 & -0.805\\
  96 &  & 18:23:06 & -36:34:26 & 357.1 & -10.578\\
\hline\end{tabular}
\end{table}

\section{Supplementary figures: King profile fittings}

This appendix contains the plot of King profile fitting over the cluster radial density profile for each investigated target (Figs.~B1$-$B4). It is available only in the supplementary material.


\section{Supplementary figures: membership}

This appendix contains the plot of isochrone fitting over the cluster most probable members for each investigated target (Figs.~C1$-$C12). It is available only in the supplementary material.


\section{Supplementary figures: isochrone fittings}

This appendix contains the plots of isochrone fitting over the cluster most probable members for each investigated target (Figs.~D1$-$D7). It is available only in the supplementary material. 


\bsp	
\label{lastpage}
\end{document}